\documentclass[aos]{imsart}

\RequirePackage{amsthm,amsmath,amsfonts,amssymb}
\RequirePackage[numbers]{natbib}
\RequirePackage{graphicx}
\RequirePackage{multirow}
\RequirePackage{mathrsfs}
\RequirePackage{xcolor}
\RequirePackage{textcomp}
\RequirePackage{manyfoot}
\RequirePackage{booktabs}
\RequirePackage{algpseudocode}
\RequirePackage{listings}
\RequirePackage{subfigure}
\RequirePackage{hyperref}

\startlocaldefs
\theoremstyle{plain}

\newcommand{\btheta}{{\mathbf\theta}}
\newcommand{\bDelta}{{\mathbf\Delta}}
\newcommand{\bGamma}{{\mathbf\Gamma}}
\newcommand{\bLambda}{{\mathbf\Lambda}}
\newcommand{\bSigma}{{\mathbf\Sigma}}
\newcommand{\bTheta}{{\mathbf\Theta}}

\newcommand{\bUpsilon}{{\mathbf\Upsilon}}

\newcommand{\bOmega}{{\mathbf\Omega}}

\theoremstyle{definition}

\newtheorem{theorem}{Theorem}

\newtheorem{assumption}{Assumption}
\newtheorem{lemma}{Lemma}

\newtheorem{remark}{Remark}
\newtheorem{corollary}{Corollary}

\def\eop{\hfill $\Box$}
\def\tr{\mbox{tr}}
\def\Diag{\mbox{Diag}}

\def\var{\mbox{var}}
\def\A{{\mathbf A}}
\def\E{{\mathbf E}}
\def\I{{\mathbf I}}
\def\U{{\mathbf U}}

\def\P{{\mathbf P}}
\def\M{{\mathbf M}}
\def\B{{\mathbf B}}
\def\g{{\mathbf g}}

\endlocaldefs

\begin{document}

\begin{frontmatter}
\title{From Simple to Composite Perturbations: A Unified Decomposition Framework for Stochastic Block Models}
\runtitle{A Unified Decomposition Framework}

\begin{aug}
\author[A]{\fnms{Jianwei}~\snm{Hu}\ead[label=e1]{jwhu@ccnu.edu.cn}}
,
\author[A]{\fnms{Ding}~\snm{Chen}\ead[label=e2]{chend@mails.ccnu.edu.cn}}
\and
\author[B]{\fnms{Ji}~\snm{Zhu}\ead[label=e3]{jizhu@umich.edu}}
\address[A]{School of Mathematics and Statistics, and Key Lab NAA--MOE, Central China Normal University, Wuhan, China \printead[presep={,\ }]{e1,e2}}

\address[B]{Department of Statistics, University of Michigan, Ann Arbor, MI, USA \printead[presep={,\ }]{e3}}
\end{aug}

\begin{abstract}
Statistical inference for stochastic block models typically relies on the spectrum of the normalized adjacency matrix $\A^*$. In practice, the true probability matrix $\mathbf{B}$ is unknown and must be replaced by a plug-in estimator $\hat{\mathbf{B}}$. This substitution introduces two distinct types of estimation error: a simple perturbation $\boldsymbol{\Delta}$, arising when $\hat{\mathbf{B}}$ replaces $\mathbf{B}$ only in the numerator, and a composite perturbation $\tilde{\boldsymbol{\Delta}}$, arising when the replacement occurs in both the numerator and the denominator.

Under both perturbation regimes, we decompose the total sum of squares into three components  and conduct a detailed analysis of their asymptotic properties. This reveals a key, and perhaps surprising, distinction between simple and composite perturbations: the cross term $\tr({\A^*}\bDelta)$ is asymptotically negligible, whereas its composite counterpart $\tr({\A^*}\tilde{\bDelta})$  is not.

 Motivated by this, we develop a unified decomposition framework, expressing the composite perturbation matrix as $\tilde{\bDelta}=\check{\A}+\bDelta+\check{\bDelta}$, where $\check{\A}$ is a bias matrix of the normalized adjacency matrix, $\bDelta$ is the simple perturbation, and $\check{\bDelta}$ is a bias matrix of $\bDelta$. This structured decomposition allows us to precisely isolate and control each source of error, leading to a refined limiting theory for two key classes of test statistics.

Concretely, for the largest  eigenvalue statistic,  we improve the existing condition from $K=O(n^{1/6-\tau})$ to the optimal rate $K=o(n^{1/6})$ under both simple and composite perturbations, with an additional community balance condition required in the composite case to control the full-rank bias matrix $\check{\A}$ introduced by the scaling factor. For the  linear spectral statistic, a major technical challenge lies in proving that all error terms induced by the perturbation, whether simple or composite, are asymptotically negligible. Our unified decomposition framework provides the necessary structure to systematically control these errors term by term, leading to a complete and rigorous proof of asymptotic normality, thereby  justifying the practical use of the linear spectral statistic when $\mathbf{B}$ is estimated.

The proposed framework provides a novel tool for analyzing plug-in estimation errors in network models, thereby establishing sharper theoretical guarantees with practical relevance. Its structured approach is readily extensible to broader inference problems like two-sample testing and directed networks, and serves as a potential principle for handling parameter estimation errors in other latent variable models and high-dimensional covariance settings that rely on normalized matrices.
\end{abstract}



\end{frontmatter}


\section{Introduction}
\label{Introduction}
Network data analysis has a wide range of applications. Networks can be divided into different communities based on specific characteristics, which helps in understanding network structures. The stochastic block model, proposed by Holland et al. \cite{Holland:1983}, is one of the most foundational and well-studied models for community structure in networks. Community detection is a central problem in the study of stochastic block models. Many popular community detection methods involve spectral clustering based on an adjacency matrix or a Laplacian matrix \cite{Amini:2013, Gao:2018, Jin:2015, Rohe:2011} and optimizing a likelihood function or its variants \cite{Amini:2013, Bickel:2009, Gao:2017, Wang:2023, Zhao:2012}.  Yet, determining the number of communities is a fundamental problem that precedes it. There are many methods to estimate the number of communities, including likelihood-based methods for model selection \cite{Hu:2020, Saldana:2017, Wang:2017},  network cross-validation methods \cite{Chen:2018, Li:2020}, and spectral methods based on the Hessian matrix \cite{Hwang:2023, Le:2022}.

In recent years, hypothesis testing has emerged as a competitive alternative framework for both community detection and estimating the number of communities in stochastic block models. For example, recursive bipartitioning algorithms for community detection \cite{Bickel:2016, Dong:2020} and sequential testing procedures for determining the number of communities \cite{Lei:2016, Wu:2024} are grounded in hypothesis testing. More broadly, hypothesis testing problems of network data can be categorized into two primary types: one-sample tests, which assess whether an observed network is derived from a specific network model, and two-sample tests, which evaluate whether two networks are generated from the same population. For one-sample settings, a variety of methods are available for hypothesis testing in network models, including the largest  eigenvalue statistics \cite{Bickel:2016, Lei:2016, Zhu:2025}, linear spectral statistics \cite{Dong:2020, Wu:2024}, maximum entry-wise deviation statistics \cite{Hu:2021, Wu:2025}, cycle count statistics \cite{Jin:2025b}. Analogous methods exist for two-sample comparisons \cite{Chen:2024, Chen:2023,  Fu:2025a, Fu:2025b, Ghoshdastidar:2020, Jin:2025a, Wu:2024a, Wu:2024b}.

The theoretical foundation for many spectral-based hypothesis tests is rooted in random matrix theory, particularly the limiting laws for Wigner-type matrices. For standard Wigner matrices, the Tracy-Widom law governing the largest eigenvalue is classical \cite{Tracy:1994}, and the asymptotic normality of linear spectral statistics is well-established \cite{Bai:2010, Bai:2005}. Importantly, the normalized adjacency matrix of a stochastic block model with known community assignments and probability matrix $\mathbf{B}$ falls into the class of generalized Wigner matrices. Consequently, the limiting distributions for the key test statistics are fully characterized in this oracle setting: the largest eigenvalue follows the Tracy-Widom law \cite{Lee:2014}, and linear spectral statistics are asymptotically normal \cite{Wang:2021}. This provides a sound theoretical basis for hypothesis testing when the model parameters are perfectly known.

However, in practice, the true probability matrix $\mathbf{B}$ is unknown and must be estimated using a plug-in estimator $\hat{\mathbf{B}}$. This substitution gives rise to two distinct types of perturbation in the plug-in normalized adjacency matrix: simple and composite. The simple perturbation matrix $\bDelta$, resulting from using the plug-in only in the numerator, has a low-rank structure. In contrast, the composite perturbation matrix $\tilde{\bDelta}$, arising from using the plug-in in both the numerator and the denominator, generally exhibits a full-rank structure. The four main points of our analysis regarding these two perturbation regimes are summarized as follows.

As a first focus of this paper, we aim to analyze the total sum of squares under both simple and composite perturbations. This can be divided into three terms: the sum of squares due to the normalized adjacency matrix, the sum of the cross term due to the normalized adjacency matrix and the perturbation matrix, and the sum of squares due to the perturbation matrix. We conduct a detailed analysis of their asymptotic properties under these two perturbation regimes. This analysis reveals their key differences: while the cross term $\tr({\A^*}\bDelta)$ is asymptotically negligible, its composite counterpart $\tr({\A^*}\tilde{\bDelta})$  is not. Furthermore, although the sum-of-squares terms $\tr(\bDelta^2)$ and $\tr(\tilde{\bDelta}^2)$ are of the same order, they converge to different limiting distributions. These results provide motivation and a theoretical foundation for the unified decomposition framework and its applications in this paper.

As a second focus of this paper, we develop the unified decomposition framework for the composite perturbation matrix. The composite perturbation matrix $\tilde{\bDelta}$ can be decomposed into three parts: a bias matrix $\check{\A}$ of the normalized adjacency matrix, the simple perturbation matrix $\bDelta$, and a bias matrix $\check{\bDelta}$ of $\bDelta$. Formally, this decomposition is expressed as: $\tilde{\bDelta}=\check{\A}+\bDelta+\check{\bDelta}$. This structured decomposition is key to analyzing cross terms involving the normalized adjacency matrix and the perturbation matrix. It reveals why the cross term  $\tr({\A^*}\tilde{\bDelta})$ is asymptotically non-negligible: the scaling factor introduces a full-rank bias matrix $\check{\A}$ that couples with $\A^*$, producing the dominant component $\tr({\A^*}\check{\A})$, which converges to a normal distribution.

As a third focus of this paper, we study the properties of the largest  eigenvalue statistic in stochastic block models.  The seminal work of Lei \cite{Lei:2016} established the Tracy-Widom limit for the largest eigenvalue statistic at the rate of $K=O(n^{1/6-\tau})$. By deriving sharper bounds on the spectral norm of the simple perturbation matrix, we improve this condition to the optimal rate $K=o(n^{1/6})$. We further apply our decomposition framework to the composite perturbation case. To reach the same optimal rate $K=o(n^{1/6})$, an additional community balance condition (Assumption \ref{assumption:0}) is required to control the full-rank bias matrix $\check{\A}$ introduced by the scaling factor. This highlights a key structural distinction between simple and composite perturbations.

As a fourth focus of this paper, we systematically study the properties of the linear spectral statistic proposed by Wu and Hu\footnote{Jiang Hu (Northeast Normal University) is a different researcher from the co-author Jianwei Hu of the present paper.} \cite{Wu:2024}. Their insightful work introduced this powerful statistic and laid the theoretical groundwork for the oracle setting with known $\mathbf{B}$. However, a complete and rigorous theoretical guarantee accommodating the plug-in estimator $\hat{\mathbf{B}}$ was still lacking. The main technical challenge lies in systematically controlling all error terms induced by the perturbation matrix. Our unified decomposition framework directly addresses this challenge, enabling a transparent, term-by-term analysis that conclusively shows  that all perturbation-induced errors are asymptotically negligible. This not only clarifies the underlying mechanism through which the perturbation impacts the linear spectral statistic but also extends and solidifies the theoretical basis of \cite{Wu:2024}, thereby establishing a more general and robust asymptotic theory for plug-in inference in stochastic block models.

\textbf{Main contributions.} In summary, our work makes four key contributions. First, we reveal key differences between simple and composite perturbations: while the cross term $\tr({\A^*}\bDelta)$ is asymptotically negligible, its composite counterpart $\tr({\A^*}\tilde{\bDelta})$  is not, and their sum-of-squares terms $\tr(\bDelta^2)$ and $\tr(\tilde{\bDelta}^2)$, despite being of the same order, converge to different limiting distributions.  Second, we develop a unified decomposition framework for the composite perturbation matrix, which reveals the mechanism underlying its non-negligible cross term. Third, we establish the optimal rate of $K=o(n^{1/6})$ for the largest eigenvalue statistic to converge to the Tracy-Widom limit  under simple perturbations, with an additional community balance condition required under composite perturbations to control the full-rank bias term $\check{\A}$ and achieve the same rate. Fourth, we provide a rigorous proof of the asymptotic normality for the linear spectral statistic, addressing the key challenge of systematically controlling all error terms induced by the perturbation matrix.

The remainder of the paper is organized as follows. Sections \ref{Simple} and \ref{Composite} decompose the total sum of squares into three components under the simple and composite perturbation regimes, respectively, and study their theoretical properties. Building on these results, a unified decomposition framework  is proposed in Section \ref{Decomposition}. In Section \ref{Applications}, this framework is applied to the largest  eigenvalue statistic and the linear spectral statistic.
Simulation studies are presented in Section \ref{Simulation}, followed by further discussion in Section \ref{Discussion}. All proofs are relegated to the Appendix.

\section{Simple Perturbation  Matrix}
\label{Simple}
In this section, we first introduce the stochastic block model, then formally define the simple perturbation matrix and examine its theoretical properties.

 Let $\A \in \{0,1\}^{n\times n}$ be the adjacency matrix  of an undirected network with no self-loops. Consequently, $\A$ is symmetric, and its diagonal elements are all zero. For all pairs $i>j$, the matrix elements $A_{ij}$ are generated independently from a Bernoulli distribution with probability $P_{ij}$. Here, the parameter $P_{ij}$ is the corresponding entry of the probability matrix $\P \in \mathbb{R}^{n\times n}$. Consider the stochastic block model with $K$ communities, where each node $i$ is associated with a community label $g_i\in [K]$. The  probability matrix $P_{ij}$ depends only on the community labels of nodes $i$ and $j$. That is, $P_{ij} = B_{g_ig_j}$, where $\B$ is a $K \times K$ symmetric probability matrix.

We define the normalized adjacency matrix $\A^*$ as follows:
\begin{equation}
	\label{eq:Astar}
	A_{ij}^* = \begin{cases}
		\dfrac{A_{ij}-P_{ij}}{\sqrt{nP_{ij}(1-P_{ij})}}, \quad & i \neq j,\\
		0, \quad & i = j.
	\end{cases}
\end{equation}

Let $\hat{\g}$ be an estimated community label vector with $K$ communities.  Throughout this paper, we assume that $\hat{\g}$ is strongly consistent, that is, $\mathbb{P}(\hat{\g}=\g)\rightarrow 1$. The maximum likelihood estimator of $\B$ is then given by
\begin{equation*}
	\widehat{B}_{uv} = \begin{cases}
		\dfrac{\sum_{i\in\mathcal{N}_{u}, j\in\mathcal{N}_{v}}A_{ij}}{n_{uv}}, \quad & u \neq v,\\
		\dfrac{\sum_{i,j\in\mathcal{N}_{u},i<j}A_{ij}}{n_{uu}}, \quad & u = v.	
	\end{cases}
\end{equation*}
where $\mathcal{N}_k=\{i:1\leq i\leq n, \hat{g}_i=k\}$ with $n_k=|\mathcal{N}_k|$ for all $1\leq k\leq K$, and $n_{uv}=n_un_v$ for $u\neq v$ and $n_{uu}=n_u(n_u-1)/2$ for $u=v$, respectively.

To establish the theoretical results, we make the following assumptions.
\begin{assumption}\label{assumption:1}
	The entries of $\B$ are uniformly bounded away from 0 and 1, and $\B$ has no identical rows.
\end{assumption}

\begin{assumption}\label{assumption:2}
	There exists a constant $c_1$, such that
	\[
	\min_{1\leq k\leq K}n_{k}\geq \frac{c_1n}{K}.
	\]
\end{assumption}

Assumption \ref{assumption:1} ensures that $\B$ is identifiable, and Assumption \ref{assumption:2} requires the size of the smallest community is at least proportional to $n/K$. Such assumptions are standard in relevant literature, including \cite{Lei:2016, Wang:2017}.

\subsection{Definition}
\label{DefinitionSimple}
We define $\bar{\A}$ as follows:
\begin{equation}\label{eq:Abar}
	\bar{A}_{ij} = \begin{cases}
		\dfrac{A_{ij}-\hat{P}_{ij}}{\sqrt{nP_{ij}(1-P_{ij})}}, \quad & i \neq j,\\
		0, \quad & i = j,
	\end{cases}
\end{equation}
where $\hat{P}_{ij}=\hat{B}_{\hat{g}_{i}\hat{g}_{j}}$.

We then define the simple perturbation matrix $\bDelta$ of the normalized adjacency matrix $\A^*$ as follows:
\begin{equation}\label{eq:Delta}
	\begin{aligned}
		\Delta_{ij} = \begin{cases}
			\dfrac{P_{ij}-\hat{P}_{ij}}{\sqrt{nP_{ij}(1-P_{ij})}}, \quad & i \neq j,\\
			0, \quad & i = j.
		\end{cases}
	\end{aligned}
\end{equation}
The simple perturbation matrix $\bDelta$ arises from replacing $P_{ij}$ with $\hat{P}_{ij}$ only in the numerator of the normalized adjacency matrix.

Then, we have
\begin{equation*}
	\begin{aligned}
		\bar{A}_{ij}&={A_{ij}^*}+\Delta_{ij}.
	\end{aligned}
\end{equation*}
That is,
\[
\bar{\A}={\A^*}+\bDelta.
\]
Therefore, $\bar{\A}$ can be decomposed into two parts: the normalized adjacency matrix ${\A^*}$ and the simple perturbation matrix $\bDelta$.

To quantify the impact of the simple perturbation on the spectral properties of $\bar{\A}$, we analyze its total sum of squares $\tr(\bar{\A}^2)$, a key metric for characterizing the global spectral behavior of the matrix. Expanding the trace of the squared matrix, we obtain the following decomposition:
\begin{equation*}
	\begin{aligned}
		\tr(\bar{\A}^2)&=\tr(({\A^*}+\bDelta)^2)\\
		&=\tr({\A^*}^2)+2\tr({\A^*}\bDelta)+\tr(\bDelta^2),
	\end{aligned}
\end{equation*}
where
\begin{equation*}
	\begin{aligned}
		\tr(\bar{\A}^2)=\sum_{i,j}\bar{A}_{ij}^2=\sum_{1\leq u, v\leq K}\sum_{i\in\mathcal{N}_{u}, j\in\mathcal{N}_{v}}\bar{A}_{ij}^2,
	\end{aligned}
\end{equation*}
\begin{equation*}
	\begin{aligned}
		\tr({\A^*}^2)=\sum_{i,j}{A_{ij}^*}^2=\sum_{1\leq u, v\leq K}\sum_{i\in\mathcal{N}_{u}, j\in\mathcal{N}_{v}}{A_{ij}^*}^2,
	\end{aligned}
\end{equation*}
\begin{equation*}
	\begin{aligned}
		\tr({\A^*}\bDelta)=\sum_{i, j}{A_{ij}^*}\Delta_{ji}=\sum_{1\leq u, v\leq K}\sum_{i\in\mathcal{N}_{u}, j\in\mathcal{N}_{v}}{A_{ij}^*}\Delta_{ji},
	\end{aligned}
\end{equation*}
and
\begin{equation*}
	\begin{aligned}
		\tr(\bDelta^2)=\sum_{i, j}\Delta_{ij}^2=\sum_{1\leq u, v\leq K}\sum_{i\in\mathcal{N}_{u}, j\in\mathcal{N}_{v}}\Delta_{ij}^2.
	\end{aligned}
\end{equation*}
This decomposition separates $\tr(\bar{\A}^2)$ into three interpretable components: $\tr({\A^*}^2)$, $\tr({\A^*}\bDelta)$ and $\tr(\bDelta^2)$,  which are respectively the sum of squares due to the normalized adjacency matrix, the sum of the cross term due to the normalized adjacency matrix and the simple perturbation matrix, and the sum of squares due to the simple perturbation matrix.

\subsection{Theoretical Properties}
We now analyze the asymptotic behavior of the three components in the decomposition of $\tr(\bar{\A}^2)$ introduced in Section \ref{DefinitionSimple}.

We first consider the sum of squares due to the normalized adjacency matrix. The following lemma, due to \cite{Wang:2021}, establishes the asymptotic distribution of $\tr({\A^*}^2)$.
\begin{lemma}
	\label{lemma:2-1}
	Let $\A^*$ be given as in (\ref{eq:Astar}). It holds that
	\begin{equation*}
		\begin{aligned}
			\frac{\tr({\A^*}^2)-(n-1)}{\sqrt{2L-2}}\stackrel{d}{\longrightarrow} N(0,1),
		\end{aligned}
	\end{equation*}
	where $L=\lim_{n\rightarrow \infty}\sum_{i,j}E|{A_{ij}^*}|^4$. Here, for $i\neq j$, $E|{A_{ij}^*}|^4=\frac{P_{ij}^3+(1-P_{ij})^3}{n^2P_{ij}(1-P_{ij})}$.
\end{lemma}

We now turn to the cross term $\tr({\A^*}\bDelta)$ due to the normalized adjacency matrix and the simple perturbation matrix.

\begin{theorem}
	\label{theorem:2-2}
	Let $\A^*$ be given as in (\ref{eq:Astar}), and let $\bDelta$ be defined as in (\ref{eq:Delta}). Suppose that Assumptions \ref{assumption:1} and \ref{assumption:2} are satisfied. For fixed $K$, we have
	\begin{equation*}
		\begin{aligned}
			-\frac{1}{2}n\tr({\A^*}\bDelta)\stackrel{d}{\longrightarrow} \chi_{\frac{K(K+1)}{2}}^2,\\
		\end{aligned}
	\end{equation*}
	and for  $K\rightarrow\infty$ and  $K=o(n^{1/2})$, we have
	\begin{equation*}
		\begin{aligned}
			\frac{-\frac{1}{2}n\tr({\A^*}\bDelta)-\frac{K(K+1)}{2}}{\sqrt{K(K+1)}}\stackrel{d}{\longrightarrow} N(0,1).
		\end{aligned}
	\end{equation*}
\end{theorem}

\begin{corollary}
	\label{corollary:2-1}
	Let $\A^*$ be given as in (\ref{eq:Astar}),  and let $\bDelta$ be defined as in (\ref{eq:Delta}). Suppose that Assumptions \ref{assumption:1} and \ref{assumption:2} are satisfied. When $K=o(n^{\frac{1}{2}})$, we have
	\begin{equation*}
		\begin{aligned}
			\tr({\A^*}\bDelta)=O_p(\frac{K^2}{n}).
		\end{aligned}
	\end{equation*}
\end{corollary}
This implies that the cross term $\tr({\A^*}\bDelta)$ is of smaller order and can be omitted in subsequent asymptotic analysis.

Next, we analyze the sum of squares due to the simple perturbation matrix. We have the following result.

\begin{lemma}
	\label{lemma:2-3}
	Let $\bDelta$ be defined as in (\ref{eq:Delta}). Suppose that Assumptions \ref{assumption:1} and \ref{assumption:2} are satisfied. For fixed $K$, we have
	\begin{equation*}
		\begin{aligned}
			\frac{1}{2}n\tr(\bDelta^2)\stackrel{d}{\longrightarrow} \chi_{\frac{K(K+1)}{2}}^2,\\
		\end{aligned}
	\end{equation*}
	and for  $K\rightarrow\infty$ and  $K=o(n^{1/2})$, we have
	\begin{equation*}
		\begin{aligned}
			\frac{\frac{1}{2}n\tr(\bDelta^2)-\frac{K(K+1)}{2}}{\sqrt{K(K+1)}}\stackrel{d}{\longrightarrow} N(0,1).
		\end{aligned}
	\end{equation*}
\end{lemma}

\begin{corollary}
	\label{corollary:2-2}
	Let $\bDelta$ be defined as in (\ref{eq:Delta}). Suppose that Assumptions \ref{assumption:1} and \ref{assumption:2} are satisfied. When $K=o(n^{1/2})$, we have
	\begin{equation*}
		\begin{aligned}
			\tr(\bDelta^2)=O_p(\frac{K^2}{n}),\,\,\, \|\bDelta\|=O_p(\frac{K}{\sqrt{n}}).
		\end{aligned}
	\end{equation*}
	where $\|\cdot\|$ denotes the spectral norm.
\end{corollary}

Finally, we combine the above results to characterize the total sum of squares $\tr(\bar{\A}^2)$.
By  Corollaries \ref{corollary:2-1} and \ref{corollary:2-2}, we have
\[
\tr(\bar{\A}^2)=\tr({\A^*}^2)+2\tr({\A^*}\bDelta)+\tr(\bDelta^2)=\tr({\A^*}^2)+O_p(\frac{K^2}{n}).
\]
By  Lemma \ref{lemma:2-1}, the following result is then immediate.

\begin{corollary}
	\label{corollary:2-3}
	Let $\bar{\A}$ be given as in (\ref{eq:Abar}). Suppose that Assumptions \ref{assumption:1} and \ref{assumption:2} are satisfied. When $K=o(n^{1/2})$, we have
	\begin{equation*}
		\begin{aligned}
			\frac{\tr(\bar{\A}^2)-(n-1)}{\sqrt{2L-2}}\stackrel{d}{\longrightarrow} N(0,1).
		\end{aligned}
	\end{equation*}
\end{corollary}

\section{Composite Perturbation  Matrix}
\label{Composite}
In this section, we formally define the composite perturbation matrix and examine its theoretical properties, revealing a striking contrast with the simple perturbation case.
\subsection{Definition}
\label{DefinitionComposite}
We define  $\hat{\A}$ as follows:
\begin{equation}\label{eq:Ahat}
	\hat{A}_{ij} = \begin{cases}
		\dfrac{A_{ij}-\hat{P}_{ij}}{\sqrt{n\hat{P}_{ij}(1-\hat{P}_{ij})}}, \quad & i \neq j,\\
		0, \quad & i = j.
	\end{cases}
\end{equation}
By  direct calculation, we have
\begin{equation*}
	\begin{aligned}
		\hat{A}_{ij}&=\bar{A}_{ij}\sqrt{\frac{nP_{ij}(1-P_{ij})}{n\hat{P}_{ij}(1-\hat{P}_{ij})}}\\
		&=A_{ij}^*+\Delta_{ij}+\gamma_{ij}\Delta_{ij},
	\end{aligned}
\end{equation*}
where $\gamma_{ij}=\frac{(A_{ij}-\hat{P}_{ij})(1-\hat{P}_{ij}-P_{ij})}{\sqrt{\hat{P}_{ij}(1-\hat{P}_{ij}})(\sqrt{P_{ij}(1-P_{ij}})+\sqrt{\hat{P}_{ij}(1-\hat{P}_{ij}}))}$.
A detailed derivation is provided in Remark \ref{remark:1} of the Appendix.

As the entries of $\P$ are uniformly bounded away from $0$ and $1$, with probability tending to $1$, we have
\begin{equation*}
	\begin{aligned}
		\gamma&\triangleq\max_{i,j}|\gamma_{ij}|\leq \frac{C}{2},
	\end{aligned}
\end{equation*}
where
\[
C=\max_{1\leq u,v\leq K}\frac{1}{B_{u,v}(1-B_{u,v})}.
\]

We then define the composite perturbation matrix $\tilde{\bDelta}$ of the normalized adjacency matrix $\A^*$ as follows:
\begin{equation}\label{eq:Deltatilde}
	\begin{aligned}
		\tilde{\Delta}_{ij} = \begin{cases}
			(1+\gamma_{ij})\Delta_{ij}, \quad & i \neq j,\\
			0, \quad & i = j.
		\end{cases}
	\end{aligned}
\end{equation}
The composite perturbation matrix $\tilde{\bDelta}$ arises from replacing $P_{ij}$ with $\hat{P}_{ij}$ in both the numerator and denominator of the normalized adjacency matrix.

Then, we have
\begin{equation*}
	\begin{aligned}
		\hat{A}_{ij}&=A^*_{ij}+\tilde{\Delta}_{ij}.
	\end{aligned}
\end{equation*}
That is
\begin{equation}
	\label{equation:3-0}
	\begin{aligned}
		\hat{\A}&=\A^*+\tilde{\bDelta}.
	\end{aligned}
\end{equation}
Therefore, $\hat{\A}$ can be decomposed into two parts: the normalized adjacency matrix ${\A^*}$ and the composite perturbation matrix $\tilde{\bDelta}$.

The total sum of squares $\tr(\hat{\A}^2)$ then admits the following decomposition:
\begin{equation*}
	\begin{aligned}
		\tr(\hat{\A}^2)&=\tr((\A^*+\tilde{\bDelta})^2)\\
		&=\tr({\A^*}^2)+2\tr(\A^*\tilde{\bDelta})+\tr(\tilde{\bDelta}^2),
	\end{aligned}
\end{equation*}
where
\begin{equation*}
	\begin{aligned}
		\tr(\hat{\A}^2)=\sum_{i,j}\hat{A}_{ij}^2=\sum_{1\leq u, v\leq K}\sum_{i\in\mathcal{N}_{u}, j\in\mathcal{N}_{v}}\hat{A}_{ij}^2,
	\end{aligned}
\end{equation*}
\begin{equation*}
	\begin{aligned}
		\tr({\A^*}\tilde{\bDelta})=\sum_{i, j}A^*_{ij}\tilde{\Delta}_{ji}=\sum_{1\leq u, v\leq K}\sum_{i\in\mathcal{N}_{u}, j\in\mathcal{N}_{v}}A^*_{ij}\tilde{\Delta}_{ji},
	\end{aligned}
\end{equation*}
and
\begin{equation*}
	\begin{aligned}
		\tr(\tilde{\bDelta}^2)=\sum_{i, j}\tilde{\Delta}_{ij}^2=\sum_{1\leq u, v\leq K}\sum_{i\in\mathcal{N}_{u}, j\in\mathcal{N}_{v}}\tilde{\Delta}_{ij}^2.
	\end{aligned}
\end{equation*}
This decomposition separates $\tr(\hat{\A}^2)$  into three interpretable components: $\tr({\A^*}^2)$, $\tr(\A^*\tilde{\bDelta})$ and $\tr(\tilde{\bDelta}^2)$,  which are respectively the sum of squares due to the normalized adjacency matrix, the sum of the cross term due to the normalized adjacency matrix and the composite perturbation matrix, and the sum of squares due to the composite perturbation matrix.

\subsection{Theoretical Properties}

We now analyze the asymptotic behavior of the three components in the decomposition of $\tr(\hat{\A}^2)$ introduced in Section \ref{DefinitionComposite}. There are three important differences between the simple and composite perturbation matrices.

First, the total sum of squares $\tr(\hat{\A}^2)$ does not converge to a normal distribution. In fact, it is equal to a constant.

\begin{lemma}
	\label{lemma:3-2}
	Let $\hat{\A}$ be given as in (\ref{eq:Ahat}). Suppose that Assumptions \ref{assumption:1} and \ref{assumption:2} are satisfied. When $K=o(n^{1/2})$, we have
	\begin{equation*}
		\begin{aligned}
			\tr(\hat{\A}^2)=n-1.
		\end{aligned}
	\end{equation*}
\end{lemma}

Second, the sum of squares due to the composite perturbation matrix does not converge in distribution  to a chi-square distribution with $K(K+1)/2$ degrees of freedom. However, we still have the following result.

\begin{lemma}
	\label{lemma:3-3}
	Let $\tilde{\bDelta}$ be defined as in (\ref{eq:Deltatilde}). Suppose that Assumptions \ref{assumption:1} and \ref{assumption:2} are satisfied. When $K=o(n^{1/2})$, we have
	\begin{equation*}
		\begin{aligned}
			\tr(\tilde{\bDelta}^2)=O_p(\frac{K^2}{n}),\,\,\,\|\tilde{\bDelta}\|=O_p(\frac{K}{\sqrt{n}}).
		\end{aligned}
	\end{equation*}
\end{lemma}

Third, the cross term $\tr({\A^*}\tilde{\bDelta})$ due to the normalized adjacency matrix and the composite perturbation matrix converges in distribution to the standard normal distribution.

The direct expansion of $\tr({\A^*}\tilde{\bDelta})$ leads to an intractable form (see Remark \ref{remark:2} in the Appendix), making it difficult to obtain its distribution. Instead, we derive its asymptotic distribution via an indirect yet concise approach.

By  Lemmas \ref{lemma:3-2} and \ref{lemma:3-3}, we have
\begin{equation*}
	\begin{aligned}
		2\tr({\A^*}\tilde{\bDelta})=\tr(\hat{\A}^2)-\tr({\A^*}^2)-\tr(\tilde{\bDelta}^2)=(n-1)-\tr({\A^*}^2)-O_p(\frac{K^2}{n}).
	\end{aligned}
\end{equation*}
By  Lemma \ref{lemma:2-1}, the following result is then immediate.
\begin{theorem}
	\label{theorem:3-2}
	Let $\A^*$ be given as in (\ref{eq:Astar}), and let $\tilde{\bDelta}$ be defined as in (\ref{eq:Deltatilde}). Suppose that Assumptions \ref{assumption:1} and \ref{assumption:2} are satisfied. When $K=o(n^{1/2})$, we have
	\begin{equation*}
		\begin{aligned}
			\frac{2\tr({\A^*}\tilde{\bDelta})}{\sqrt{2L-2}}\stackrel{d}{\longrightarrow} N(0,1).
		\end{aligned}
	\end{equation*}
\end{theorem}

\subsection{Structural Comparison with Simple Perturbation}
Theorem \ref{theorem:3-2} implies that cross term $\tr({\A^*}\tilde{\bDelta})$ due to the normalized adjacency matrix and the composite perturbation matrix cannot be omitted. In comparison with Theorem \ref{theorem:2-2},  there is an essential difference between the simple and composite perturbation  matrices. To explain this phenomenon, we define $\bar{\bDelta}$ as follows:
\begin{equation}\label{eq:Deltabar}
	\begin{aligned}
		\bar{\Delta}_{ij} = \begin{cases}
			\Delta_{ij}, \quad & i \neq j,\\
			\Upsilon_i, \quad & i = j,
		\end{cases}
	\end{aligned}
\end{equation}
where $\Upsilon_i=\frac{P_{ii}-\hat{P}_{ii}}{\sqrt{nP_{ii}(1-P_{ii})}}$. That is,  $\bar{\bDelta}=\bDelta+\bUpsilon$, where $\bUpsilon=\Diag\{\Upsilon_1,\ldots,\Upsilon_n\}$. Then, $\bar{\bDelta}$ is a $K\times K$ block-wise constant symmetric matrix, its rank is at most $K$.

To exploit the structural property of low-rank matrices, the simple perturbation matrix $\bDelta$ is first reformulated as the low-rank matrix $\bar{\bDelta}$.  After accounting for and removing the effect of $\bUpsilon$, $\bDelta$ retains its low-rank nature. Conversely, the composite perturbation matrix $\tilde{\bDelta}$ is generally of full rank.

Under simple perturbation, by Corollary \ref{corollary:2-1}, we have $\operatorname{tr}(\mathbf{A}^* \Delta) = O_p(K^2/n) = o_p(1)$ for $K = o(n^{1/2})$, which is asymptotically negligible. In stark contrast, in the composite case, by Theorem \ref{theorem:3-2}, we have $\operatorname{tr}(\mathbf{A}^* \tilde{\Delta}) = O_p(1)$, which is not negligible. This intrinsic discrepancy highlights a key structural distinction between  the two perturbations: $\Delta$ is inherently low-rank, while $\tilde{\Delta}$ typically exhibits a full-rank structure.

More importantly, although $\tilde{\Delta}_{ij}=(1+\gamma_{ij})\Delta_{ij}$, we cannot write it as $\tilde{\bDelta}=(1+O_p(\gamma))\bDelta$. If we did so, we would obtain
\begin{equation*}
	\begin{aligned}
		\tr({\A^*}\tilde{\bDelta})=(1+O_p(\gamma))\tr({\A^*}\bDelta).
	\end{aligned}
\end{equation*}
This leads to a contradiction: $O_p(1) = o_p(1)$. Thus, when we extract a scalar factor from a matrix, we must pay particular attention, as this may destroy the inherent structure of the matrix.

\section{A Unified Decomposition Framework}
\label{Decomposition}
In this section, we develop a unified decomposition framework that breaks the composite perturbation matrix into interpretable components with distinct spectral properties.
\subsection{Decomposition of the Composite Perturbation Matrix}
There exists a key distinction between the asymptotic behaviors of $\tr({\A^*}\bDelta)$ and $\tr({\A^*}\tilde{\bDelta})$. In Section \ref{Composite}, we explain this phenomenon through the distinct structural properties of $\bDelta$ and $\tilde{\bDelta}$. In this section, we further investigate the underlying mechanism by examining the composition of the composite perturbation matrix. We propose a unified decomposition framework, which breaks the composite perturbation down into several structurally explicit components. To this end, we start from the relationship between $\hat{A}_{ij}$ and $\bar{A}_{ij}$, which shows how the re-scaling factor introduces additional bias terms.

Specifically, we have
\begin{equation}
	\label{equation:4-9}
	\begin{aligned}
		\hat{A}_{ij}&=\bar{A}_{ij}\sqrt{\frac{nP_{ij}(1-P_{ij})}{n\hat{P}_{ij}(1-\hat{P}_{ij})}}\\
		&=A_{ij}^*+\alpha_{ij}{A_{ij}^*}+\Delta_{ij}+\alpha_{ij}\Delta_{ij},
	\end{aligned}
\end{equation}
where \begin{equation}
	\label{equation:4-0}
\begin{aligned}
\alpha_{ij}=\frac{(P_{ij}-\hat{P}_{ij})(1-\hat{P}_{ij}-P_{ij})}{\sqrt{\hat{P}_{ij}(1-\hat{P}_{ij}})(\sqrt{P_{ij}(1-P_{ij}})+\sqrt{\hat{P}_{ij}(1-\hat{P}_{ij}}))}.
	\end{aligned}
\end{equation}
A detailed derivation is provided in Remark \ref{remark:3} of the Appendix.

As the entries of $\P$ are uniformly bounded away from $0$ and $1$, by Hoeffding's  inequality, we have
\[
\max_{i,j}|P_{ij}-\hat{P}_{ij}|=O_p(\frac{K\log^{\frac{1}{2}}K}{n}).
\]
Thus
\begin{equation*}
	\begin{aligned}
		\alpha&\triangleq\max_{i,j}|\alpha_{ij}|= O_p(\frac{K\log^{\frac{1}{2}}K}{n}).
	\end{aligned}
\end{equation*}

We define the bias matrix ${\check{\A}}$ of the normalized adjacency matrix ${\A^*}$ as follows:
\begin{equation}\label{eq:Acheck}
	\begin{aligned}
		\check{A}_{ij} = \begin{cases}
			\alpha_{ij}{A_{ij}^*}, \quad & i \neq j,\\
			0, \quad & i = j.
		\end{cases}
	\end{aligned}
\end{equation}

We then define  $\check{\bDelta}$ as the bias matrix of the simple perturbation matrix $\bDelta$:
\begin{equation}\label{eq:Deltacheck}
	\begin{aligned}
		\check{\Delta}_{ij} = \begin{cases}
			\alpha_{ij}\Delta_{ij}, \quad & i \neq j,\\
			0, \quad & i = j.
		\end{cases}
	\end{aligned}
\end{equation}
Then \eqref{equation:4-9} can be rewritten as
\begin{equation}
	\label{equation:4-1}
	\begin{aligned}
		\hat{\A}={\A^*}+\check{\A}+\bDelta+\check{\bDelta}.
	\end{aligned}
\end{equation}
In comparison with \eqref{equation:3-0}, $\hat{\A}$ can be further decomposed into four parts: the normalized adjacency matrix ${\A^*}$ and its bias matrix $\check{\A}$, the simple perturbation matrix $\bDelta$ and its bias matrix $\check{\bDelta}$. This decomposition allows us to systematically analyze how each component contributes to the spectral properties of the overall matrix:
\begin{itemize}
	\item ${\A^*}$ captures the theoretical Wigner matrix structure;
	\item $\check{\A}$  reflects the direct correction to ${\A^*}$ due to re-scaling;
	\item $\bDelta$  accounts for the perturbation due to the estimation of $\P$;
	\item $\check{\bDelta}$  shows the further adjustment to $\bDelta$ due to re-scaling.
\end{itemize}
Here, $\check{\A}$ is a full-rank matrix that directly couples estimation error with the random fluctuations of ${\A^*}$, while $\bDelta$ and $\check{\bDelta}$ are low-rank and structured. This structural distinction is key to our unified analysis.

By \eqref{equation:3-0} and \eqref{equation:4-1}, the composite perturbation matrix $\tilde{\bDelta}$ can be expressed as:
\begin{equation}
	\label{equation:4-3}
	\begin{aligned}
		\tilde{\bDelta}=\check{\A}+\bDelta+\check{\bDelta}.
	\end{aligned}
\end{equation}
We refer to this decomposition as the unified decomposition framework for the composite perturbation matrix. The framework reveals that the re-scaling factor introduces an additional full-rank bias matrix  ${\check{\A}}$, which is the key source of the full-rank nature of the composite perturbation matrix $\tilde{\bDelta}$. This structured decomposition allows us to precisely isolate and control the distinct sources of perturbation, thereby establishing a clear mathematical framework for deriving the limiting distributions.

\subsection{Asymptotic Analysis of the Cross Term}
By the decomposition in \eqref{equation:4-3}, we can write
\begin{equation*}
	\begin{aligned}
		\tr({\A^*}\tilde{\bDelta})=\tr({\A^*}\check{\A})+\tr({\A^*}\bDelta)+\tr({\A^*}\check{\bDelta}).
	\end{aligned}
\end{equation*}

Under the condition $K=o(n^{1/2})$, Corollary \ref{corollary:2-1} and Lemma \ref{lemma:4-4} imply that both $\tr({\A^*}\bDelta)$ and $\tr({\A^*}\check{\bDelta})$ are of order
$O_p(K^2/n)$, which is negligible compared to $\tr({\A^*}\tilde{\bDelta})$. Consequently,
\begin{equation}
	\label{equation:4-8}
	\begin{aligned}
		\tr({\A^*}\tilde{\bDelta})=\tr({\A^*}\check{\A})+O_p(\frac{K^2}{n}).
	\end{aligned}
\end{equation}
This reveals that the cross term in the composite perturbation setting is essentially governed by the bias matrix $\check{\A}$.

Our unified decomposition framework in \eqref{equation:4-8} provides a direct route to analyze $\tr({\A^*}\tilde{\bDelta})$. Unlike the indirect derivation in Section \ref{Composite}, which relies on Lemma \ref{lemma:2-1}  to handle $\tr({\A^*}\tilde{\bDelta})$, we can now exploit the structural decomposition \eqref{equation:4-8}. This allows us to isolate the dominant term $\tr({\A^*}\check{\A})$ and obtain its asymptotic distribution directly, thereby offering a more transparent and self-contained proof.
\begin{theorem}
	\label{theorem:3-3}
	Let $\A^*$ be given as in (\ref{eq:Astar}), and let $\check{\A}$ be defined as in (\ref{eq:Acheck}). Suppose that Assumptions \ref{assumption:1} and \ref{assumption:2} are satisfied. When $K=o(n^{1/2})$, we have
	\begin{equation*}
		\begin{aligned}
			\frac{2\tr({\A^*}\check{\A})}{\sqrt{2L-2}}\stackrel{d}{\longrightarrow} N(0,1).
		\end{aligned}
	\end{equation*}
\end{theorem}
Combining \eqref{equation:4-8} with Theorem \ref{theorem:3-3}, we immediately recover the limiting distribution of $\tr({\A^*}\tilde{\bDelta})$, as previously stated in Theorem \ref{theorem:3-2}, without invoking Lemma \ref{lemma:2-1}.

Thus, the transition from simple to  composite perturbations is not merely a re-scaling. It introduces a new, structurally distinct component, the bias matrix $\check{\A}$,  that substantially alters the perturbation's composition and its asymptotic impact on test statistics.

By distinguishing these components, we can more precisely control the order and asymptotic behavior of each part in subsequent analyses. This yields a unified framework for comparing the asymptotic behaviors under simple and composite perturbations.

\section{Applications}
\label{Applications}
In this section, we first apply the results in Sections \ref{Simple}, \ref{Composite} and \ref{Decomposition} to the largest  eigenvalue statistic $\lambda_1(\hat{\A})$ in \cite{Lei:2016}, then to the linear spectral statistic $\tr(\hat{\A}^{3})$ in \cite{Wu:2024}.

\subsection{Largest  Eigenvalue Statistic}
We begin with the largest  eigenvalue statistic, whose asymptotic theory for stochastic block models was first established in the influential  work of Lei \cite{Lei:2016}. Our decomposition framework clarifies the distinct impacts of simple and composite perturbations on the extremal spectrum, allowing us to sharpen the existing condition from $K=O(n^{1/6-\tau})$ to the optimal rate $K=o(n^{1/6})$.
\subsubsection{Under Simple Perturbation}
Recall from Section \ref{Simple} that $\bar{\A}={\A^*}+\bDelta,$  where the simple perturbation matrix $\bDelta$ can be reformulated as the low-rank matrix $\bar{\bDelta}=\bDelta+\bUpsilon$. Exploiting this low-rank structure is essential for precisely bounding the difference between the largest eigenvalue of $\bar{\A}+\bUpsilon$ and that of the normalized adjacency matrix $\A^*$. After accounting for and removing the effect of $\bUpsilon$, we can establish the limiting distribution of $\lambda_1(\bar{\A})$.

Given that $\bar{\bDelta}$ is a $K\times K$ block-wise constant symmetric matrix, it admits a low-rank factorization. Let $\bar{\bDelta}=\bTheta\bSigma\bTheta^T$, where $\bTheta=(\btheta_1,\dots,\btheta_K)$ and $\bSigma$ is a $K\times K$ symmetric matrix. Since the rank of $\bar{\bDelta}$ is at most $K$,
the corresponding principal subspace is spanned by $\btheta_1,\dots,\btheta_K$, where $\btheta_k\in \mathbb{R}^n$
is the unit norm indicator of the $k$-th community in $\mathbf{g}$. That is, ${(\btheta_k)}_i=\frac{1}{\sqrt{n_k}}\mathbf{1}_{\{\mathbf{g}_i=k\}}$.

This allows us to precisely control the spectral norms of $\bar{\bDelta}$, $\bSigma$, and $\|\bUpsilon\|$.
\begin{lemma}
	\label{lemma:4-1}
	Let $\bar{\bDelta}$ be given as in (\ref{eq:Deltabar}). Suppose that Assumptions \ref{assumption:1} and \ref{assumption:2} are satisfied.  When $K=o(n^{1/2})$, we have
	\begin{equation*}
		\begin{aligned}
			\|\bar{\bDelta}\|=O_p(\frac{K}{\sqrt{n}}),\,\,\,\|\bSigma\|=O_p(\frac{K}{\sqrt{n}}),\|\bUpsilon\|=O_p(\frac{K}{n}).
		\end{aligned}
	\end{equation*}
\end{lemma}

With this tighter bound in hand, we can now follow the perturbation argument of Lei \cite{Lei:2016} but under the optimal growth condition on $K$.

\begin{corollary}
	\label{corollary:4-0}
	Let $\bar{\A}$ be given as in \eqref{eq:Abar}. Suppose that Assumptions \ref{assumption:1} and \ref{assumption:2} are satisfied. When $K=o(n^{1/6})$, we have
	\begin{equation*}
		\begin{aligned}
			n^{2/3}(\lambda_1(\bar{\A})-2)\stackrel{d}{\rightarrow} TW_1,
		\end{aligned}
	\end{equation*}
where $TW_1$ denotes the Tracy-Widom distribution with index 1 and
$\lambda_i(\bar{\A})$ denotes the $i$-th largest eigenvalue of the matrix $\bar{\A}$.
\end{corollary}

\subsubsection{Under Composite Perturbation}

When $K=O(n^{1/6-\tau})$, where $\tau$ is a positive constant, Lei \cite{Lei:2016} showed  that,
\begin{equation*}
	\begin{aligned}
		n^{2/3}(\lambda_1(\hat{\A})-2)\stackrel{d}{\rightarrow} TW_1,
	\end{aligned}
\end{equation*}
where $\lambda_i(\hat{\A})$ denotes the $i$-th largest eigenvalue of the matrix $\hat{\A}$. However, there exists a noticeable theoretical gap between $K=O(n^{1/6-\tau})$ and the optimal rate $K=o(n^{1/6})$. A natural question then arises: What additional condition would be necessary to achieve the optimal rate $K=o(n^{1/6})$?

In the proof of Corollary \ref{corollary:4-0}, we exploit the low-rank property of $\bar{\bDelta}=\bDelta+\bUpsilon$, whose rank is at most $K$. Motivated by this, we construct $\hat{\bDelta}$ as follows to ensure it is also of rank at most $K$:
\begin{equation}\label{eq:Deltahat}
	\begin{aligned}
		\hat{\Delta}_{ij} = \begin{cases}
			(1+\alpha_{ij})\Delta_{ij}, \quad & i \neq j,\\
			(1+\alpha_{ii})\Upsilon_i, \quad & i = j.
		\end{cases}
	\end{aligned}
\end{equation}
Let $\check{\Upsilon}_i=(1+\alpha_{ii})\Upsilon_i$ and $\check{\bUpsilon}=\Diag\{\check{\Upsilon}_1,\ldots,\check{\Upsilon}_n\}$.
This construction yields the decomposition $\hat{\bDelta}=\bDelta+\check{\bDelta}+\check{\bUpsilon}$.

Using the unified decomposition framework in Section \ref{Decomposition}, we have $\hat{\A}={\A^*}+\check{\A}+\bDelta+\check{\bDelta}$. Exploiting the low-rank structure of $\hat{\bDelta}$ is essential for precisely bounding the difference between the largest eigenvalue of $\hat{\A}+\check{\bUpsilon}$ and that of the normalized adjacency matrix  $\A^*$. After accounting for and removing the effect of $\check{\bUpsilon}$, we can establish the limiting distribution of $\lambda_1(\hat{\A})$.

Let $\hat{\bDelta}=\bTheta\hat{\bSigma}\bTheta^T$, where $\hat{\bSigma}$ is a $K\times K$ symmetric matrix. The spectral norms of $\hat{\bDelta}$, $\hat{\bSigma}$, and $\|\check{\bUpsilon}\|$ can be precisely controlled as follows.

\begin{lemma}
	\label{lemma:4-01}
	Let $\hat{\bDelta}$ be given as in \eqref{eq:Deltahat}. Suppose that Assumptions \ref{assumption:1} and \ref{assumption:2} are satisfied.  When $K=o(n^{1/2})$, we have
	\begin{equation*}
		\begin{aligned}
			\|\hat{\bDelta}\|=O_p(\frac{K}{\sqrt{n}}),\,\,\,\|\hat{\bSigma}\|=O_p(\frac{K}{\sqrt{n}}),\|\check{\bUpsilon}\|=O_p(\frac{K}{n}).
		\end{aligned}
	\end{equation*}
\end{lemma}
In comparison with the simple case, the composite perturbation matrix $\tilde{\bDelta}$ contains an additional full-rank matrix $\check{\A}$ introduced by the re-scaling factor. To control the influence of $\check{\A}$, we make the following assumption.

\begin{assumption}\label{assumption:0}
	There exists a constant $c_2$, such that
	\[
	\max_{1\leq k\leq K}n_{k}\leq \frac{c_2n}{\log K}.
	\]
\end{assumption}
Assumption \ref{assumption:0} requires the size of the largest community is at most proportional to $n/\log K$. This condition plays a pivotal role in ensuring the spectral norm of the bias matrix $\check{\A}$ remains asymptotically negligible. In combination with Assumption \ref{assumption:2}, it imposes a balanced community structure, which is realistic in many real-world networks (e.g., social or biological networks) and prevents pathological cases where estimation and inference become unreliable.
\begin{lemma}
	Let $\check{\A}$ be defined as in \eqref{eq:Acheck}. Suppose that Assumptions \ref{assumption:1}, \ref{assumption:2} and \ref{assumption:0} are satisfied.  When $K=o(n^{1/2})$, we have
	\label{lemma:4-7}
	\begin{equation*}
		\begin{aligned}
			\|\check{\A}\|=O_p(\frac{K^{2}}{n}).
		\end{aligned}
	\end{equation*}
\end{lemma}

This implies that when $K=o(n^{1/6})$, the spectral norm of the bias matrix $\check{\A}$ of the normalized adjacency matrix converges to zero in probability at a rate faster than $n^{-2/3}$. Consequently, under the $n^{2/3}$-scaling, the influence of $\check{\A}$ on the extreme eigenvalue becomes asymptotically negligible, ensuring that the limiting distribution of $\lambda_1(\hat{\A})$ is dominated by $\lambda_1({\A^*})$.

By Lemmas \ref{lemma:4-01} and \ref{lemma:4-7}, the rate $K=O(n^{1/6-\tau})$ can be improved to $K=o(n^{1/6})$, which is the optimal rate.
\begin{theorem}\label{theorem:4-2}
	Let $\hat{\A}$ be given as in \eqref{eq:Ahat}. Suppose that Assumptions \ref{assumption:1}, \ref{assumption:2} and \ref{assumption:0} are satisfied. When $K=o(n^{1/6})$, we have
	\begin{equation*}
		\begin{aligned}
			n^{2/3}(\lambda_1(\hat{\A})-2)\stackrel{d}{\rightarrow} TW_1,
		\end{aligned}
	\end{equation*}
\end{theorem}

\subsubsection{Comparison}
It is instructive to compare the conditions required for the largest  eigenvalue statistic under simple perturbation (Corollary \ref{corollary:4-0}) and composite perturbation (Theorem \ref{theorem:4-2}). While both achieve the optimal rate $K=o(n^{1/6})$ for the Tracy-Widom limit of $\lambda_1(\bar{\A})$ and $\lambda_1(\hat{\A})$, respectively,  the composite perturbation requires an extra balance condition (Assumption \ref{assumption:0}) to control the full-rank bias matrix $\check{\A}$ introduced by the scaling factor. This reflects the higher sensitivity of extreme eigenvalues to unstructured, full-rank perturbations compared to low-rank ones.

\subsection{Linear Spectral Statistic}
We now turn to the linear spectral statistic proposed by Wu and Hu\footnotemark[\value{footnote}] \cite{Wu:2024}
\footnotetext[\value{footnote}]{Jiang Hu (Northeast Normal University) is a different researcher from the co-author Jianwei Hu of the present paper.},  which they demonstrated to be highly effective for goodness-of-fit testing in stochastic block models. However, a key technical challenge in establishing its asymptotic normality lies in proving that all error terms induced by the perturbation, whether simple or composite, are asymptotically negligible. Our unified decomposition framework provides the necessary analytical structure to address this challenge, enabling a rigorous, term-by-term analysis that yields a conclusive proof of asymptotic normality.
\subsubsection{Under Simple Perturbation}

Recall that $\bar{\A}={\A^*}+\bDelta$. Then, we have
\begin{equation}
\label{equation:4-02}
	\begin{aligned}
		\tr(\bar{\A}^3)&=\tr(({\A^*}+\bDelta)^3)\\
		&=\tr({\A^*}^3)+3\tr({\A^*}^2\bDelta)+3\tr({\A^*}\bDelta^2)+\tr(\bDelta^3).
	\end{aligned}
\end{equation}

In the proof of the following Corollary \ref{corollary:4-1}, the last two terms are shown to tend to zero in probability. We therefore focus on the term $\tr({\A^*}^2\bDelta)$. We have the following result.
\begin{lemma}
	\label{lemma:4-8}
	Let $\A^*$ be given as in \eqref{eq:Astar}, and let $\bDelta$ be defined as in (\ref{eq:Delta}). Suppose that Assumptions \ref{assumption:1} and \ref{assumption:2} are satisfied. When $K=o(n^{1/2})$, we have
	\begin{equation*}
		\begin{aligned}
			\tr({\A^*}^{2}\bDelta)&=O_p(\frac{K^2}{n}).
		\end{aligned}
	\end{equation*}
\end{lemma}

This implies that the cross term due to the two powers of the normalized adjacency matrix and the simple perturbation  matrix can be omitted.

In summary, under the condition $K=o(n^{1/2})$, all error terms induced by the estimation in \eqref{equation:4-02} can be omitted.  Consequently, the asymptotic behaviour of $\tr(\bar{\A}^3)$
is entirely governed by the leading term $\tr({\A^*}^3)$. By Lemma \ref{lemma:appendix-1}, we have the following result.
\begin{corollary}
	\label{corollary:4-1}
	Let $\bar{\A}$ be given as in (\ref{eq:Abar}). Suppose that Assumptions \ref{assumption:1} and \ref{assumption:2} are satisfied. When $K=o(n^{1/2})$, we have
	\begin{equation*}
		\begin{aligned}
			\frac{1}{\sqrt{6}}\tr(\bar{\A}^{3})\stackrel{d}{\longrightarrow} N(0,1).
		\end{aligned}
	\end{equation*}
\end{corollary}

\subsubsection{Under Composite Perturbation}

A major technical challenge arises under the composite perturbation setting: rigorously establishing the asymptotic normality of $\tr(\hat{\A}^3)$ is non-trivial. To address this, we employ the unified decomposition framework in Section \ref{Decomposition}.

Recall that $\hat{\A}={\A^*}+\tilde{\bDelta}$. The linear spectral statistic can be first decomposed as:
\begin{equation}
	\label{equation:4-2}
	\begin{aligned}
		\tr(\hat{\A}^3)&=\tr(({\A^*}+\tilde{\bDelta})^3)\\
		&=\tr({\A^*}^3)+3\tr({\A^*}^2\tilde{\bDelta})+3\tr({\A^*}\tilde{\bDelta}^2)+\tr(\tilde{\bDelta}^3).
	\end{aligned}
\end{equation}
Our goal is to prove that, under suitable conditions, all terms except the leading term $\tr({\A^*})$ are asymptotically negligible error terms, ensuring that
$\tr(\hat{\A}^3)$  shares the same asymptotic normal distribution as $\tr({\A^*})$. In the proof of Theorem \ref{theorem:4-3}, the last two terms are shown to tend to zero in probability.  We therefore focus on the cross term $\tr({\A^*}^2\tilde{\bDelta})$.

Using the decomposition \eqref{equation:4-3}, the cross term $\tr({\A^*}^2\tilde{\bDelta})$ can be further split as:
\begin{equation}
	\label{equation:4-5}
	\begin{aligned}
		\tr({\A^*}^2\tilde{\bDelta})&=\tr({\A^*}^2\check{\A})+\tr({\A^*}^2\bDelta)+\tr({\A^*}^2\check{\bDelta}).
	\end{aligned}
\end{equation}

We now control each term in \eqref{equation:4-5} individually. For the first term $\tr({\A^*}^2\check{\A})$, we have the following result.
\begin{lemma}
	\label{lemma:4-5}
	Let $\A^*$ be given as in \eqref{eq:Astar}, and let $\check{\A}$ be defined as in \eqref{eq:Acheck}. Suppose that Assumptions \ref{assumption:1} and \ref{assumption:2} are satisfied. When $K=o(n^{1/2})$, we have
	\begin{equation*}
		\begin{aligned}
			\tr({\A^*}^2\check{\A})&= O_p(\frac{K^2}{n}).
		\end{aligned}
	\end{equation*}
\end{lemma}
That is, the cross term due to the bias matrix $\check{\A}$ and two powers of the normalized adjacency matrix can  be omitted.

The second term $\tr({\A^*}^2\bDelta)$ can be omitted by Lemma \ref{lemma:4-8}. For the third term $\tr({\A^*}^2\check{\bDelta})$, we have the following result.
\begin{lemma}
	\label{lemma:4-4}
	Let $\A^*$ be given as in \eqref{eq:Astar}, and let $\check{\bDelta}$ be defined as in \eqref{eq:Deltacheck}. Suppose that Assumptions \ref{assumption:1} and \ref{assumption:2} are satisfied. When $K=o(n^{1/2})$, for any fixed $k$, we have
	\begin{equation*}
		\begin{aligned}
			\tr({\A^*}^{k}\check{\bDelta})&=O_p(\frac{K^2}{n}).
		\end{aligned}
	\end{equation*}
\end{lemma}
That is, the cross term due to the bias matrix $\check{\bDelta}$ of the simple perturbation  matrix and powers of the normalized adjacency matrix can also be omitted.

In summary, under the condition $K=o(n^{1/2})$, all error terms induced by the estimation in \eqref{equation:4-2} can be omitted.  Consequently, the asymptotic behaviour of $\tr(\hat{\A}^3)$
is entirely governed by the leading term $\tr({\A^*}^3)$. By Lemma \ref{lemma:appendix-1}, we have the following result.
\begin{theorem}
	\label{theorem:4-3}
	Let $\hat{\A}$ be given as in \eqref{eq:Ahat}. Suppose that Assumptions \ref{assumption:1} and \ref{assumption:2} are satisfied. When $K=o(n^{1/2})$, we have
	\begin{equation*}
		\begin{aligned}
			\frac{1}{\sqrt{6}}\tr(\hat{\A}^{3})\stackrel{d}{\longrightarrow} N(0,1).
		\end{aligned}
	\end{equation*}
\end{theorem}

\subsubsection{Comparison}
A noteworthy feature of the linear spectral statistic is the identical condition $K=o(n^{1/2})$ required for both the simple perturbation setting (Corollary \ref{corollary:4-1}) and the composite perturbation setting (Theorem \ref{theorem:4-3}). That is, for the linear spectral statistic, the transition from simple to composite perturbations does not tighten the asymptotic condition on $K$. This contrasts with the largest  eigenvalue statistic, where the composite perturbation requires an extra balance assumption (Assumption \ref{assumption:0}).

\section{Simulation}\label{Simulation}
This section validates the properties of simple and composite perturbations via simulation studies, focusing on the limiting distributions of both the cross terms and sum-of-squares terms. In our experiments, community labels are estimated using the profile-pseudo likelihood method \cite{Wang:2023}, and the probability matrix is estimated via maximum likelihood.

\subsection{The Distribution of the Cross Terms}
\label{sec:6-1}
\begin{figure}[http]
	\centering
	\subfigure{\includegraphics[width=0.32\textwidth]{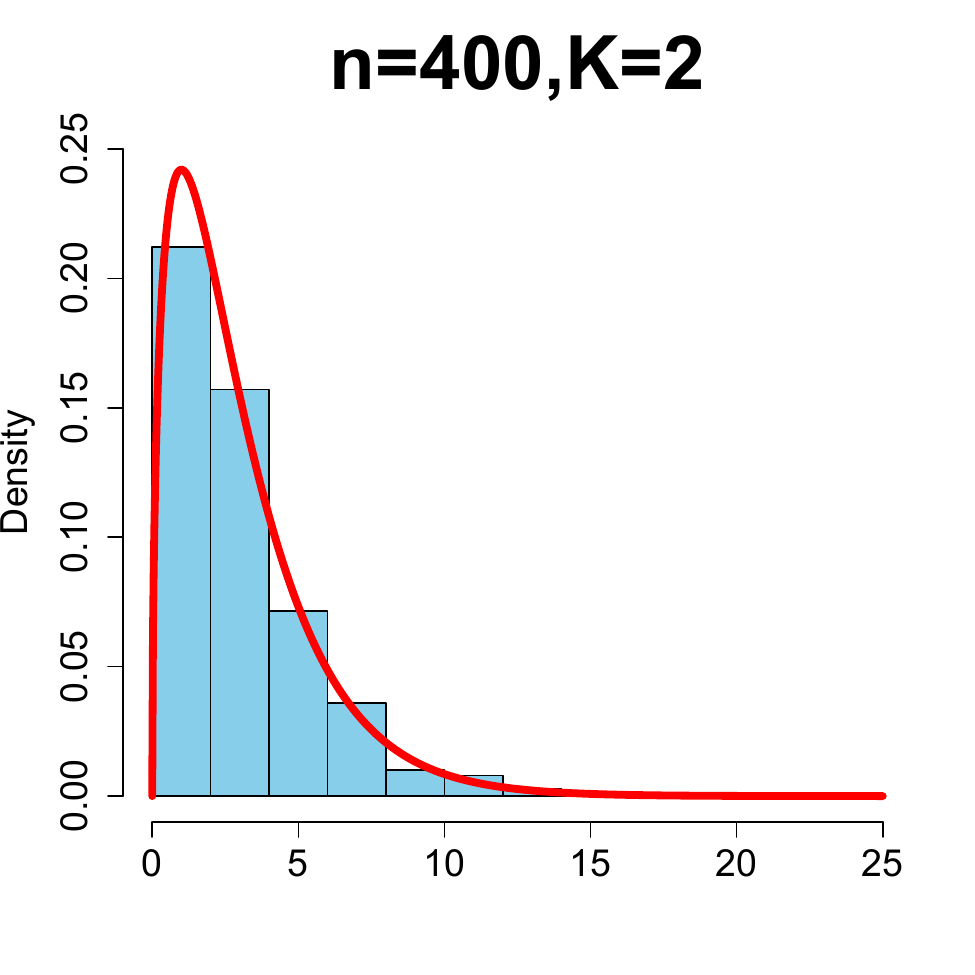}}
	\subfigure{\includegraphics[width=0.32\textwidth]{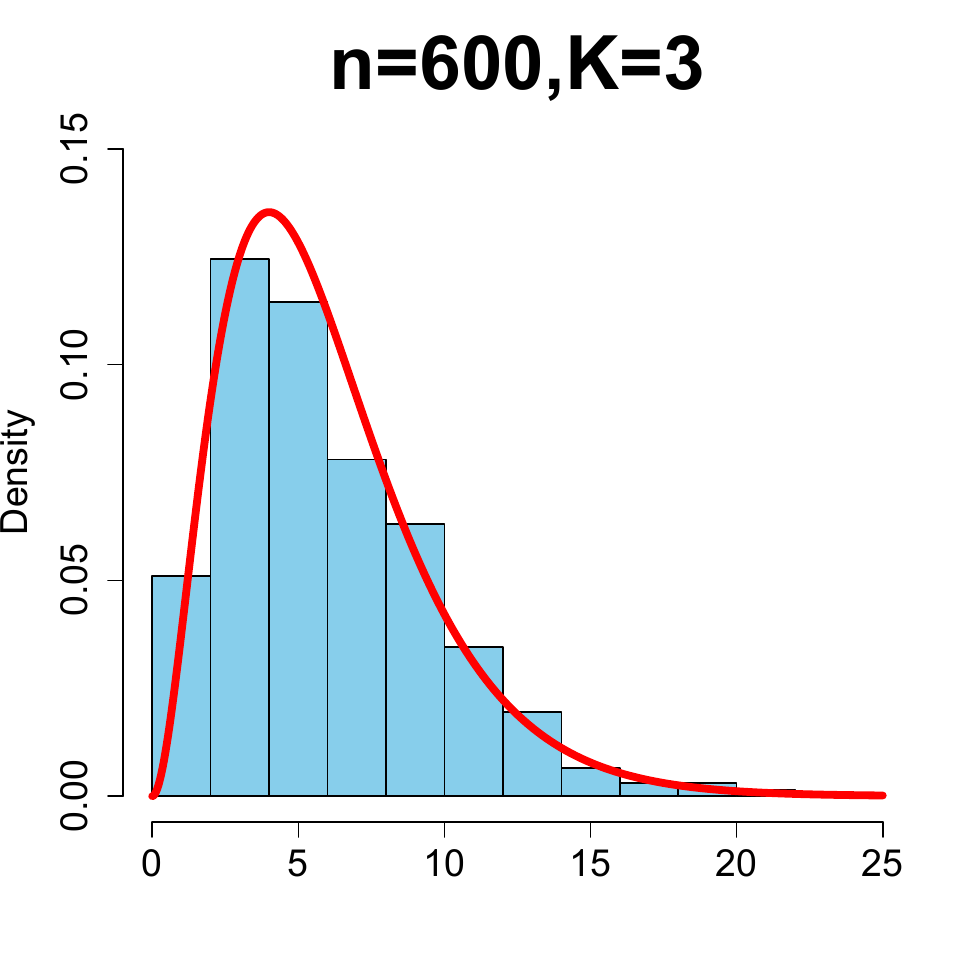}}
	\subfigure{\includegraphics[width=0.32\textwidth]{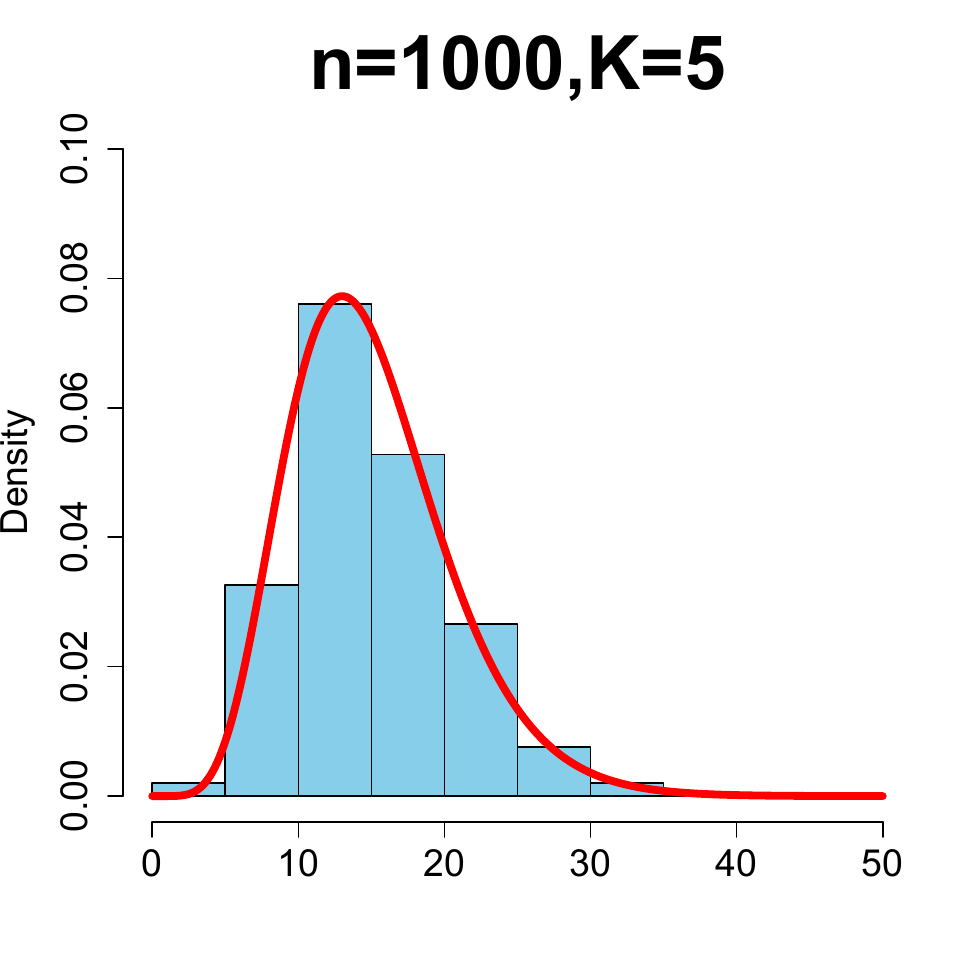}}
	\caption{Histograms of $-\frac{n}{2}\tr({\A^*}{\bDelta})$ from $1000$ data replications. The red solid line represents the densities of a chi-square distribution with $K(K+1)/2$ degrees of freedom.}
	\label{fig:theorem:2-2}
\end{figure}

\begin{figure}[http]
	\centering
	\subfigure{\includegraphics[width=0.32\textwidth]{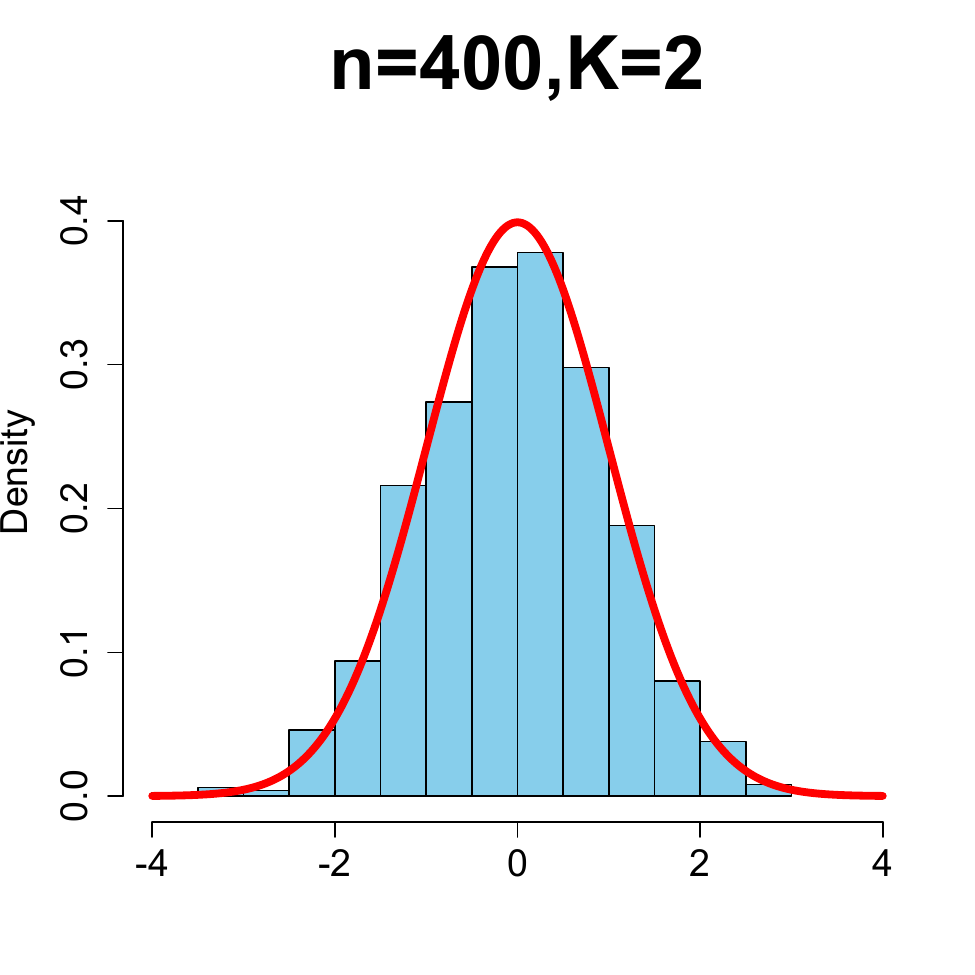}}
	\subfigure{\includegraphics[width=0.32\textwidth]{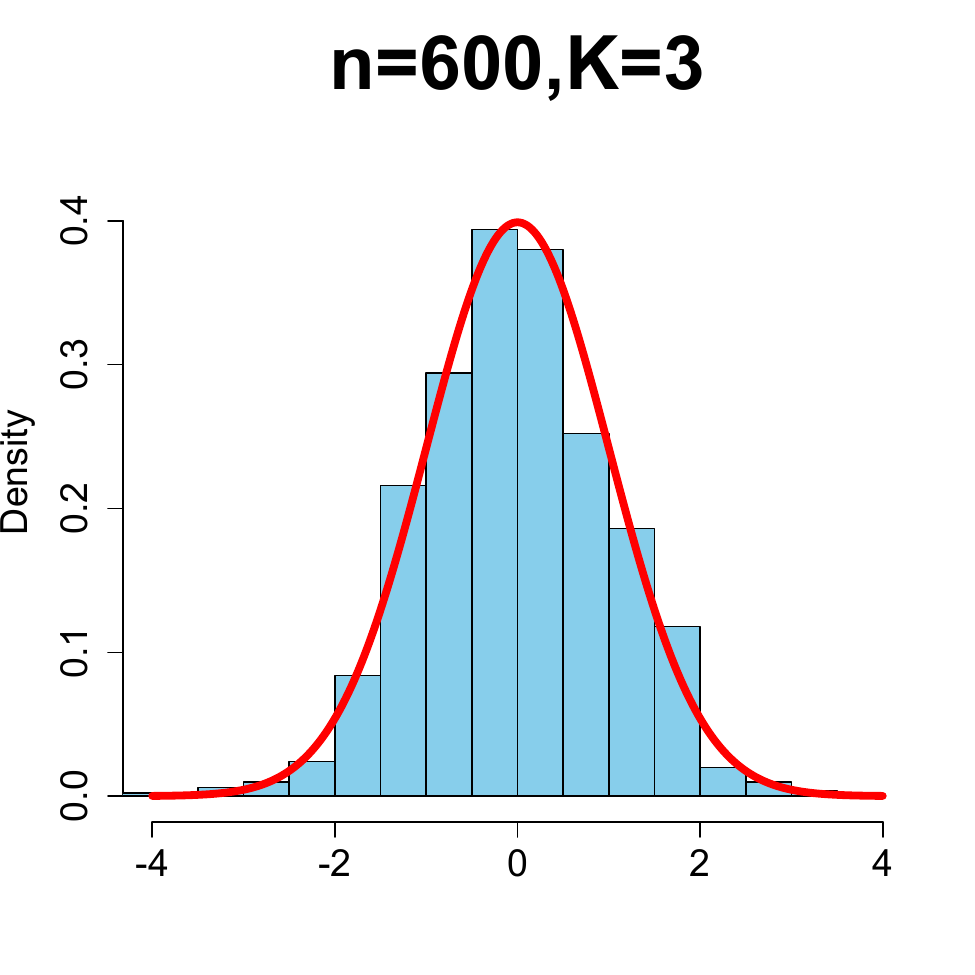}}
	\subfigure{\includegraphics[width=0.32\textwidth]{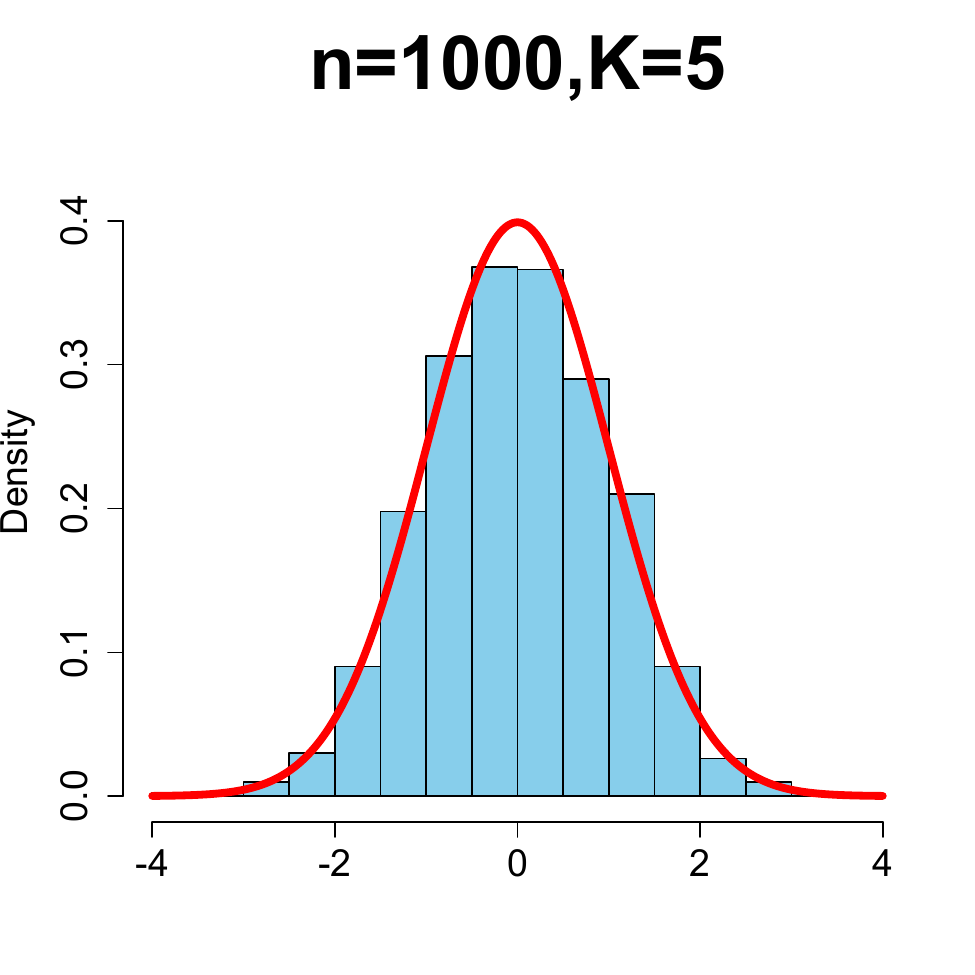}}
	\caption{Histograms of $\frac{2\tr({\A^*}\tilde{\bDelta})}{\sqrt{2L-2}}$ from $1000$ data replications. The red solid line represents the densities of the standard normal distribution.}
	\label{fig:theorem:3-2}
\end{figure}

\begin{figure}[http]
	\centering
	\subfigure{\includegraphics[width=0.32\textwidth]{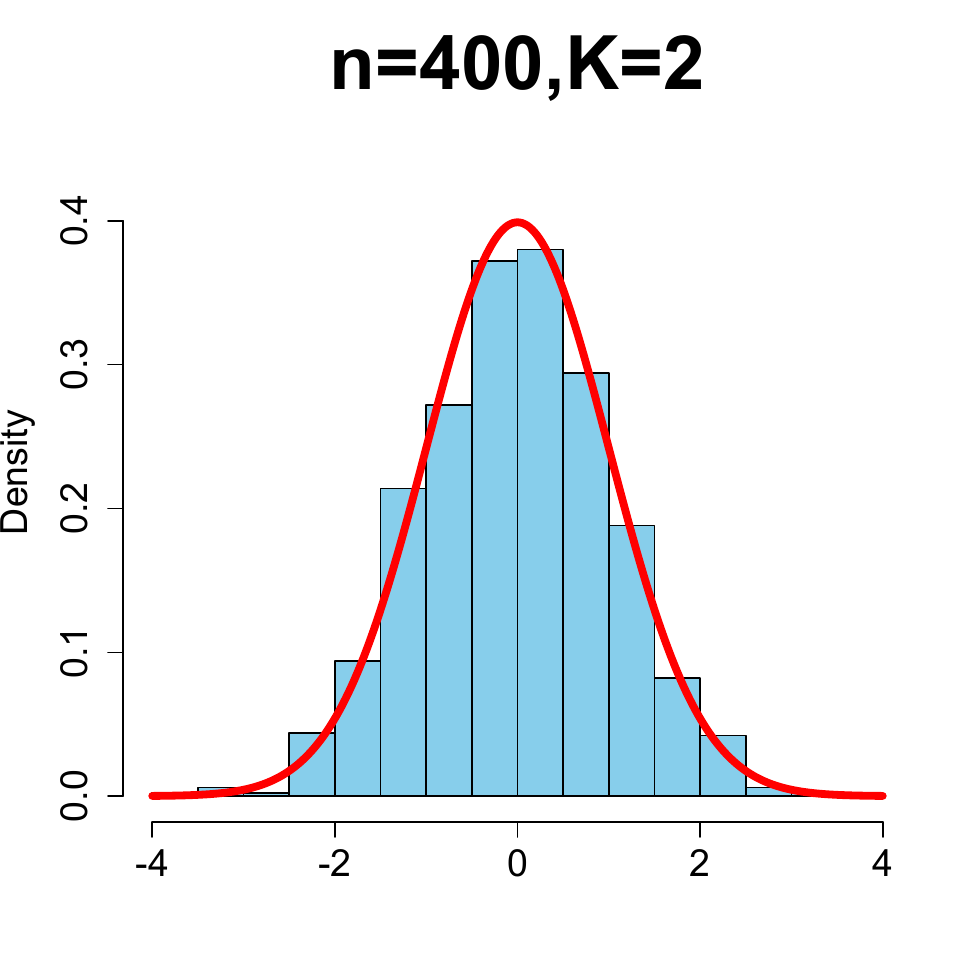}}
	\subfigure{\includegraphics[width=0.32\textwidth]{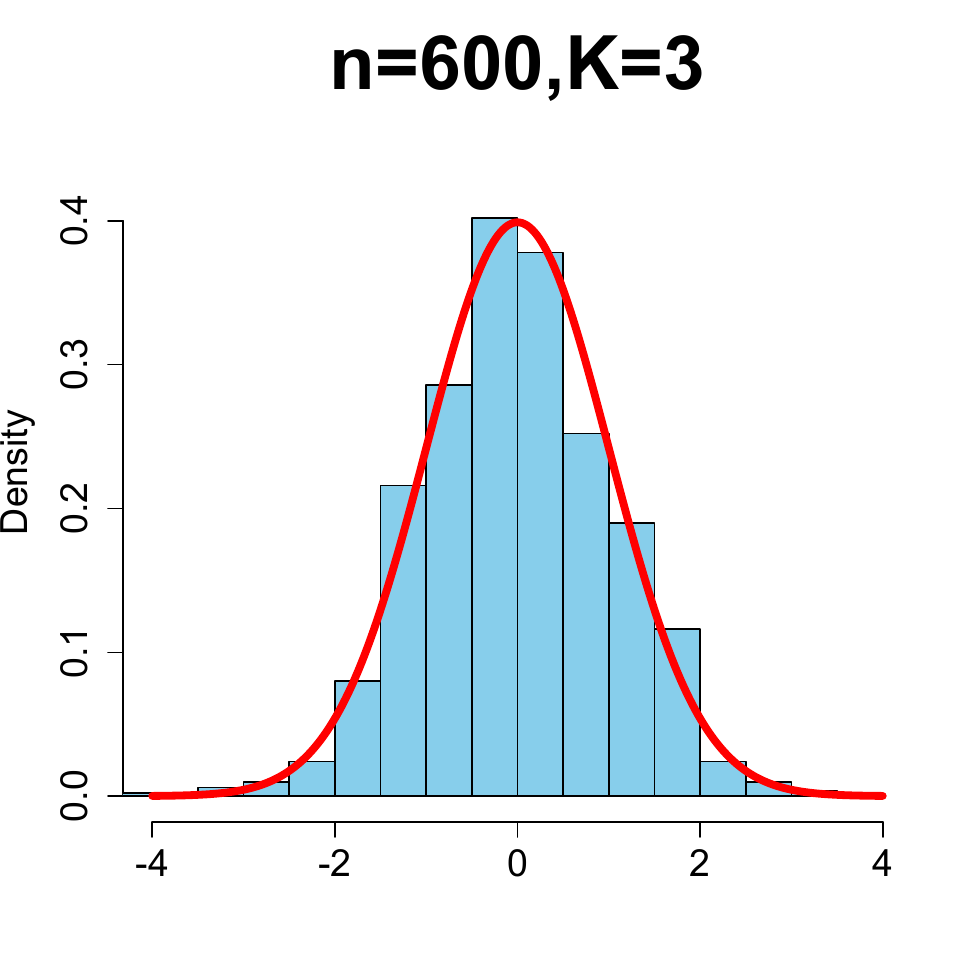}}
	\subfigure{\includegraphics[width=0.32\textwidth]{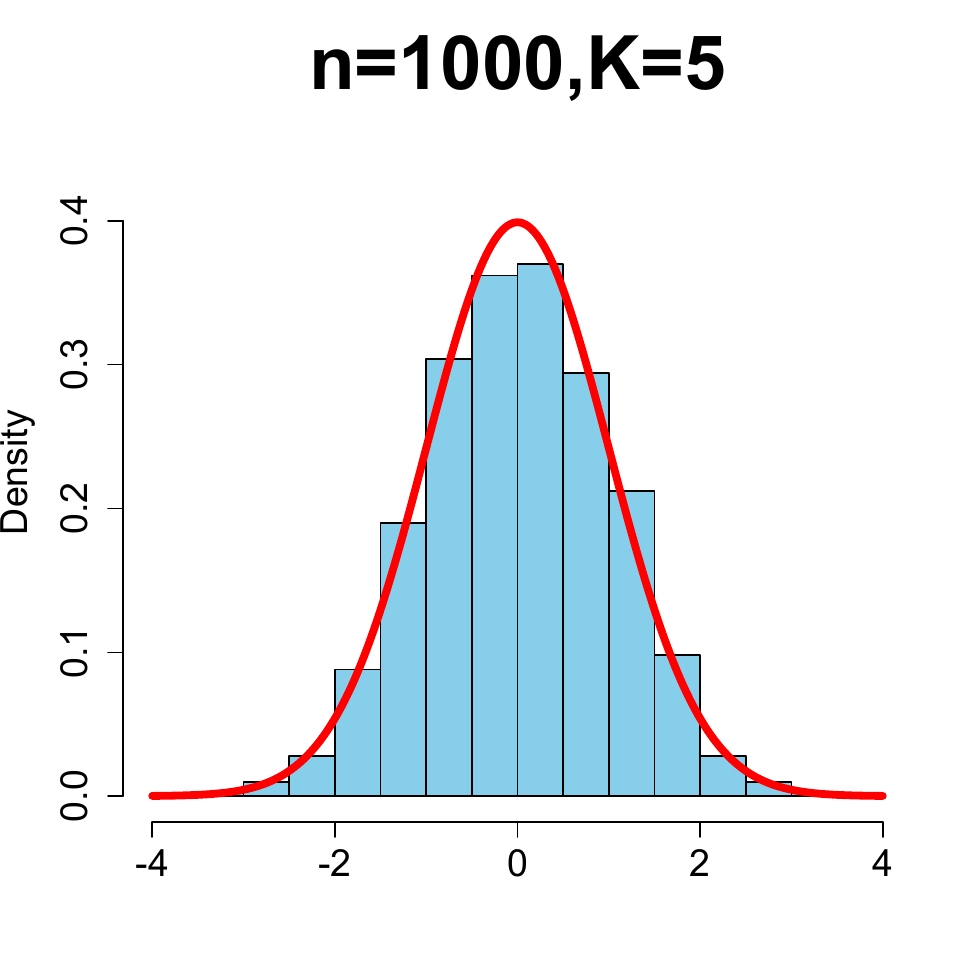}}
	\caption{Histograms of  $\frac{2\tr({\A^*}\check{\A})}{\sqrt{2L-2}}$ from $1000$ data replications. The red solid line represents the densities of the standard normal distribution.}
	\label{fig:theorem:3-3}
\end{figure}

In this subsection, we examine the limiting distributions of the cross terms induced by the normalized adjacency matrix under simple and composite perturbations.  The asymptotic properties of the cross terms $\tr({\A^*}{\bDelta})$, $\tr({\A^*}\tilde{\bDelta})$, and $\tr({\A^*}\check{\A})$ are established in Theorems \ref{theorem:2-2}-\ref{theorem:3-3}, respectively. We set the network size to $n =200K$ and consider $K \in \{2,3,5\}$  communities of equal size, i.e., $\pi_1 = \pi_2 = \ldots = \pi_K = 1/K$. The probability matrix is specified as
$\B = 0.1(1+3\times\mathbb{I}(u=v))$. Each configuration is simulated $1000$ times. Figure \ref{fig:theorem:2-2} shows the empirical distribution of $-\frac{n}{2}\tr({\A^*}{\bDelta})$, which is well approximated by a chi-square distribution with $K(K+1)/2$ degrees of freedom. Figures \ref{fig:theorem:3-2}-\ref{fig:theorem:3-3} display the empirical distributions of $2\tr({\A^*}\tilde{\bDelta})/\sqrt{2L-2}$ and $2\tr({\A^*}\check{\A})/\sqrt{2L-2}$, both of which align closely with the standard normal distribution.

These results highlight a fundamental shift in the asymptotic behavior of the cross term when transitioning from a simple to a composite perturbation.   Furthermore, under the composite perturbation setting, $\tr({\A^*}\tilde{\bDelta})$ and $\tr({\A^*}\check{\A})$ share the same limiting distribution.

\subsection{The Distribution of the Sum-of-Squares Terms}
\begin{figure}[http]
	\centering
	\subfigure{\includegraphics[width=0.32\textwidth]{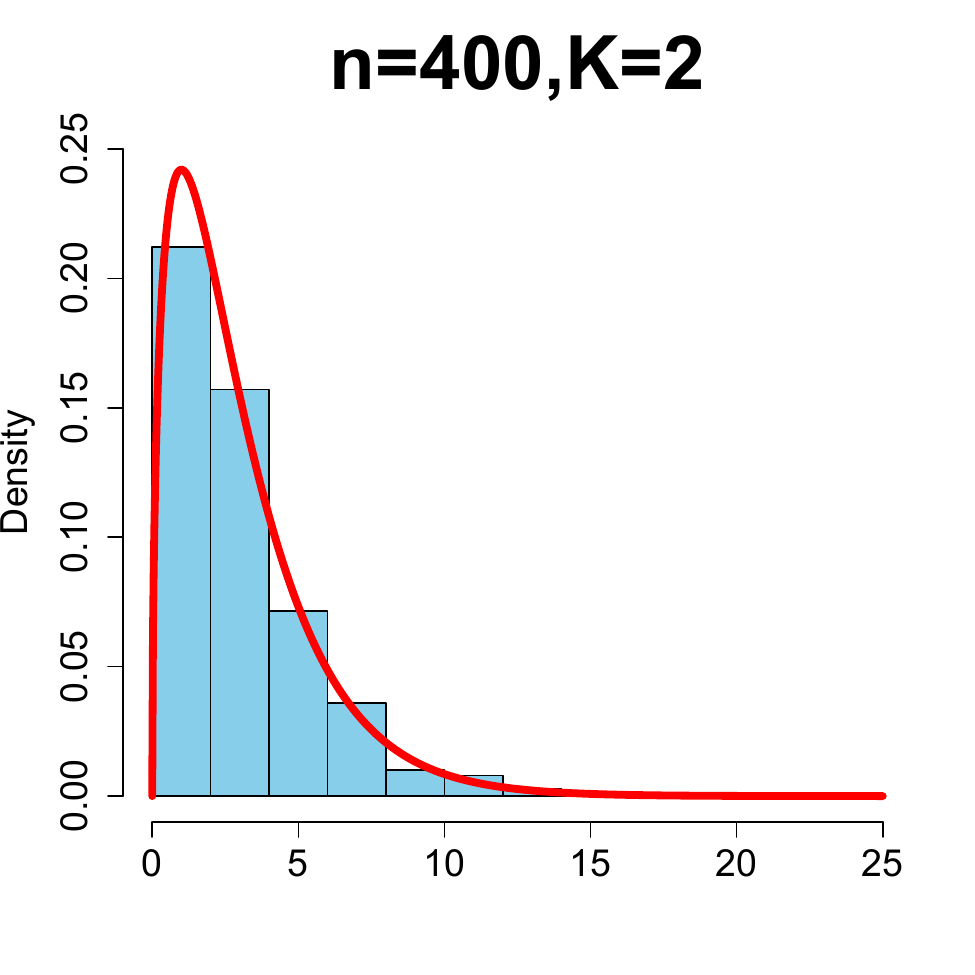}}
	\subfigure{\includegraphics[width=0.32\textwidth]{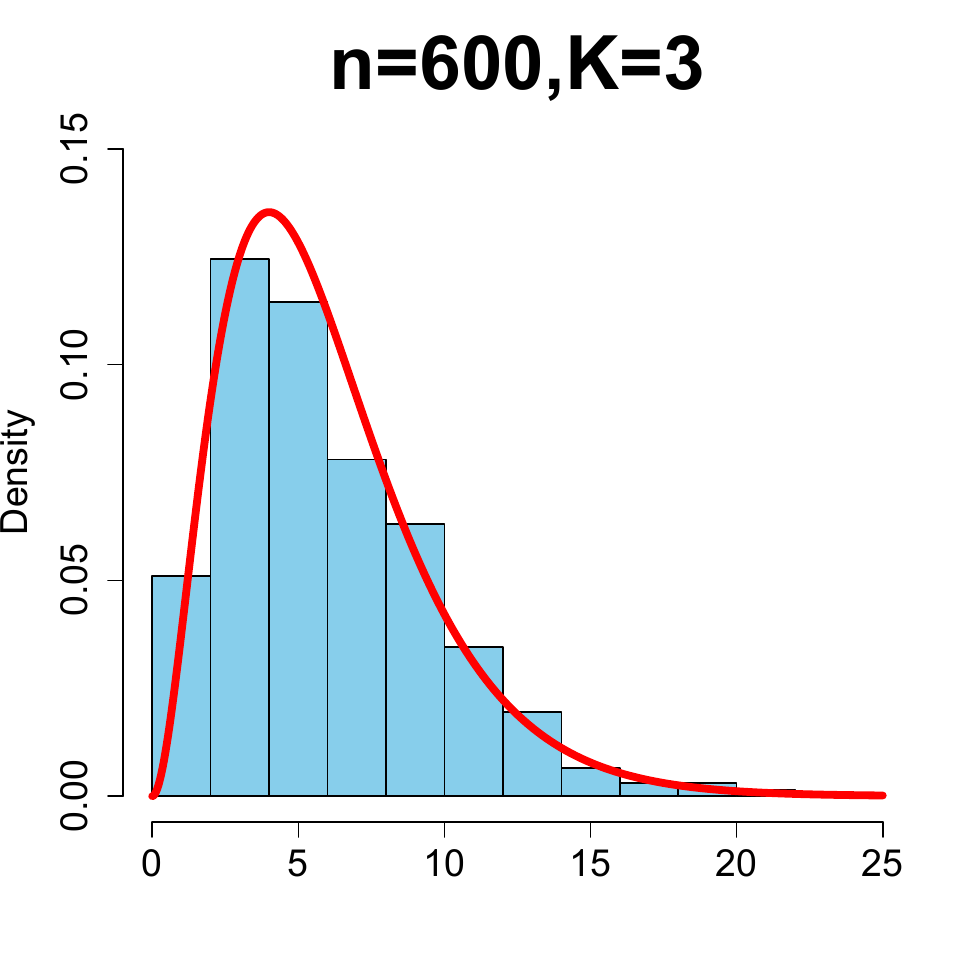}}
	\subfigure{\includegraphics[width=0.32\textwidth]{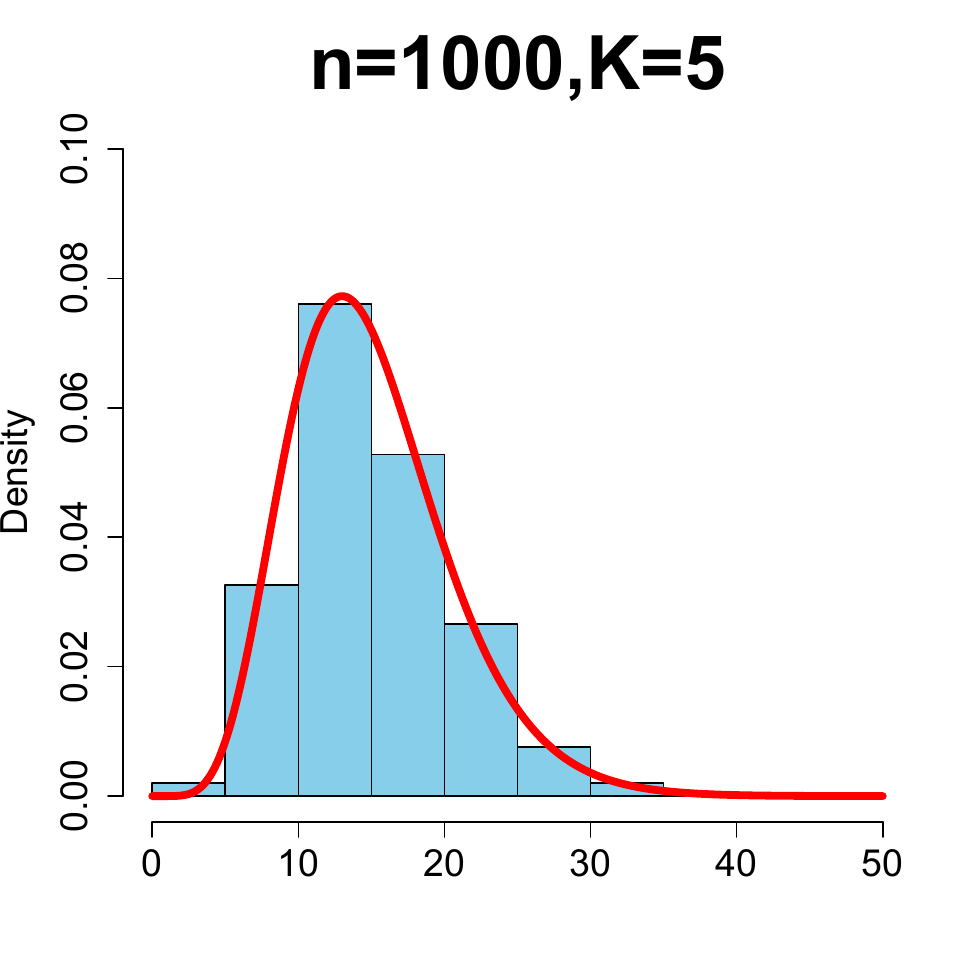}}
	\caption{Histograms of $\frac{n}{2}\tr({\bDelta}^2)$ from 1000 simulation replications. The red solid line represents the density of a chi-square distribution with $K(K+1)/2$ degrees of freedom.}
	\label{fig:lemma:2-3}
\end{figure}

\begin{figure}[http]
	\centering
	\subfigure{\includegraphics[width=0.32\textwidth]{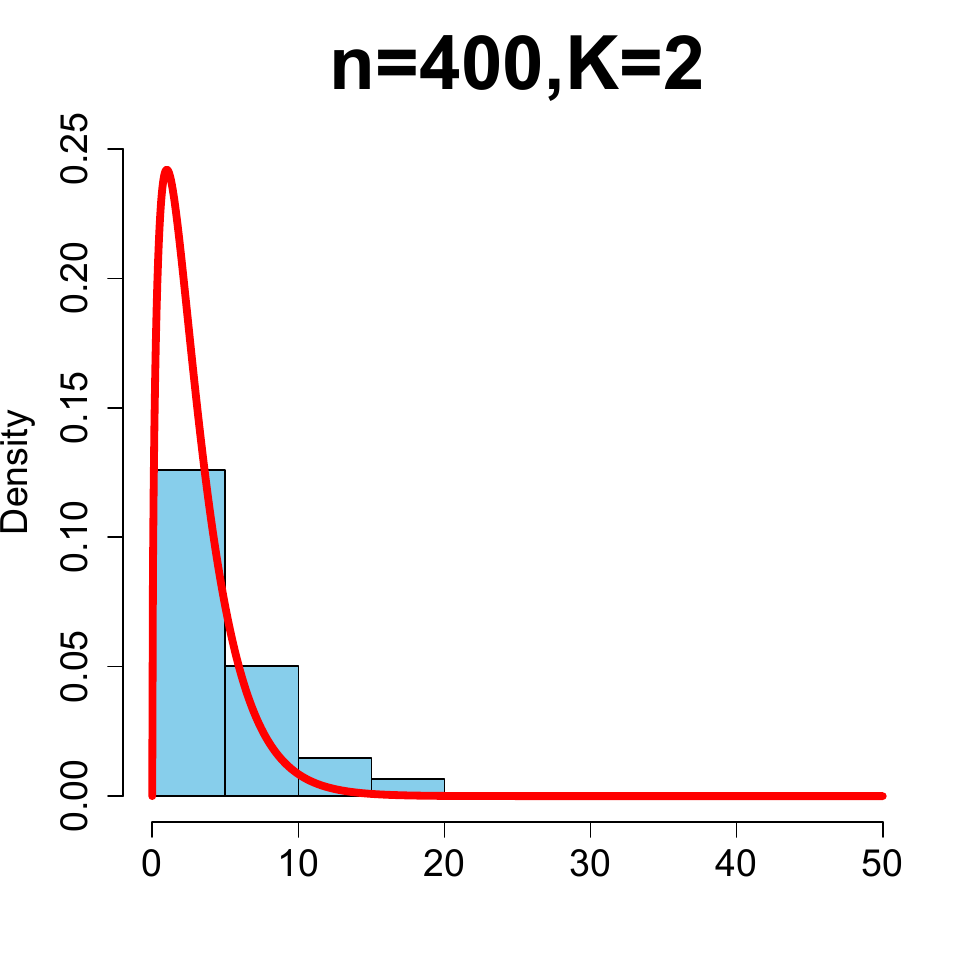}}
	\subfigure{\includegraphics[width=0.32\textwidth]{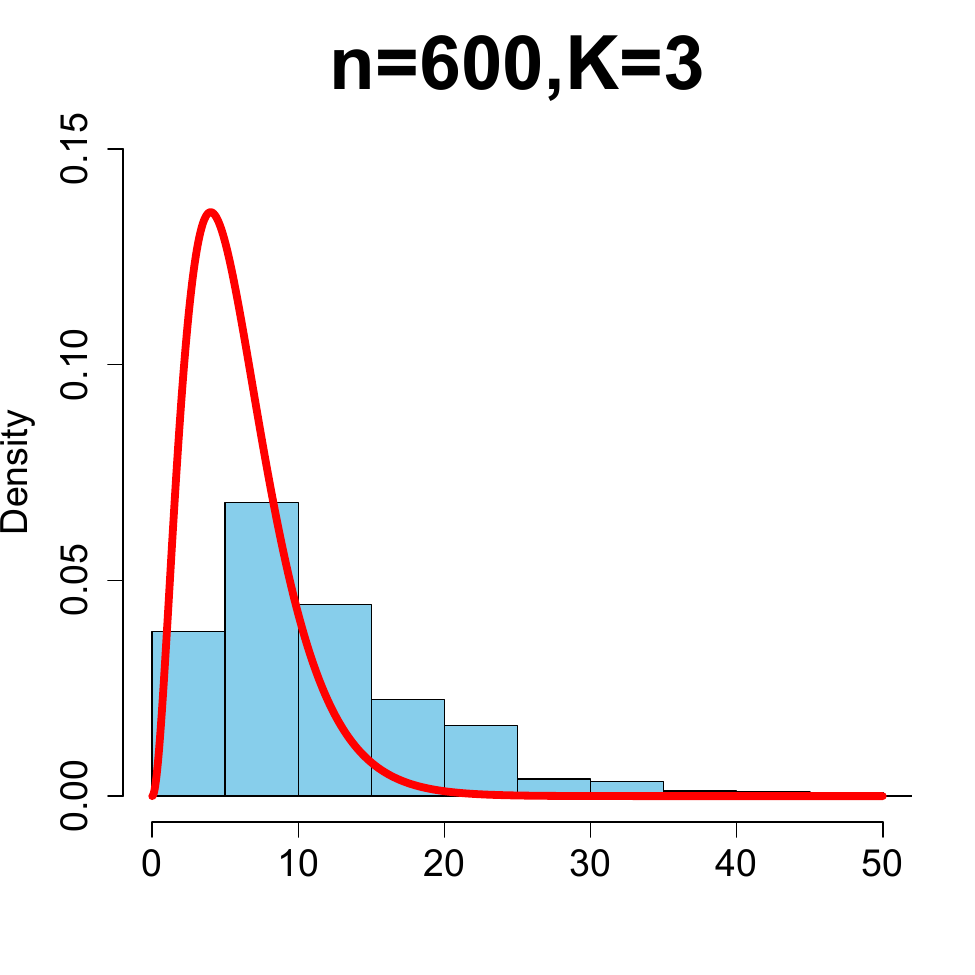}}
	\subfigure{\includegraphics[width=0.32\textwidth]{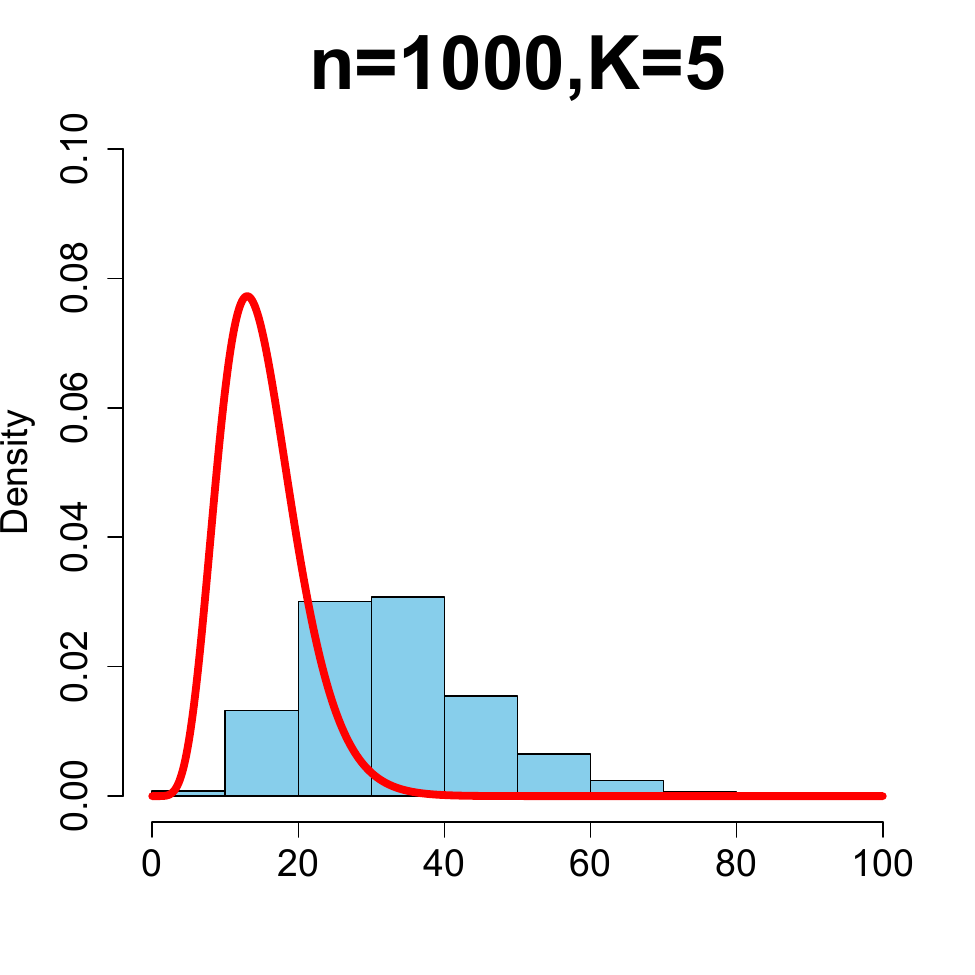}}
	\caption{Histograms of $\frac{n}{2}\tr(\tilde{\bDelta}^2)$ from 1000 simulation replications. The red solid line represents the density of a chi-square distribution with $K(K+1)/2$ degrees of freedom.}
	\label{fig:lemma:3-3}
\end{figure}

In this subsection, we compare the limiting distributions of the sum of squares induced by  the simple and composite perturbation matrices. The asymptotic distribution of $\tr({\bDelta}^2)$ is provided in Lemma \ref{lemma:2-3}. Network parameters follow the settings of Section \ref{sec:6-1}. Figures \ref{fig:lemma:2-3}
and \ref{fig:lemma:3-3} present the empirical distributions of $\frac{n}{2}\tr({\bDelta}^2)$ and $\frac{n}{2}\tr(\tilde{\bDelta}^2)$,  based on 1000 simulation replications, respectively.

 In contrast to $\tr({\bDelta}^2)$, the statistic $\tr(\tilde{\bDelta}^2)$ does not converge to a chi-square distribution with $K(K+1)/2$ degrees of freedom. This result demonstrates a fundamental divergence in the limiting behavior of the two sum-of-squares terms.

\section{Discussion}\label{Discussion}
In this paper, we have systematically investigated the impact of parameter estimation error on spectral-based inference for stochastic block models, focusing on the distinction between simple and composite perturbations in the normalized adjacency matrix. A key and perhaps surprising finding is that the cross term involving the composite perturbation matrix is asymptotically non-negligible, in stark contrast to its simple perturbation counterpart.  This motivated us to develop a unified decomposition framework for the composite perturbation matrix in stochastic block models. Applying this framework to the largest  eigenvalue statistic \cite{Lei:2016} and the linear spectral statistic \cite{Wu:2024},
we have derived more precise conclusions, which strengthens the theoretical guarantees for their use in practice.

Although a thorough power analysis was beyond the scope of \cite{Wu:2024}, the decomposition framework developed in this paper can be naturally extended to incorporate such an analysis. This provides a promising direction for future research, leading to a more complete theoretical understanding of the test's performance. The approach involves introducing a signal term into the framework, derived from the probability matrix under the alternative hypothesis, which is structurally analogous to constructing a term like $\mathbf{\Delta}$.

Furthermore, the unified decomposition framework developed in this paper is not confined to one-sample tests in undirected stochastic block models. It offers a principled approach to handle plug-in estimation errors in a variety of network inference settings, including two-sample testing problems \cite{Chen:2024} and directed networks \cite{Zhu:2025}.

Beyond stochastic block models,  our decomposition framework suggests a general principle for statistical inference with estimated parameters: composite perturbations introduce structurally distinct bias terms that must be isolated and controlled separately. This principle offers a possible path for analyzing plug-in estimation errors in a broader class of latent variable models and high-dimensional covariance estimation problems that rely on normalized matrices for inference.

\begin{appendix}

\section{Auxiliary Lemmas}

\begin{lemma}[Lei \cite{Lei:2016}]
	\label{lemma:appendix-2}
	Let $\A^*$ be given as in (\ref{eq:Astar}). Then, we have
	\begin{equation*}
		n^{2/3}(\lambda_1(\A^*)-2)\stackrel{d}{\rightarrow} TW_1,
	\end{equation*}
	where $TW_1$ denotes the Tracy-Widom distribution with index 1 and $\lambda_i(\A^*)$ denotes the $i$-th largest eigenvalue of the matrix $\A^*$.
\end{lemma}

\begin{lemma}[Wu and Hu \cite{Wu:2024}]
	\label{lemma:appendix-1}
	Let $\A^*$ be given as in (\ref{eq:Astar}). Then, we have
	\begin{equation*}
		\frac{1}{\sqrt{6}}\tr({\A^*}^{3})\stackrel{d}{\longrightarrow} N(0,1).
	\end{equation*}
\end{lemma}

\begin{lemma}[Vershynin \cite{Vershynin:2018}]
	\label{Vershynin:2018}
	Let $\M$ be an $p\times q$ random matrix with independent, mean-zero, sub-gaussian entries $M_{ij}$. Then, for any $t>0$, we have
	\[
	\|\M\|\leq C\sigma(\sqrt{p}+\sqrt{q}+t),
	\]
	with probability at least $1-2\exp(-t^2)$. Here $\sigma=\max_{i,j}\|M_{ij}\|_{\psi_2}$.
\end{lemma}

\section{Remarks}

\begin{remark}
	\label{remark:1}
	In fact, we have
	\begin{equation*}
		\begin{aligned}
			\hat{A}_{ij}&=\bar{A}_{ij}\sqrt{\frac{nP_{ij}(1-P_{ij})}{n\hat{P}_{ij}(1-\hat{P}_{ij})}}\\
			&=\bar{A}_{ij}\big(\sqrt{\frac{P_{ij}(1-P_{ij})}{\hat{P}_{ij}(1-\hat{P}_{ij})}}-1\big)+\bar{A}_{ij}\\
			&=\bar{A}_{ij}\frac{\frac{P_{ij}(1-P_{ij})}{\hat{P}_{ij}(1-\hat{P}_{ij})}-1}{\sqrt{\frac{P_{ij}(1-P_{ij})}{\hat{P}_{ij}(1-\hat{P}_{ij})}}+1}+\bar{A}_{ij}\\
			&=\bar{A}_{ij}\frac{\frac{(P_{ij}-\hat{P}_{ij})(1-\hat{P}_{ij}-P_{ij})}{\hat{P}_{ij}(1-\hat{P}_{ij})}}{\sqrt{\frac{P_{ij}(1-P_{ij})}{\hat{P}_{ij}(1-\hat{P}_{ij})}}+1}+\bar{A}_{ij}\\
			&=\dfrac{P_{ij}-\hat{P}_{ij}}{\sqrt{nP_{ij}(1-P_{ij})}}\frac{(A_{ij}-\hat{P}_{ij})(1-\hat{P}_{ij}-P_{ij})}{\hat{P}_{ij}(1-\hat{P}_{ij})(\sqrt{\frac{P_{ij}(1-P_{ij})}{\hat{P}_{ij}(1-\hat{P}_{ij})}}+1)}+\bar{A}_{ij}\\
			&=A^*_{ij}+\Delta_{ij}+\dfrac{P_{ij}-\hat{P}_{ij}}{\sqrt{nP_{ij}(1-P_{ij})}}\frac{(A_{ij}-\hat{P}_{ij})(1-\hat{P}_{ij}-P_{ij})}{\sqrt{\hat{P}_{ij}(1-\hat{P}_{ij}})(\sqrt{P_{ij}(1-P_{ij}})+\sqrt{\hat{P}_{ij}(1-\hat{P}_{ij}}))}\\
			&=A^*_{ij}+\Delta_{ij}+\gamma_{ij}\Delta_{ij},
		\end{aligned}
	\end{equation*}
	where $\gamma_{ij}=\frac{(A_{ij}-\hat{P}_{ij})(1-\hat{P}_{ij}-P_{ij})}{\sqrt{\hat{P}_{ij}(1-\hat{P}_{ij}})(\sqrt{P_{ij}(1-P_{ij}})+\sqrt{\hat{P}_{ij}(1-\hat{P}_{ij}}))}$.
\end{remark}

\eop

\begin{remark}
	\label{remark:2}
	In fact, we have
\begin{equation*}
	\begin{aligned}
		\tr({\A^*}\tilde{\bDelta})&=\sum_{i}({\A^*}\tilde{\bDelta})_{ii}\\
		&=\sum_{i}\sum_{j}A^*_{ij}\tilde{\Delta}_{ji}\\
		&=\sum_{i}\sum_{j}A^*_{ij}(1+\gamma_{ji})\Delta_{ji}\\
		&=\sum_{1\leq u, v\leq K}\sum_{i\in\mathcal{N}_{u}, j\in\mathcal{N}_{v}}A^*_{ij}\gamma_{ji}\Delta_{ji}+\tr({\A^*}\bDelta)\\
		&=\sum_{1\leq u, v\leq K}\frac{B_{uv}-\hat{B}_{uv}}{\sqrt{nB_{uv}(1-B_{uv})}}\sum_{i\in\mathcal{N}_{u}, j\in\mathcal{N}_{v}}\gamma_{ji}A^*_{ij}+\tr({\A^*}\bDelta).
	\end{aligned}
\end{equation*}
\end{remark}

\begin{remark}
	\label{remark:3}
	In fact, we have
	\begin{equation*}
		\begin{aligned}
			\hat{A}_{ij}&=\bar{A}_{ij}\sqrt{\frac{nP_{ij}(1-P_{ij})}{n\hat{P}_{ij}(1-\hat{P}_{ij})}}\\
			&=\bar{A}_{ij}\big(\sqrt{\frac{P_{ij}(1-P_{ij})}{\hat{P}_{ij}(1-\hat{P}_{ij})}}-1\big)+\bar{A}_{ij}\\
			&=\bar{A}_{ij}\frac{\frac{P_{ij}(1-P_{ij})}{\hat{P}_{ij}(1-\hat{P}_{ij})}-1}{\sqrt{\frac{P_{ij}(1-P_{ij})}{\hat{P}_{ij}(1-\hat{P}_{ij})}}+1}+\bar{A}_{ij}\\
			&=\bar{A}_{ij}\frac{\frac{(P_{ij}-\hat{P}_{ij})(1-\hat{P}_{ij}-P_{ij})}{\hat{P}_{ij}(1-\hat{P}_{ij})}}{\sqrt{\frac{P_{ij}(1-P_{ij})}{\hat{P}_{ij}(1-\hat{P}_{ij})}}+1}+\bar{A}_{ij}\\
			&=\bar{A}_{ij}\frac{(P_{ij}-\hat{P}_{ij})(1-\hat{P}_{ij}-P_{ij})}{\hat{P}_{ij}(1-\hat{P}_{ij})(\sqrt{\frac{P_{ij}(1-P_{ij})}{\hat{P}_{ij}(1-\hat{P}_{ij})}}+1)}+\bar{A}_{ij}\\
			&=\bar{A}_{ij}\frac{(P_{ij}-\hat{P}_{ij})(1-\hat{P}_{ij}-P_{ij})}{\sqrt{\hat{P}_{ij}(1-\hat{P}_{ij}})(\sqrt{P_{ij}(1-P_{ij}})+\sqrt{\hat{P}_{ij}(1-\hat{P}_{ij}}))}+\bar{A}_{ij}\\
			&=(1+\alpha_{ij})\bar{A}_{ij}\\
			&=(1+\alpha_{ij})(A_{ij}^*+\Delta_{ij})\\
			&=A_{ij}^*+\alpha_{ij}{A_{ij}^*}+\Delta_{ij}+\alpha_{ij}\Delta_{ij},
		\end{aligned}
	\end{equation*}
	where \[
	\alpha_{ij}=\frac{(P_{ij}-\hat{P}_{ij})(1-\hat{P}_{ij}-P_{ij})}{\sqrt{\hat{P}_{ij}(1-\hat{P}_{ij}})(\sqrt{P_{ij}(1-P_{ij}})+\sqrt{\hat{P}_{ij}(1-\hat{P}_{ij}}))}.
	\]
\end{remark}
\eop

\section{Proofs }

\subsection{Proof of Theorem \ref{theorem:2-2}}

By  direct calculation, we have
\begin{equation*}
	\begin{aligned}
		\tr\left(\A^*\bDelta\right)&=\sum_{i}\left(\A^*\bDelta\right)_{ii}\\
		&=\sum_{i}\sum_{j}{A_{ij}^*}\Delta_{ji}\\
		&=\sum_{1\leq u, v\leq K}\sum_{i\in\mathcal{N}_{u}, j\in\mathcal{N}_{v}}{A_{ij}^*}\Delta_{ji}\\
		&=\sum_{1\leq u, v\leq K}\frac{B_{uv}-\hat{B}_{uv}}{\sqrt{nB_{uv}(1-B_{uv})}}\sum_{i\in\mathcal{N}_{u}, j\in\mathcal{N}_{v}}{A_{ij}^*}\\
		&=\sum_{1\leq u, v\leq K}\frac{B_{uv}-\hat{B}_{uv}}{\sqrt{nB_{uv}(1-B_{uv})}}\sum_{i\in\mathcal{N}_{u}, j\in\mathcal{N}_{v}}{A_{ij}^*}\\
		&=\frac{1}{n}\sum_{1\leq u,v\leq K}\dfrac{(B_{uv}-\hat{B}_{uv})}{\sqrt{B_{uv}(1-B_{uv})}}\dfrac{\sum_{i\in\mathcal{N}_{u}, j\in\mathcal{N}_{v}}(A_{ij}-B_{uv})}{\sqrt{B_{uv}(1-B_{uv})}}\\
		&=-\frac{2}{n}\sum_{1\leq u\leq v\leq K}\dfrac{n_{uv}(B_{uv}-\hat{B}_{uv})^2}{B_{uv}(1-B_{uv})}\\
		&\triangleq -\frac{2}{n}\eta.
	\end{aligned}
\end{equation*}
For fixed $K$, by the central limit theory, we have
\begin{equation}
	\label{theorem:2-2-equation-1}
	\begin{aligned}
		\dfrac{\sqrt{n_{uv}}(B_{uv}-\hat{B}_{uv})}{\sqrt{B_{uv}(1-B_{uv})}}\stackrel{d}{\longrightarrow} N(0,1).
	\end{aligned}
\end{equation}
Thus, $\eta$ converges in distribution to a chi-square distribution with $K(K+1)/2$ degrees of freedom.

\eop

\subsection{Proof of Lemma \ref{lemma:2-3}}
By  direct calculation, we have
\begin{equation}
	\label{equation:3-3}
	\begin{aligned}
		\tr(\bDelta^2)&=\sum_{i, j}\Delta_{ij}^2\\
		&=\sum_{1\leq u, v\leq K}\sum_{i\in\mathcal{N}_{u}, j\in\mathcal{N}_{v}}\Delta_{ij}^2\\
		&=\frac{2}{n}\sum_{1\leq u\leq v\leq K}\dfrac{n_{uv}(B_{uv}-\hat{B}_{uv})^2}{B_{uv}(1-B_{uv})}\\
		&\triangleq \frac{2}{n}\eta.
	\end{aligned}
\end{equation}

By the central limit theory, $\eta$ converges in distribution to a chi-square distribution with $K(K+1)/2$ degrees of freedom.
\eop

\subsection{Proof of Corollary \ref{corollary:2-2}}
By spectral decomposition, we may express $\bDelta$ as
\begin{equation*}
	\begin{aligned}
		\bDelta&=\bGamma\bLambda\bGamma^\top,
	\end{aligned}
\end{equation*}
where $\bLambda=\Diag\{\lambda_1,\ldots,\lambda_n\}$, and $\bGamma=(\bGamma_1^\top,\ldots,\bGamma_n^\top)^\top$ is an $n\times n$ orthogonal matrix collecting the eigenvectors of $\bDelta$.

By Lemma \ref{lemma:2-3}, we have
\[
\tr(\bDelta^2)=\sum_{i}\lambda_i^2=O_p(\frac{K^2}{n}).
\]
Then, we have
\[
\|\bDelta\|=\max_{i}|\lambda_i|\leq \sqrt{\sum_{i}\lambda_i^2}=O_p(\frac{K}{\sqrt{n}}),
\]
where $\|\cdot\|$ denotes the spectral norm.

\eop

\subsection{Proof of Lemma \ref{lemma:3-2}}

By  direct calculation, we have
\begin{equation*}
	\begin{aligned}
		\tr(\hat{\A}^2)&=\sum_{i}\sum_{j}\hat{A}_{ij}^2\\
		&=\sum_{1\leq u, v\leq K}\sum_{i\in\mathcal{N}_{u}, j\in\mathcal{N}_{v}}\hat{A}_{ij}^2\\
		&=\sum_{1\leq u, v\leq K}\frac{1}{n\hat{B}_{uv}(1-\hat{B}_{uv})}\sum_{i\in\mathcal{N}_{u}, j\in\mathcal{N}_{v}}(A_{ij}-\hat{B}_{uv})^2\\
		&=\sum_{1\leq u, v\leq K}\frac{1}{n\hat{B}_{uv}(1-\hat{B}_{uv})}\sum_{i\in\mathcal{N}_{u}, j\in\mathcal{N}_{v}}(A_{ij}^2-2A_{ij}\hat{B}_{uv}+\hat{B}_{uv}^2)\\
		&=\sum_{1\leq u, v\leq K}\frac{1}{n\hat{B}_{uv}(1-\hat{B}_{uv})}\sum_{i\in\mathcal{N}_{u}, j\in\mathcal{N}_{v}}(A_{ij}-2A_{ij}\hat{B}_{uv}+\hat{B}_{uv}^2)\\
		&=\sum_{1\leq u\leq v\leq K}\frac{2n_{uv}\hat{B}_{uv}(1-\hat{B}_{uv})}{n\hat{B}_{uv}(1-\hat{B}_{uv})}\\
		&=\sum_{1\leq u\leq v\leq K}\frac{2n_{uv}}{n}=n-1.
	\end{aligned}
\end{equation*}

\eop

\subsection{Proof of Lemma \ref{lemma:3-3}}

By spectral decomposition, we may express $\tilde{\bDelta}$ as
\begin{equation*}
	\begin{aligned}
		\tilde{\bDelta}&=\tilde{\bGamma}\tilde{\bLambda}\tilde{\bGamma}^\top,
	\end{aligned}
\end{equation*}
where $\tilde{\bLambda}=\Diag\{\tilde{\lambda}_1,\ldots,\tilde{\lambda}_n\}$, and $\tilde{\bGamma}=(\tilde{\bGamma}_1^\top,\ldots,\tilde{\bGamma}_n^\top)^\top$ is an $n\times n$ orthogonal matrix collecting the eigenvectors of $\tilde{\bDelta}$.

By Corollary \ref{corollary:2-2}, we have
\begin{equation*}
	\begin{aligned}
		\sum_{i}\tilde{\lambda}_i^2&=\tr(\tilde{\bDelta}^2)\\
		&=\sum_{i, j}(1+\gamma_{ij})^2\Delta_{ij}^2\\
		&\leq (1+\gamma)^2\sum_{1\leq u, v\leq K}\sum_{i\in\mathcal{N}_{u}, j\in\mathcal{N}_{v}}\Delta_{ij}^2\\
		&=\frac{(1+\gamma)^2}{n}\sum_{1\leq u\leq v\leq K}\dfrac{2n_{uv}(\hat{B}_{uv}-B_{uv})^2}{B_{uv}(1-B_{uv})}\\
		&=(1+\gamma)^2\tr(\bDelta^2)\\
		&=O_p(\frac{K^2}{n}).
	\end{aligned}
\end{equation*}
Then, we have
\begin{equation*}
	\begin{aligned}
		\|\tilde{\bDelta}\|=\max_{i}|\tilde{\lambda}_i|\leq \sqrt{\sum_{i}\tilde{\lambda}_i^2}=O_p(\frac{K}{\sqrt{n}}),
	\end{aligned}
\end{equation*}
where $\|\cdot\|$ denotes the spectral norm.

\eop

\subsection{Proof of Theorem \ref{theorem:3-3}}

First, for $u\neq v$, we have
\begin{equation}
	\label{theorem:3-3-equation-2}
	\begin{aligned}
		\E(\sum_{i\in\mathcal{N}_{u}, j\in\mathcal{N}_{v}}{\A^*_{ij}}^2)=\frac{n_{uv}}{n},
	\end{aligned}
\end{equation}
and
\begin{equation}
	\label{theorem:3-3-equation-3}
	\begin{aligned}
		\sigma_{uv}^2&\triangleq\var(\sum_{i\in\mathcal{N}_{u}, j\in\mathcal{N}_{v}}{\A^*_{ij}}^2)\\
		&=\sum_{i\in\mathcal{N}_{u}, j\in\mathcal{N}_{v}}\big(\E{A^*_{ij}}^4-(\E{A^*_{ij}}^2)^2\big)\\
		&=\sum_{i\in\mathcal{N}_{u}, j\in\mathcal{N}_{v}}\big(\E{A^*_{ij}}^4-\frac{1}{n^2}\big)\\
		&=\sum_{i\in\mathcal{N}_{u}, j\in\mathcal{N}_{v}}\E{A^*_{ij}}^4-\frac{n_{uv}}{n^2}\\
		&=O(\frac{n_{uv}}{n^2}).
	\end{aligned}
\end{equation}
Combining \eqref{theorem:3-3-equation-2} and  \eqref{theorem:3-3-equation-3}, by the central limit theory, we have
\begin{equation*}
	\begin{aligned}
		\frac{\sum_{i\in\mathcal{N}_{u}, j\in\mathcal{N}_{v}}{\A^*_{ij}}^2-n_{uv}/n}{\sigma_{uv}}\stackrel{d}{\longrightarrow} N(0,1).
	\end{aligned}
\end{equation*}
That is
\begin{equation}
	\label{theorem:3-3-equation-4}
	\begin{aligned}
		\sum_{i\in\mathcal{N}_{u}, j\in\mathcal{N}_{v}}{\A^*_{ij}}^2&=\frac{n_{uv}}{n}+O_p(\sqrt{\frac{n_{uv}}{n^2}}).
	\end{aligned}
\end{equation}
Similarly, for $u=v$, we have
\begin{equation}
	\label{theorem:3-3-equation-04}
	\begin{aligned}
		\sum_{i\in\mathcal{N}_{u}, j\in\mathcal{N}_{u}}{\A^*_{ij}}^2&=\frac{2n_{uu}}{n}+O_p(\sqrt{\frac{n_{uu}}{n^2}}).
	\end{aligned}
\end{equation}

Next, by \eqref{theorem:3-3-equation-4} and \eqref{theorem:3-3-equation-04}, we have
\begin{equation}
	\label{theorem:3-3-equation-5}
	\begin{aligned}
		\tr({\A^*}\check{\A})&=\sum_{i}(\check{\A}{\A^*})_{ii}=\sum_{i}\sum_{j}\alpha_{ij}A^*_{ij}A^*_{ji}\\
		&=\sum_{i}\sum_{j}\alpha_{ij}{A^*_{ij}}^2=\sum_{1\leq u,v\leq K}\alpha_{uv}\sum_{i\in\mathcal{N}_{u}, j\in\mathcal{N}_{v}}{A^*_{ij}}^2\\
		&=2\sum_{1\leq u\leq v\leq K}\alpha_{uv}\big(\frac{n_{uv}}{n}+O_p(\sqrt{\frac{n_{uv}}{n^2}})\big)\\
		&=2\sum_{1\leq u\leq v\leq K}\frac{n_{uv}}{n}\alpha_{uv}+\sum_{1\leq u\leq v\leq K}\alpha_{uv}O_p(\sqrt{\frac{n_{uv}}{n^2}})\big).
	\end{aligned}
\end{equation}
By \eqref{equation:4-0}, \eqref{equation:3-3} and Corollary \ref{corollary:2-2}, we have
\begin{equation}
	\label{lemma:4-3-equation-3}
	\begin{aligned}
		2\sum_{1\leq u\leq v\leq K}n_{uv}\alpha_{uv}^2&\leq 2C_1\sum_{1\leq u\leq v\leq K}\dfrac{n_{uv}(B_{uv}-\hat{B}_{uv})^2}{B_{uv}(1-B_{uv})}\\
		&=C_1n\tr(\bDelta^2)\\
		&=O_p(K^2),
	\end{aligned}
\end{equation}
where
\[
C_1=\max_{u,v}\frac{B_{uv}(1-B_{uv})}{\hat{B}_{uv}(1-\hat{B}_{uv})\big(\sqrt{(B_{uv}(1-B_{uv})}+\sqrt{\hat{B}_{uv}(1-\hat{B}_{uv})}\big)^2}.
\]
By \eqref{lemma:4-3-equation-3}, we have
\begin{equation}
	\label{theorem:3-3-equation-6}
	\begin{aligned}
		\sum_{1\leq u\leq v\leq K}\alpha_{uv}O_p(\sqrt{\frac{n_{uv}}{n^2}})&\leq \sqrt{K^2\sum_{1\leq u\leq v\leq K}\alpha_{uv}^2O_p(\frac{n_{uv}}{n^2})}\\
		&\leq \sqrt{O_p(\frac{K^2}{n^2})\sum_{1\leq u\leq v\leq K}n_{uv}\alpha_{uv}^2}\\
		&=\sqrt{O_p(\frac{K^2}{n^2})O_p(K^2)}\\
		&=O_p(\frac{K^2}{n}).
	\end{aligned}
\end{equation}
By \eqref{theorem:3-3-equation-5} and \eqref{theorem:3-3-equation-6}, we have
\begin{equation}
	\label{theorem:3-3-equation-7}
	\begin{aligned}
		2\tr({\A^*}\check{\A})&=4\sum_{1\leq u\leq v\leq K}\frac{n_{uv}}{n}\alpha_{uv}+O_p(\frac{K^2}{n}).
	\end{aligned}
\end{equation}
By Taylor's expansion, we have
\begin{equation*}
	\begin{aligned}
		\frac{1-\hat{B}_{uv}-B_{uv}}{\sqrt{\hat{B}_{uv}(1-\hat{B}_{uv})}\big(\sqrt{(B_{uv}(1-B_{uv})}+\sqrt{\hat{B}_{uv}(1-\hat{B}_{uv})}\big)}-\frac{1-2B_{uv}}{2B_{uv}(1-B_{uv})}&=O_p(\hat{B}_{uv}-B_{uv}).
	\end{aligned}
\end{equation*}
Then, we have
\begin{equation}
	\label{theorem:3-3-equation-8}
	\begin{aligned}
		\alpha_{uv}&=(B_{uv}-\hat{B}_{uv})\frac{1-2B_{uv}}{2B_{uv}(1-B_{uv})}+O_p((B_{uv}-\hat{B}_{uv})^2).
	\end{aligned}
\end{equation}
By \eqref{theorem:2-2-equation-1} and \eqref{theorem:3-3-equation-8}  , we have
\begin{equation}
	\label{theorem:3-3-equation-9}
	\begin{aligned}
		2\tr({\A^*}\check{\A})&=4\sum_{1\leq u\leq v\leq K}\frac{n_{uv}}{n}\alpha_{uv}\\
		&=\frac{2}{n}\sum_{1\leq u\leq v\leq K}n_{uv}(B_{uv}-\hat{B}_{uv})\frac{1-2B_{uv}}{B_{uv}(1-B_{uv})}\\
		&+O_p(\frac{1}{n}\sum_{1\leq u\leq v\leq K}n_{uv}(B_{uv}-\hat{B}_{uv})^2)\\
		&=\frac{2}{n}\sum_{1\leq u\leq v\leq K}n_{uv}(B_{uv}-\hat{B}_{uv})\frac{1-2B_{uv}}{B_{uv}(1-B_{uv})}+O_p(\frac{K^2}{n})\\
		&\triangleq\eta'+O_p(\frac{K^2}{n}).
	\end{aligned}
\end{equation}

Third, by \eqref{theorem:3-3-equation-9}, we have
\begin{equation}
	\label{theorem:3-3-equation-10}
	\begin{aligned}
		\E(\eta')=0.
	\end{aligned}
\end{equation}
We also have
\begin{equation}
	\label{theorem:3-3-equation-11}
	\begin{aligned}
		\var(\eta')&=\frac{4}{n^2}\sum_{1\leq u\leq v\leq K}\var\big(n_{uv}(B_{uv}-\hat{B}_{uv})\big)\big(\frac{1-2B_{uv}}{B_{uv}(1-B_{uv})}\big)^2\\
		&=\frac{4}{n^2}\sum_{1\leq u\leq v\leq K}n_{uv}\frac{(1-2B_{uv})^2}{B_{uv}(1-B_{uv})}\\
		&=\frac{4}{n^2}\sum_{1\leq u\leq v\leq K}n_{uv}\big(\frac{3B_{uv}^2-3B_{uv}+1}{B_{uv}(1-B_{uv})}-1\big)\\
		&=\frac{4}{n^2}\sum_{1\leq u\leq v\leq K}n_{uv}\frac{3B_{uv}^2-3B_{uv}+1}{B_{uv}(1-B_{uv})}-\frac{4}{n^2}\sum_{1\leq u\leq v\leq K}n_{uv}\\
		&=\frac{4}{n^2}\sum_{1\leq u\leq v\leq K}n_{uv}\frac{B_{uv}^3+(1-B_{uv})^3}{B_{uv}(1-B_{uv})}-2+\frac{2}{n}\\
		&\rightarrow 2L-2.
	\end{aligned}
\end{equation}

Finally, by \eqref{theorem:3-3-equation-7}, \eqref{theorem:3-3-equation-9}, \eqref{theorem:3-3-equation-10} and \eqref{theorem:3-3-equation-11},  by the central limit theory, we have
\[
\frac{2\tr({\A^*}\check{\A})}{\sqrt{2L-2}}\stackrel{d}{\longrightarrow} N(0,1).
\]

\eop

\subsection{Proof of Lemma \ref{lemma:4-1}}

It is known that
\begin{equation}
	\label{lemma:4-1-equation-1}
	\begin{aligned}
		\tr(\bUpsilon^2)&=\sum_{i}\frac{(P_{ii}-\hat{P}_{ii})^2}{nP_{ii}(1-P_{ii})}\\
		&=\sum_{1\leq u\leq K}\sum_{i\in\mathcal{N}_{u}}\frac{(P_{ii}-\hat{P}_{ii})^2}{nP_{ii}(1-P_{ii})}\\
		&=\frac{1}{n}\sum_{1\leq u\leq K}\frac{n_{u}(B_{uu}-\hat{B}_{uu})^2}{B_{uu}(1-B_{uu})}.
	\end{aligned}
\end{equation}
By Assumption  \ref{assumption:2} and \eqref{theorem:2-2-equation-1}, we have
\begin{equation}
	\label{lemma:4-1-equation-3}
	\begin{aligned}
		\frac{n_{u}(B_{uu}-\hat{B}_{uu})^2}{B_{uu}(1-B_{uu})}=\frac{n_{u}}{n_{uu}}\frac{n_{uu}(B_{uu}-\hat{B}_{uu})^2}{B_{uu}(1-B_{uu})}=O_p(\frac{K}{n}).
	\end{aligned}
\end{equation}
By \eqref{lemma:4-1-equation-1} and \eqref{lemma:4-1-equation-3}, we have
\begin{equation}
	\label{lemma:4-1-equation-4}
	\begin{aligned}
		\tr(\bUpsilon^2)=\sum_{i}\lambda_i^2(\bUpsilon)=O_p(\frac{K^2}{n^2}).
	\end{aligned}
\end{equation}

Thus,  we have
\[
\max_{i}|\lambda_i(\bUpsilon)|\leq \sqrt{\sum_{i}\lambda_i^2(\bUpsilon)}=O_p(\frac{K}{n}).
\]

Next, by Corollary \ref{corollary:2-2} and \eqref{lemma:4-1-equation-4}, we have
\begin{equation}
	\label{lemma:4-1-equation-6}
	\begin{aligned}
		\sum_{i}\bar{\lambda}_i^2&=\tr(\bar{\bDelta}^2)\\
		&=\sum_{i, j}\Delta_{ij}^2+\sum_{1\leq u\leq K}\sum_{i\in\mathcal{N}_{u}}\Upsilon_{i}^2\\
		&=\sum_{1\leq u, v\leq K}\sum_{i\in\mathcal{N}_{u}, j\in\mathcal{N}_{v}}\Delta_{ij}^2+\sum_{1\leq u\leq K}\sum_{i\in\mathcal{N}_{u}}\Upsilon_{i}^2\\
		&={\frac{1}{n}\sum_{1\leq u\leq v\leq K}\dfrac{2n_{uv}(B_{uv}-\hat{B}_{uv})^2}{B_{uv}(1-B_{uv})}}+\frac{1}{n}\sum_{1\leq u\leq K}\frac{n_{u}(B_{uu}-\hat{B}_{uu})^2}{B_{uu}(1-B_{uu})}\\
		&=\tr(\bDelta^2)+\tr(\bUpsilon^2)\\
		&=O_p(\frac{K^2}{n})+O_p(\frac{K^2}{n^2})\\
		&=O_p(\frac{K^2}{n}).
	\end{aligned}
\end{equation}
By \eqref{lemma:4-1-equation-6}, we have
\begin{equation*}
	\begin{aligned}
		\|\bar{\bDelta}\|=\max_{i}|\bar{\lambda}_i|\leq \sqrt{\sum_{i}\bar{\lambda}_i^2}=O_p(\frac{K}{\sqrt{n}}).
	\end{aligned}
\end{equation*}

Since $\bTheta^T\bTheta=\I_{K}$, we have
\begin{equation*}
	\begin{aligned}
		\bSigma=\bTheta^T\bar{\bDelta}\bTheta.
	\end{aligned}
\end{equation*}
That is
\begin{equation*}
	\begin{aligned}
		\|\bSigma\|\leq \|\bar{\bDelta}\|=O_p(\frac{K}{\sqrt{n}}).
	\end{aligned}
\end{equation*}

\eop

\subsection{Proof of Corollary \ref{corollary:4-0}}
By spectral decomposition, we may express $\bar{\bDelta}$ as
\begin{equation*}
	\begin{aligned}
		\bar{\bDelta}&=\bar{\bGamma}\bar{\bLambda}\bar{\bGamma}^\top,
	\end{aligned}
\end{equation*}
where $\bar{\bLambda}=\Diag\{\bar{\lambda}_1,\ldots,\bar{\lambda}_n\}$, and $\bar{\bGamma}=(\bar{\bGamma}_1^\top,\ldots,\bar{\bGamma}_n^\top)^\top$ is an $n\times n$ orthogonal matrix collecting the eigenvectors of $\bar{\bLambda}$. Let $|\bar{\bDelta}|=\bar{\bGamma}|\bar{\bLambda}|\bar{\bGamma}^\top=\sum_{i}|\bar{\lambda}_i|\bar{\bGamma}_i\bar{\bGamma}_i^\top$. By spectral decomposition, we may express ${\A^*}$ as
\begin{equation*}
	\label{corollary:4-0-equation-5}
	\begin{aligned}
		{\A^*}&=\U\bOmega\U^\top,
	\end{aligned}
\end{equation*}
where $\bOmega=\Diag\{\omega_1,\ldots,\omega_n\}$, and $\U=(\U_1^\top,\ldots,\U_n^\top)^\top$ is an $n\times n$ orthogonal matrix collecting the eigenvectors of ${\A^*}$.

First, similar to the proof of (19) in \cite{Lei:2016}, by  Lemma \ref{lemma:4-1}, we have
\begin{equation}
	\label{corollary:4-0-equation-1}
	\begin{aligned}
		\max_{j}\U_j^\top|\bar{\bDelta}|\U_j=O_p(K^2n^{-\frac{3}{2}}).
	\end{aligned}
\end{equation}

Next, similar to the proof of (20) in \cite{Lei:2016}, by \eqref{corollary:4-0-equation-1}, we have
\begin{equation}
	\label{corollary:4-0-equation-2}
	\begin{aligned}
		\lambda_1(\bar{\A}+\bUpsilon)&\geq\lambda_1({\A^*})+\U_1^\top\bar{\bDelta}\U_1\\
		&\geq\lambda_1({\A^*})+\U_1^\top|\bar{\bDelta}|\U_1\\
		&\geq\lambda_1({\A^*})-\max_{j}\U_j^\top|\bar{\bDelta}|\U_j\\
		&\geq\lambda_1({\A^*})-O_p(K^2n^{-\frac{3}{2}})\\
		&=\lambda_1({\A^*})-o_p(n^{-\frac{3}{2}}).
	\end{aligned}
\end{equation}
Similar to the proof of (21) in \cite{Lei:2016}, we have
\begin{equation}
	\label{corollary:4-0-equation-3}
	\begin{aligned}
		\lambda_1(\bar{\A}+\bUpsilon)&\leq\lambda_1({\A^*})+O_p(K^{\frac{7}{2}}n^{-\frac{5}{4}})\\
		&=\lambda_1({\A^*})+o_p(n^{-\frac{2}{3}}).
	\end{aligned}
\end{equation}
By \eqref{corollary:4-0-equation-2} and \eqref{corollary:4-0-equation-3}, we have
\begin{equation*}
	\begin{aligned}
		\lambda_1(\bar{\A}+\bUpsilon)&=\lambda_1({\A^*})+o_p(n^{-\frac{2}{3}}).
	\end{aligned}
\end{equation*}
By  Lemma \ref{lemma:4-1},  we have
\begin{equation}
	\label{corollary:4-0-equation-4}
	\begin{aligned}
		\lambda_1(\bar{\A})=\lambda_1({\A^*})+o_p(n^{-\frac{3}{2}}).
	\end{aligned}
\end{equation}

Finally, by \eqref{corollary:4-0-equation-4} and   Lemma \ref{lemma:appendix-2},  we get the result.

\eop

\subsection{Proof of Lemma \ref{lemma:4-01}}
By \eqref{lemma:4-1-equation-4}, we have
\begin{equation}
	\label{lemma:4-01-equation-1}
	\begin{aligned}
		\tr(\check{\bUpsilon}^2)&=\sum_{i} (1+\alpha_{ii})^2\frac{(P_{ii}-\hat{P}_{ii})^2}{nP_{ii}(1-P_{ii})}\\
		&\leq (1+\alpha)^2\sum_{1\leq u\leq K}\sum_{i\in\mathcal{N}_{u}}\frac{(P_{ii}-\hat{P}_{ii})^2}{nP_{ii}(1-P_{ii})}\\
		&=\frac{(1+\alpha)^2}{n}\sum_{1\leq u\leq K}\frac{n_{u}(B_{uu}-\hat{B}_{uu})^2}{B_{uu}(1-B_{uu})}\\
		&=(1+\alpha)^2\tr(\bUpsilon^2)\\
		&=O_p(\frac{K^2}{n^2}).
	\end{aligned}
\end{equation}
By \eqref{lemma:4-01-equation-1},  we have
\begin{equation}
	\label{lemma:4-01-equation-2}
	\begin{aligned}
		\max_{i}|\lambda_i(\check{\bUpsilon})|\leq \sqrt{\sum_{i}\lambda_i^2(\check{\bUpsilon})}=O_p(\frac{K}{n}).
	\end{aligned}
\end{equation}

Next, by Corollary \ref{corollary:2-2} and  \eqref{lemma:4-1-equation-4}, we have
\begin{equation*}
	\begin{aligned}
		\sum_{i}\hat{\lambda}_i^2&=\tr(\hat{\bDelta}^2)\\
		&=\sum_{i, j}(1+\alpha_{ij})^2\Delta_{ij}^2+\sum_{1\leq u\leq K}\sum_{i\in\mathcal{N}_{u}}(1+\alpha_{ii})^2\Upsilon_{i}^2\\
		&\leq(1+\alpha)^2\big(\sum_{1\leq u, v\leq K}\sum_{i\in\mathcal{N}_{u}, j\in\mathcal{N}_{v}}\Delta_{ij}^2+\sum_{1\leq u\leq K}\sum_{i\in\mathcal{N}_{u}}\Upsilon_{i}^2\big)\\
		&=(1+\alpha)^2\big(\frac{2}{n}\sum_{1\leq u\leq v\leq K}\dfrac{n_{uv}(B_{uv}-\hat{B}_{uv})^2}{B_{uv}(1-B_{uv})}+\frac{1}{n}\sum_{1\leq u\leq K}\frac{n_{u}(B_{uu}-\hat{B}_{uu})^2}{B_{uu}(1-B_{uu})}\big)\\
		&=(1+\alpha)^2\big(\tr(\bDelta^2)+\tr(\bUpsilon^2)\big)\\
		&=O_p(\frac{K^2}{n})+O_p(\frac{K^2}{n^2})\\
		&=O_p(\frac{K^2}{n}).
	\end{aligned}
\end{equation*}
Thus, we have
\begin{equation*}
	\begin{aligned}
		\|\hat{\bDelta}\|=\max_{i}|\hat{\lambda}_i|\leq \sqrt{\sum_{i}\hat{\lambda}_i^2}=O_p(\frac{K}{\sqrt{n}}).
	\end{aligned}
\end{equation*}

Since $\bTheta^T\bTheta=\I_{K}$,  we have
\begin{equation*}
	\begin{aligned}
		\hat{\bSigma}=\bTheta^T\hat{\bDelta}\bTheta.
	\end{aligned}
\end{equation*}
That is
\begin{equation*}
	\begin{aligned}
		\|\hat{\bSigma}\|\leq \|\hat{\bDelta}\|=O_p(\frac{K}{\sqrt{n}}).
	\end{aligned}
\end{equation*}

\eop

\subsection{Proof of Lemma \ref{lemma:4-7}}

Consider the block representation of ${\A^*}$:
\[
{\A^*}=({\A^*}_{uv})_{1\leq u,v\leq K},
\]
where ${\A^*}_{uv}$ is the  sub-matrix corresponding to the rows in the $u$-th community  and
columns in the $v$-th community. For ${\A^*}_{uv}$, noting that $p=n_u$, $q=n_v$, $\sigma=O(\frac{1}{\sqrt{n}})$, by Lemma \ref{Vershynin:2018}, for any $t>0$, we have
\begin{equation*}
	\begin{aligned}
		\mathbb{P}\left(\|{\A^*}_{uv}\|>C(\sqrt{\frac{n_u}{n}}+\sqrt{\frac{n_v}{n}}+\frac{t}{\sqrt{n}})\right)&\leq 2\exp\{-t^2\}.
	\end{aligned}
\end{equation*}
Then, for any $t>0$, we have
\begin{equation*}
	\begin{aligned}
		\mathbb{P}\left(\max_{u,v}\|{\A^*}_{uv}\|>C\sqrt{\frac{\max_kn_k}{n}}+t\right)&\leq \sum_{u,v}\mathbb{P}\left(\|{\A^*}_{uv}\|>C\sqrt{\frac{\max_kn_k}{n}}+t\right)\\
		&\leq 2K^2\exp\{-cnt^2\}\\
		&\leq 2\exp\{2\log K-cnt^2\}.
	\end{aligned}
\end{equation*}
That is
\begin{equation*}
	\begin{aligned}
		\max_{u,v}\|{\A^*}_{uv}\|=O_p(\sqrt{\frac{\max_kn_k}{n}}+\sqrt{\frac{\log K}{n}})\\
	\end{aligned}
\end{equation*}
By Assumption \ref{assumption:0}, we have

\begin{equation*}
	\begin{aligned}
		\max_{u,v}\|{\A^*}_{uv}\|=O_p(\log^{-1/2} K).
	\end{aligned}
\end{equation*}
Thus, we have
\begin{equation*}
	\begin{aligned}
		\|\check{\A}\|&=\|(\alpha_{ij}{\A^*}_{ij})_{n\times n}\|\\
		&\leq K\max_{u,v}|\alpha_{uv}|\|{\A^*}_{uv}\|\\
		&=O_p(\frac{K^2\log^{1/2} K}{n})\max_{u,v}\|{\A^*}_{uv}\|\\
		&\leq O_p(\frac{K^2\log^{1/2} K}{n})O_p(\log^{-1/2} K)\\
		&=O_p(\frac{K^{2}}{n}).
	\end{aligned}
\end{equation*}

\eop

\subsection{Proof of Theorem \ref{theorem:4-2}}
By spectral decomposition, we may express $\hat{\bDelta}$ as
\begin{equation*}
	\begin{aligned}
		\hat{\bDelta}&=\hat{\bGamma}\hat{\bLambda}\hat{\bGamma}^\top,
	\end{aligned}
\end{equation*}
where $\hat{\bLambda}=\Diag\{\hat{\lambda}_1,\ldots,\hat{\lambda}_n\}$, and $\hat{\bGamma}=(\hat{\bGamma}_1^\top,\ldots,\hat{\bGamma}_n^\top)^\top$ is an $n\times n$ orthogonal matrix collecting the eigenvectors of $\hat{\bDelta}$. Let $|\hat{\bDelta}|=\hat{\bGamma}|\hat{\bLambda}|\hat{\bGamma}^\top=\sum_{i}|\hat{\lambda}_i|\hat{\bGamma}_i\hat{\bGamma}_i^\top$. Recall that
\begin{equation*}
	\begin{aligned}
		{\A^*}&=\U\bOmega\U^\top,
	\end{aligned}
\end{equation*}
where $\bOmega=\Diag\{\omega_1,\ldots,\omega_n\}$, and $\U=(\U_1^\top,\ldots,\U_n^\top)^\top$ is an $n\times n$ orthogonal matrix collecting the eigenvectors of ${\A^*}$.

Similar to the proof of (19) in \cite{Lei:2016}, by Lemma \ref{lemma:4-01}, we have
\begin{equation}
	\label{theorem:4-2-equation-6}
	\begin{aligned}
		\max_{j}\U_j^\top|\hat{\bDelta}|\U_j=O_p(K^2n^{-\frac{3}{2}}).
	\end{aligned}
\end{equation}
Similar to the proof of (20) in \cite{Lei:2016}, by \eqref{lemma:4-01-equation-2}, \eqref{theorem:4-2-equation-6} and Lemma \ref{lemma:4-7}, we have
\begin{equation}
	\label{theorem:4-2-equation-7}
	\begin{aligned}
		\lambda_1(\hat{\A})&\geq\lambda_1({\A^*})+\U_1^\top\check{\A}\U_1+\U_1^\top(\bDelta+\check{\bDelta})\U_1\\
		&=\lambda_1({\A^*})+\U_1^\top\check{\A}\U_1+\U_1^\top(\bDelta+\check{\bDelta}+\check{\bUpsilon})\U_1-\U_1^\top\check{\bUpsilon}\U_1\\
		&\geq\lambda_1({\A^*})+\U_1^\top\check{\A}\U_1-\max_{j}|\U_j^\top(\bDelta+\check{\bDelta}+\check{\bUpsilon})\U_j|-\U_1^\top\check{\bUpsilon}\U_1\\
		&\geq\lambda_1({\A^*})+\U_1^\top\check{\A}\U_1-\max_{j}\U_j^\top|\hat{\bDelta}|\U_j-\U_1^\top\check{\bUpsilon}\U_1\\
		&\geq\lambda_1({\A^*})-o_p(n^{-\frac{2}{3}})-O_p(K^2n^{-\frac{3}{2}})-O_p(\frac{K}{n})\\
		&=\lambda_1({\A^*})-o_p(n^{-\frac{2}{3}}).
	\end{aligned}
\end{equation}
Similar to the proof of (21) in \cite{Lei:2016} and \eqref{theorem:4-2-equation-7}, by \eqref{lemma:4-01-equation-2}, \eqref{theorem:4-2-equation-6} and Lemma \ref{lemma:4-7}, we have
\begin{equation}
	\label{theorem:4-2-equation-8}
	\begin{aligned}
		\lambda_1(\hat{\A})&\leq\lambda_1({\A^*})+O_p(K^{\frac{7}{2}}n^{-\frac{5}{4}})\\
		&=\lambda_1({\A^*})+o_p(n^{-\frac{2}{3}}).
	\end{aligned}
\end{equation}
Recall that
\begin{equation}
	\label{theorem:4-2-equation-0}
	\begin{aligned}
		\hat{\A}={\A^*}+\check{\A}+\bDelta+\check{\bDelta}.
	\end{aligned}
\end{equation}
By \eqref{theorem:4-2-equation-7}, \eqref{theorem:4-2-equation-8}, \eqref{theorem:4-2-equation-0} and  Lemma \ref{lemma:appendix-2}, we have
\[
\lambda_1(\hat{\A})=\lambda_1({\A^*})+o_p(n^{-\frac{2}{3}}).
\]

\eop

\subsection{Proof of Lemma \ref{lemma:4-8}}

By  direct calculation, we have
\begin{equation}
\label{lemma:8-3-equation-02}
	\begin{aligned}
		\tr({\A^*}^2\bDelta)&=\sum_{i}({\A^*}^2\bDelta)_{ii}\\
		&=\sum_{i}\sum_{j}\sum_{k}{A_{ij}^*}A^*_{jk}\Delta_{ki}\\
		&=\sum_{i}\sum_{k}\Delta_{ki}\sum_{j}{A_{ij}^*}A^*_{jk}\\
		&=\sum_{1\leq u, v\leq K}\sum_{k\in\mathcal{N}_{u}, i\in\mathcal{N}_{v}}\Delta_{ki}\sum_{j}{A_{ij}^*}A^*_{jk}\\
		&=\sum_{1\leq u, v\leq K}\frac{B_{uv}-\hat{B}_{uv}}{\sqrt{nB_{uv}(1-B_{uv})}}\sum_{k\in\mathcal{N}_{u}, i\in\mathcal{N}_{v}}\sum_{j}{A_{ij}^*}A^*_{jk}\\
		&\leq \sqrt{\sum_{1\leq u,v\leq K}\frac{(B_{uv}-\hat{B}_{uv})^2}{nB_{uv}(1-B_{uv})}\sum_{1\leq u,v\leq K}(\sum_{k\in\mathcal{N}_{u}, i\in\mathcal{N}_{v}}\sum_{j}{A_{ij}^*}A^*_{jk})^2}\\
		&= \sqrt{\sum_{1\leq u,v\leq K}\frac{(B_{uv}-\hat{B}_{uv})^2}{B_{uv}(1-B_{uv})}\sum_{1\leq u,v\leq K}\frac{1}{n}(\sum_{k\in\mathcal{N}_{u}, i\in\mathcal{N}_{v}}\sum_{j}{A_{ij}^*}A^*_{jk})^2}.
	\end{aligned}
\end{equation}

Note that
\begin{equation*}
	\begin{aligned}
		(\sum_{k\in\mathcal{N}_{u},i\in\mathcal{N}_{v}}\sum_{j}{A_{ij}^*}A^*_{jk})^2&=\sum_{k\in\mathcal{N}_{u}}(\sum_{i\in\mathcal{N}_{v}}\sum_{j}{A_{ij}^*}A^*_{jk})^2\\
		&+\sum_{k\in\mathcal{N}_{u},l\in\mathcal{N}_{u},k\neq l}\sum_{i\in\mathcal{N}_{v}}(\sum_{j}{A_{ij}^*}A^*_{jk})\sum_{i'\in\mathcal{N}_{v}}(\sum_{j'}{A_{i'j'}^*}A^*_{j'l})\\
		&=\sum_{k\in\mathcal{N}_{u}}\sum_{i\in\mathcal{N}_{v}}(\sum_{j}{A_{ij}^*}A^*_{jk})^2\\
		&+\sum_{k\in\mathcal{N}_{u}}\sum_{i\in\mathcal{N}_{v},s\in\mathcal{N}_{v},i\neq s}(\sum_{j}{A_{ij}^*}A^*_{jk})(\sum_{j'}{A_{sj'}^*}A^*_{j'k})\\
		&+\sum_{k\in\mathcal{N}_{u},l\in\mathcal{N}_{u},k\neq l}\sum_{i\in\mathcal{N}_{v}}(\sum_{j}{A_{ij}^*}A^*_{jk})\sum_{i'\in\mathcal{N}_{v}}(\sum_{j'}{A_{i'j'}^*}A^*_{j'l})\\
		&=\sum_{k\in\mathcal{N}_{u}}\sum_{i\in\mathcal{N}_{v}}\sum_{j}{A_{ij}^*}^2{A^*_{jk}}^2\\
		&+\sum_{k\in\mathcal{N}_{u}}\sum_{i\in\mathcal{N}_{v}}\sum_{j\neq l}{A_{ij}^*}A^*_{jk}{A_{il}^*}A^*_{lk}\\
		&+\sum_{k\in\mathcal{N}_{u}}\sum_{i\in\mathcal{N}_{v},s\in\mathcal{N}_{v},i\neq s}{A^*_{is}}^2{A^*_{sk}}^2\\
		&+\sum_{k\in\mathcal{N}_{u}}\sum_{i\in\mathcal{N}_{v},s\in\mathcal{N}_{v},i\neq s}(\sum_{j \neq s}{A_{ij}^*}A^*_{jk})(\sum_{j' \neq i}{A_{sj'}^*}A^*_{j'k})\\
		&+\sum_{k\in\mathcal{N}_{u},l\in\mathcal{N}_{u},k\neq l}\sum_{i\in\mathcal{N}_{v}}{A_{il}^*}^2{A^*_{lk}}^2\\
		&+\sum_{k\in\mathcal{N}_{u},l\in\mathcal{N}_{u},k\neq l}\sum_{i\in\mathcal{N}_{v}}{A_{il}^*}A^*_{lk}\sum_{i'\in\mathcal{N}_{v}, i'\neq i}{A_{i'k}^*}A^*_{kl},\\
		&+\sum_{k\in\mathcal{N}_{u},l\in\mathcal{N}_{u},k\neq l}\sum_{i\in\mathcal{N}_{v}}(\sum_{j \neq l}{A_{ij}^*}A^*_{jk})\sum_{i'\in\mathcal{N}_{v}}(\sum_{j' \neq k}{A_{i'j'}^*}A^*_{j'l}).\\
	\end{aligned}
\end{equation*}
For $u\neq v$, we have
\begin{equation*}
	\begin{aligned}
		\E\big(\sum_{k\in\mathcal{N}_{u},i\in\mathcal{N}_{v}}\sum_{j}{A_{ij}^*}A^*_{jk})^2\big)
		&=\E\big(\sum_{k\in\mathcal{N}_{u}}\sum_{i\in\mathcal{N}_{v}}\sum_{j}{A_{ij}^*}^2{A^*_{jk}}^2\big)\\
		&+\E\big(\sum_{k\in\mathcal{N}_{u}}\sum_{i\in\mathcal{N}_{v}}\sum_{j\neq l}{A_{ij}^*}A^*_{jk}{A_{il}^*}A^*_{lk}\big)\\
		&+\E\big(\sum_{k\in\mathcal{N}_{u}}\sum_{i\in\mathcal{N}_{v},s\in\mathcal{N}_{v},i\neq s}{A^*_{is}}^2{A^*_{sk}}^2\big)\\
		&+\E\big(\sum_{k\in\mathcal{N}_{u}}\sum_{i\in\mathcal{N}_{v},s\in\mathcal{N}_{v},i\neq s}(\sum_{j \neq s}{A_{ij}^*}A^*_{jk})(\sum_{j' \neq i}{A_{sj'}^*}A^*_{j'k})\big)\\
		&+\E\big(\sum_{k\in\mathcal{N}_{u},l\in\mathcal{N}_{u},k\neq l}\sum_{i\in\mathcal{N}_{v}}{A_{il}^*}^2{A^*_{lk}}^2\big)\\
		&+\E\big(\sum_{k\in\mathcal{N}_{u},l\in\mathcal{N}_{u},k\neq l}\sum_{i\in\mathcal{N}_{v}}{A_{il}^*}A^*_{lk}\sum_{i'\in\mathcal{N}_{v}, i'\neq i}{A_{i'k}^*}A^*_{kl}\big)\\
		&+\E\big(\sum_{k\in\mathcal{N}_{u},l\in\mathcal{N}_{u},k\neq l}\sum_{i\in\mathcal{N}_{v}}(\sum_{j \neq l}{A_{ij}^*}A^*_{jk})\sum_{i'\in\mathcal{N}_{v}}(\sum_{j' \neq k}{A_{i'j'}^*}A^*_{j'l})\big)\\
		&=\E\big(\sum_{k\in\mathcal{N}_{u}}\sum_{i\in\mathcal{N}_{v}}\sum_{j}{A_{ij}^*}^2{A^*_{jk}}^2\big)\\
		&+\E\big(\sum_{k\in\mathcal{N}_{u}}\sum_{i\in\mathcal{N}_{v},s\in\mathcal{N}_{v},i\neq s}{A^*_{is}}^2{A^*_{sk}}^2\big)\\
		&+\E\big(\sum_{k\in\mathcal{N}_{u},l\in\mathcal{N}_{u},k\neq l}\sum_{i\in\mathcal{N}_{v}}{A_{il}^*}^2{A^*_{lk}}^2\big)\\
		&\leq\frac{3n_{uv}}{n}.
	\end{aligned}
\end{equation*}

Similarly, for $u=v$, we have
\begin{equation*}
	\begin{aligned}
		\E\big(\sum_{k\in\mathcal{N}_{u},i\in\mathcal{N}_{u}}\sum_{j}{A_{ij}^*}A^*_{jk}\big)^2&\leq\frac{6n_{uu}}{n}.
	\end{aligned}
\end{equation*}
Thus, we have
\[
\E\big(\sum_{1\leq u,v\leq K}\frac{1}{n}\big(\sum_{k\in\mathcal{N}_{u},i\in\mathcal{N}_{v}}\sum_{j}{A_{ij}^*}A^*_{jk}\big)^2\big)\leq 3.
\]
That is
\begin{equation}
	\label{lemma:8-3-equation-2}
	\begin{aligned}
		\sum_{1\leq u,v\leq K}\frac{1}{n}\big(\sum_{k\in\mathcal{N}_{u},i\in\mathcal{N}_{v}}\sum_{j}{A_{ij}^*}A^*_{jk}\big)^2&=O_p(1).
	\end{aligned}
\end{equation}

By Assumption \ref{assumption:2} and Corollary \ref{corollary:2-2}, we have
\begin{equation}
	\label{lemma:4-3-equation-4}
	\begin{aligned}
		\sum_{1\leq u,v\leq K}\frac{(B_{uv}-\hat{B}_{uv})^2}{B_{uv}(1-B_{uv})}&=2\sum_{1\leq u\leq v\leq K}\frac{1}{n_{uv}}\frac{n_{uv}(B_{uv}-\hat{B}_{uv})^2}{B_{uv}(1-B_{uv})}\\
       &\leq \max_{u,v}\frac{1}{n_{uv}}2\sum_{1\leq u\leq v\leq K}\frac{n_{uv}(B_{uv}-\hat{B}_{uv})^2}{B_{uv}(1-B_{uv})}\\
		&\leq c_1^2\frac{K^2}{n^2}O_p(K^2)=O_p(\frac{K^4}{n^2}).
	\end{aligned}
\end{equation}
By  \eqref{lemma:8-3-equation-02}, \eqref{lemma:8-3-equation-2} and \eqref{lemma:4-3-equation-4}, we have
\begin{equation*}
	\begin{aligned}
		\tr({\A^*}^2\bDelta)
		&\leq \sqrt{\sum_{1\leq u,v\leq K}\frac{(B_{uv}-\hat{B}_{uv})^2}{B_{uv}(1-B_{uv})}\sum_{1\leq u,v\leq K}\frac{1}{n}(\sum_{k\in\mathcal{N}_{u}, i\in\mathcal{N}_{v}}\sum_{j}{A_{ij}^*}A^*_{jk})^2}\\
		&=\sqrt{O_p(\frac{K^4}{n^2})O_p(1)}\\
		&= O_p(\frac{K^2}{n}).
	\end{aligned}
\end{equation*}

\eop

\subsection{Proof of Corollary \ref{corollary:4-1}}
By Corollary \ref{corollary:2-2}, we have
\begin{equation}
\label{corollary:4-1-equation-1}
\begin{aligned}
\tr({\A^*}\bDelta^2)&=\tr(\bDelta^2{\A^*})\\
&=\tr(\bGamma\bLambda^2\bGamma^\top{\A^*})\\
&=\tr(\bLambda^2\bGamma^\top{\A^*}\bGamma)=\sum_{i}\lambda_i^2\tr(\bGamma_i^\top{\A^*}\bGamma_i)\\
&\leq \sum_{i}2\lambda_i^2\tr(\bGamma_i^\top\bGamma_i)=2\sum_{i}\lambda_i^2\\
&=O_p(\frac{K^2}{n}).
\end{aligned}
\end{equation}
We also have
\begin{equation}
\label{corollary:4-1-equation-2}
\begin{aligned}
\tr(\bDelta^3)&=\tr(\bGamma\bLambda^2\bGamma^\top\bDelta)\\
&=\tr(\bLambda^2\bGamma^\top\bDelta\bGamma)=\sum_{i}\lambda_i^2\tr(\bGamma_i^\top\bDelta\bGamma_i)\\
&\leq O_p(\frac{K}{\sqrt{n}})\sum_{i}\lambda_i^2\tr(\bGamma_i^\top\bGamma_i)=O_p(\frac{K}{\sqrt{n}})\sum_{i}\lambda_i^2\\
&=O_p((\frac{K}{\sqrt{n}})^3).
\end{aligned}
\end{equation}

Combining \eqref{equation:4-02}, \eqref{corollary:4-1-equation-1}, \eqref{corollary:4-1-equation-2} and Lemmas \ref{lemma:4-8}, \ref{lemma:appendix-1}, we get the result.

\eop

\subsection{Proof of Lemma \ref{lemma:4-5}}
By  direct calculation, we have
\begin{equation}
	\label{lemma:4-3-equation-1}
	\begin{aligned}
		\tr({\A^*}^2\check{\A})&=\sum_{i}({\A^*}^2\check{\A})_{ii}\\
		&=\sum_{i}\sum_{j}\sum_{k}{A_{ij}^*}A^*_{jk}\alpha_{ki}A_{ki}\\
		&=\sum_{i}\sum_{k}\alpha_{ki}A^*_{ki}\sum_{j}{A_{ij}^*}A^*_{jk}\\
		&=\sum_{1\leq u,v\leq K}\alpha_{uv}\sum_{k\in\mathcal{N}_{u}, i\in\mathcal{N}_{v}}A^*_{ki}\sum_{j}{A_{ij}^*}A^*_{jk}\\
		&\leq \sqrt{\sum_{1\leq u,v\leq K}\alpha_{uv}^2\sum_{1\leq u,v\leq K}(\sum_{k\in\mathcal{N}_{u}, i\in\mathcal{N}_{v}}A^*_{ki}\sum_{j}{A_{ij}^*}A^*_{jk})^2}.
	\end{aligned}
\end{equation}
Note that
\begin{equation*}
	\begin{aligned}
		(\sum_{k\in\mathcal{N}_{u},i\in\mathcal{N}_{v}}A^*_{ki}\sum_{j}{A_{ij}^*}A^*_{jk})^2&=\sum_{k\in\mathcal{N}_{u}}(\sum_{i\in\mathcal{N}_{v}}{A^*_{ki}}\sum_{j}{A_{ij}^*}A^*_{jk})^2\\
		&+\sum_{k\in\mathcal{N}_{u},l\in\mathcal{N}_{u},k\neq l}\sum_{i\in\mathcal{N}_{v}}A^*_{ki}(\sum_{j}{A_{ij}^*}A^*_{jk})\sum_{i\in\mathcal{N}_{v}}A^*_{li}(\sum_{j'}{A_{ij'}^*}A^*_{j'l})\\
		&=\sum_{k\in\mathcal{N}_{u}}\sum_{i\in\mathcal{N}_{v}}{A^*_{ki}}^2(\sum_{j}{A_{ij}^*}A^*_{jk})^2\\
		&+\sum_{k\in\mathcal{N}_{u}}\sum_{i\in\mathcal{N}_{v},s\in\mathcal{N}_{v},i\neq s}{A^*_{ki}}(\sum_{j}{A_{ij}^*}A^*_{jk}){A^*_{ks}}(\sum_{j'}{A_{sj'}^*}A^*_{j'k})\\
		&+\sum_{k\in\mathcal{N}_{u},l\in\mathcal{N}_{u},k\neq l}\sum_{i\in\mathcal{N}_{v}}A^*_{ki}(\sum_{j}{A_{ij}^*}A^*_{jk})\sum_{i'\in\mathcal{N}_{v}}A^*_{li'}(\sum_{j'}{A_{i'j'}^*}A^*_{j'l})\\
		&=\sum_{k\in\mathcal{N}_{u}}\sum_{i\in\mathcal{N}_{v}}{A^*_{ki}}^2\sum_{j}{A_{ij}^*}^2{A^*_{jk}}^2\\
		&+\sum_{k\in\mathcal{N}_{u}}\sum_{i\in\mathcal{N}_{v}}{A^*_{ki}}^2\sum_{j\neq l}{A_{ij}^*}A^*_{jk}{A_{il}^*}A^*_{lk}\\
		&+\sum_{k\in\mathcal{N}_{u}}\sum_{i\in\mathcal{N}_{v},s\in\mathcal{N}_{v},i\neq s}{A^*_{ki}}^2{A^*_{is}}^2{A^*_{sk}}^2\\
		&+\sum_{k\in\mathcal{N}_{u}}\sum_{i\in\mathcal{N}_{v},s\in\mathcal{N}_{v},i\neq s}{A^*_{ki}}(\sum_{j \neq s}{A_{ij}^*}A^*_{jk}){A^*_{ks}}(\sum_{j' \neq i}{A_{sj'}^*}A^*_{j'k})\\
		&+\sum_{k\in\mathcal{N}_{u},l\in\mathcal{N}_{u},k\neq l}\sum_{i\in\mathcal{N}_{v}}{A^*_{ki}}^2{A_{il}^*}^2{A^*_{lk}}^2\\
		&+\sum_{k\in\mathcal{N}_{u},l\in\mathcal{N}_{u},k\neq l}\sum_{i\in\mathcal{N}_{v}}A^*_{ki}{A_{il}^*}A^*_{lk}\sum_{i'\in\mathcal{N}_{v}, i\neq i'}A^*_{li'}{A_{i'k}^*}A^*_{kl}\\
		&+\sum_{k\in\mathcal{N}_{u},l\in\mathcal{N}_{u},k\neq l}\sum_{i\in\mathcal{N}_{v}}A^*_{ki}(\sum_{j \neq l}{A_{ij}^*}A^*_{jk})\sum_{i'\in\mathcal{N}_{v}}A^*_{li'}(\sum_{j' \neq k}{A_{i'j'}^*}A^*_{j'l}).
	\end{aligned}
\end{equation*}

For $u\neq v$, we have
\begin{equation*}
	\begin{aligned}
		\E\big(\sum_{k\in\mathcal{N}_{u},i\in\mathcal{N}_{v}}{A^*_{ki}}\sum_{j}{A_{ij}^*}A^*_{jk}\big)^2&=\E\big(\sum_{k\in\mathcal{N}_{u}}\sum_{i\in\mathcal{N}_{v}}{A^*_{ki}}^2\sum_{j}{A_{ij}^*}^2{A^*_{jk}}^2\big)\\
		&+\E\big(\sum_{k\in\mathcal{N}_{u}}\sum_{i\in\mathcal{N}_{v}}{A^*_{ki}}^2\sum_{j\neq l}{A_{ij}^*}A^*_{jk}{A_{il}^*}A^*_{lk}\big)\\
		&+\E\big(\sum_{k\in\mathcal{N}_{u}}\sum_{i\in\mathcal{N}_{v},s\in\mathcal{N}_{v},i\neq s}{A^*_{ki}}^2{A^*_{is}}^2{A^*_{sk}}^2\big)\\
		&+\E\big(\sum_{k\in\mathcal{N}_{u}}\sum_{i\in\mathcal{N}_{v},s\in\mathcal{N}_{v},i\neq s}{A^*_{ki}}(\sum_{j \neq s}{A_{ij}^*}A^*_{jk}){A^*_{ks}}(\sum_{j' \neq i}{A_{sj'}^*}A^*_{j'k})\big)\\
		&+\E\big(\sum_{k\in\mathcal{N}_{u},l\in\mathcal{N}_{u},k\neq l}\sum_{i\in\mathcal{N}_{v}}{A^*_{ki}}^2{A_{il}^*}^2{A^*_{lk}}^2\big)\\
		&+\E\big(\sum_{k\in\mathcal{N}_{u},l\in\mathcal{N}_{u},k\neq l}\sum_{i\in\mathcal{N}_{v}}A^*_{ki}{A_{il}^*}A^*_{lk}\sum_{i'\in\mathcal{N}_{v}, i\neq i'}A^*_{li'}{A_{i'k}^*}A^*_{kl}\big)\\
		&+\E\big(\sum_{k\in\mathcal{N}_{u},l\in\mathcal{N}_{u},k\neq l}\sum_{i\in\mathcal{N}_{v}}A^*_{ki}(\sum_{j \neq l}{A_{ij}^*}A^*_{jk})\sum_{i'\in\mathcal{N}_{v}}A^*_{li'}(\sum_{j' \neq k}{A_{i'j'}^*}A^*_{j'l})\big)\\
		&=\sum_{k\in\mathcal{N}_{u}}\sum_{i\in\mathcal{N}_{v}}\E\big({A^*_{ki}}^2\big)\sum_{j}\E\big({A_{ij}^*}^2{A^*_{jk}}^2\big)\\
		&+\sum_{k\in\mathcal{N}_{u}}\sum_{i\in\mathcal{N}_{v},s\in\mathcal{N}_{v},i\neq s}\E\big({A^*_{ki}}^2{A^*_{is}}^2{A^*_{sk}}^2\big)\\
		&+\sum_{k\in\mathcal{N}_{u},l\in\mathcal{N}_{u},k\neq l}\sum_{i\in\mathcal{N}_{v}}\E\big({A^*_{ki}}^2{A_{il}^*}^2{A^*_{lk}}^2\big)\\
		&\leq\frac{3n_{uv}}{n^2}.\\
	\end{aligned}
\end{equation*}

Similarly, for $u=v$, we have
\begin{equation*}
	\begin{aligned}
		\E\big(\sum_{k\in\mathcal{N}_{u},i\in\mathcal{N}_{u}}{A^*_{ki}}\sum_{j}{A_{ij}^*}A^*_{jk}\big)^2&\leq\frac{6n_{uu}}{n^2}.
	\end{aligned}
\end{equation*}
Thus, we have
\[
\E\big(\sum_{1\leq u,v\leq K}\big(\sum_{k\in\mathcal{N}_{u},i\in\mathcal{N}_{v}}{A^*_{ki}}\sum_{j}{A_{ij}^*}A^*_{jk}\big)^2\big)\leq 3.
\]
That is
\begin{equation}
	\label{lemma:4-3-equation-2}
	\begin{aligned}
		\sum_{1\leq u,v\leq K}\big(\sum_{k\in\mathcal{N}_{u},i\in\mathcal{N}_{v}}{A^*_{ki}}\sum_{j}{A_{ij}^*}A^*_{jk}\big)^2&=O_p(1).
	\end{aligned}
\end{equation}
By Assumption \ref{assumption:2} and \eqref{lemma:4-3-equation-3}, we have
\begin{equation}
	\label{lemma:4-3-equation-5}
	\begin{aligned}
		\sum_{1\leq u,v\leq K}\alpha_{uv}^2&=2\sum_{1\leq u\leq v\leq K}\frac{1}{n_{uv}}n_{uv}\alpha_{uv}^2\\
        &\leq \max_{u,v}\frac{1}{n_{uv}}2\sum_{1\leq u\leq v\leq K}n_{uv}\alpha_{uv}^2\\
		&\leq c_1^2\frac{K^2}{n^2}O_p(K^2)=O_p(\frac{K^4}{n^2}).
	\end{aligned}
\end{equation}
By  \eqref{lemma:4-3-equation-1}, \eqref{lemma:4-3-equation-2} and \eqref{lemma:4-3-equation-5}, we have
\begin{equation*}
	\begin{aligned}
		\tr({\A^*}^2\check{\A})&\leq \sqrt{\sum_{1\leq u,v\leq K}\alpha_{uv}^2\sum_{1\leq u,v\leq K}(\sum_{k\in\mathcal{N}_{u}, i\in\mathcal{N}_{v}}A^*_{ki}\sum_{j}{A_{ij}^*}A^*_{jk})^2}\\
		&=\sqrt{O_p(\frac{K^4}{n^2})O_p(1)}\\
		&= O_p(\frac{K^2}{n}).
	\end{aligned}
\end{equation*}

\eop

\subsection{Proof of Lemma \ref{lemma:4-4}}

By spectral decomposition, we may express $\check{\bDelta}$ as
\begin{equation*}
	\begin{aligned}
		\check{\bDelta}&=\check{\bGamma}\check{\bLambda}\check{\bGamma}^\top,
	\end{aligned}
\end{equation*}
where $\check{\bLambda}=\Diag\{\check{\lambda}_1,\ldots,\check{\lambda}_n\}$, and $\check{\bGamma}=(\check{\bGamma}_1^\top,\ldots,\check{\bGamma}_n^\top)^\top$ is an $n\times n$ orthogonal matrix collecting the eigenvectors of $\check{\bDelta}$.

By \eqref{equation:4-0} and \eqref{eq:Deltacheck}, we have
\begin{equation}
	\label{lemma:4-4-equation-2}
	\begin{aligned}
		\sum_{i}\check{\lambda}_i^2&=\tr(\check{\bDelta}^2)=\sum_{i,j}\alpha_{ij}^2\Delta_{ij}^2\\
		&=2\sum_{1\leq u\leq v\leq K}n_{uv}\alpha_{uv}^2\Delta_{uv}^2\leq\frac{C_2}{n}\max_{u,v}\frac{1}{n_{uv}}2\sum_{1\leq u\leq v\leq K}\dfrac{n_{uv}^2(B_{uv}-\hat{B}_{uv})^4}{B_{uv}^2(1-B_{uv})^2},
	\end{aligned}
\end{equation}
where
\[
C_2=\max_{u,v}\frac{B_{uv}(1-B_{uv})(1-B_{uv}-\hat{B}_{uv})^2}{\hat{B}_{uv}(1-\hat{B}_{uv})\big(\sqrt{(B_{uv}(1-B_{uv})}+\sqrt{\hat{B}_{uv}(1-\hat{B}_{uv})}\big)^2}.
\]
By  \eqref{theorem:2-2-equation-1}, we have
\begin{equation}
	\label{lemma:4-4-equation-3}
	2\sum_{1\leq u\leq v\leq K}\dfrac{n_{uv}^2(B_{uv}-\hat{B}_{uv})^4}{B_{uv}^2(1-B_{uv})^2}=O_p(K^2).
\end{equation}
By Assumption \ref{assumption:2}, \eqref{lemma:4-4-equation-2} and \eqref{lemma:4-4-equation-3}, we have
\begin{equation}
	\label{lemma:4-4-equation-5}
	\begin{aligned}
		\sum_{i}\check{\lambda}_i^2&=O_p(\frac{K^4}{n^3}).
	\end{aligned}
\end{equation}
For any fixed $k$, by \eqref{lemma:4-4-equation-5},  we have
\begin{equation}
	\label{lemma:4-4-equation-4}
	\begin{aligned}
		\tr\big(({\A^*}^k\check{\bDelta})^T({\A^*}^k\check{\bDelta})\big)&=\tr\big(({\A^*}^k\check{\bDelta})^T({\A^*}^k\check{\bDelta})\big)\\
		&=\tr\big(\check{\bDelta}^T({\A^*}^k)^T{\A^*}^k\check{\bDelta}\big)\\
		&=\tr(\check{\bDelta}^2{\A^*}^{2k})=\tr(\check{\bGamma}\check{\bLambda}^2\check{\bGamma}^\top{\A^*}^{2k})\\
		&=\tr(\check{\bLambda}^2\check{\bGamma}^\top{\A^*}^{2k}\check{\bGamma})=\sum_{i}\check{\lambda}_i^2\tr(\check{\bGamma}_i^\top{\A^*}^{2k}\check{\bGamma}_i)\\
		&\leq \sum_{i}2^{2k}\check{\lambda}_i^2\tr(\check{\bGamma}_i^\top\check{\bGamma}_i)=2^{2k}\sum_{i}\check{\lambda}_i^2\\
		&=O_p(\frac{K^4}{n^3}).
	\end{aligned}
\end{equation}
By \eqref{lemma:4-4-equation-4}, we have
\begin{equation*}
	\begin{aligned}
		\tr\big({\A^*}^k{\check{\bDelta}}\big)&=\sum_{i}\lambda_i\big({\A^*}^k{\check{\bDelta}}\big)\\
		&\leq \sqrt{n\sum_{i}\lambda_i^2\big({\A^*}^k\check{\bDelta}\big)}\\
		&=\sqrt{n\sum_{i}\lambda_i\big(({\A^*}^k\check{\bDelta})^T({\A^*}^k\check{\bDelta})\big)}\\
		&=\sqrt{n\tr\big(({\A^*}^k\check{\bDelta})^T({\A^*}^k\check{\bDelta})\big)}\\
		&=O_p(\frac{K^2}{n}).
	\end{aligned}
\end{equation*}

\eop

\subsection{Proof of Theorem \ref{theorem:4-3}}

By \eqref{equation:4-5}, Lemmas \ref{lemma:4-8}, \ref{lemma:4-5} and \ref{lemma:4-4}, we have
\begin{equation}
	\label{theorem:4-3-equation-0}
	\begin{aligned}
		\tr({\A^*}^{2}\tilde{\bDelta})&=O_p(\frac{K^2}{n}).
	\end{aligned}
\end{equation}
By Lemma \ref{lemma:3-3}, we have
\begin{equation}
	\label{theorem:4-3-equation-1}
	\begin{aligned}
		\tr({\A^*}\tilde{\bDelta}^2)&=\tr(\tilde{\bDelta}^2{\A^*})=\tr(\tilde{\bGamma}\tilde{\bLambda}^2\tilde{\bGamma}^\top{\A^*})\\
		&=\tr(\tilde{\bLambda}^2\tilde{\bGamma}^\top{\A^*}\tilde{\bGamma})=\sum_{i}\tilde{\lambda}_i^2\tr(\tilde{\bGamma}_i^\top{\A^*}\tilde{\bGamma}_i)\\
		&\leq \sum_{i}2\tilde{\lambda}_i^2\tr(\tilde{\bGamma}_i^\top\tilde{\bGamma}_i)=2\sum_{i}\tilde{\lambda}_i^2\\
		&=O_p(\frac{K^2}{n}).
	\end{aligned}
\end{equation}
By Lemma \ref{lemma:3-3}, we also have
\begin{equation}
	\label{theorem:4-3-equation-2}
	\begin{aligned}
		\tr(\tilde{\bDelta}^3)&=\tr(\tilde{\bGamma}\tilde{\bLambda}^2\tilde{\bGamma}^\top\tilde{\bDelta})\\
		&=\tr(\tilde{\bLambda}^2\tilde{\bGamma}^\top\tilde{\bDelta}\tilde{\bGamma})=\sum_{i}\tilde{\lambda}_i^2\tr(\tilde{\bGamma}_i^\top\tilde{\bDelta}\tilde{\bGamma}_i)\\
		&\leq O_p(\frac{K}{\sqrt{n}})\sum_{i}\tilde{\lambda}_i^2\tr(\tilde{\bGamma}_i^\top\tilde{\bGamma}_i)=O_p(\frac{K}{\sqrt{n}})\sum_{i}\tilde{\lambda}_i^2\\
		&=O_p((\frac{K}{\sqrt{n}})^3).
	\end{aligned}
\end{equation}

Combining \eqref{equation:4-2}, \eqref{theorem:4-3-equation-0}, \eqref{theorem:4-3-equation-1}, \eqref{theorem:4-3-equation-2} and Lemma \ref{lemma:appendix-1}, we get the result.

\eop

\end{appendix}

\bibliographystyle{imsart-number}
\bibliography{bibliography}

\begin{thebibliography}{38}

\bibitem{Amini:2013}
\begin{barticle}[author]
\bauthor{\bsnm{Amini},~\bfnm{Arash~A}\binits{A.~A.}},
  \bauthor{\bsnm{Chen},~\bfnm{Aiyou}\binits{A.}},
  \bauthor{\bsnm{Bickel},~\bfnm{Peter~J}\binits{P.~J.}} \AND
  \bauthor{\bsnm{Levina},~\bfnm{Elizaveta}\binits{E.}}
(\byear{2013}).
\btitle{Pseudo-likelihood methods for community detection in large sparse
  networks}.
\bjournal{The Annals of Statistics}
\bvolume{41}
\bpages{2097-2122}.
\end{barticle}
\endbibitem

\bibitem{Bai:2010}
\begin{bbook}[author]
\bauthor{\bsnm{Bai},~\bfnm{Zhidong}\binits{Z.}} \AND
  \bauthor{\bsnm{Silverstein},~\bfnm{Jack~W.}\binits{J.~W.}}
(\byear{2010}).
\btitle{Spectral Analysis of Large Dimensional Random Matrices},
\bedition{2} ed.
\bpublisher{Springer}, \baddress{New York}.
\end{bbook}
\endbibitem

\bibitem{Bai:2005}
\begin{barticle}[author]
\bauthor{\bsnm{Bai},~\bfnm{ZhiDong}\binits{Z.}} \AND
  \bauthor{\bsnm{Yao},~\bfnm{Jianfeng}\binits{J.}}
(\byear{2005}).
\btitle{On the convergence of the spectral empirical process of Wigner
  matrices}.
\bjournal{Bernoulli}
\bvolume{11}
\bpages{1059-1092}.
\end{barticle}
\endbibitem

\bibitem{Bickel:2009}
\begin{barticle}[author]
\bauthor{\bsnm{Bickel},~\bfnm{Peter~J.}\binits{P.~J.}} \AND
  \bauthor{\bsnm{Chen},~\bfnm{Aiyou}\binits{A.}}
(\byear{2009}).
\btitle{A nonparametric view of network models and Newman-Girvan and other
  modularities}.
\bjournal{Proceedings of the National Academy of Sciences of the United States
  of America}
\bvolume{106}
\bpages{21068-21073}.
\end{barticle}
\endbibitem

\bibitem{Bickel:2016}
\begin{barticle}[author]
\bauthor{\bsnm{Bickel},~\bfnm{Peter~J.}\binits{P.~J.}} \AND
  \bauthor{\bsnm{Sarkar},~\bfnm{Purnamrita}\binits{P.}}
(\byear{2016}).
\btitle{Hypothesis testing for automated community detection in networks}.
\bjournal{Journal of the Royal Statistical Society: Series B (Statistical
  Methodology)}
\bvolume{78}
\bpages{253-273}.
\end{barticle}
\endbibitem

\bibitem{Chen:2018}
\begin{barticle}[author]
\bauthor{\bsnm{Chen},~\bfnm{Kehui}\binits{K.}} \AND
  \bauthor{\bsnm{Lei},~\bfnm{Jing}\binits{J.}}
(\byear{2018}).
\btitle{Network cross-validation for determining the number of communities in
  network data}.
\bjournal{Journal of the American Statistical Association}
\bvolume{113}
\bpages{241-251}.
\end{barticle}
\endbibitem

\bibitem{Chen:2024}
\begin{barticle}[author]
\bauthor{\bsnm{Chen},~\bfnm{Li}\binits{L.}},
  \bauthor{\bsnm{Josephs},~\bfnm{Nathaniel}\binits{N.}},
  \bauthor{\bsnm{Lin},~\bfnm{Lizhen}\binits{L.}},
  \bauthor{\bsnm{Zhou},~\bfnm{Jie}\binits{J.}} \AND
  \bauthor{\bsnm{Kolaczyk},~\bfnm{Eric~D.}\binits{E.~D.}}
(\byear{2024}).
\btitle{A spectral-based framework for hypothesis testing in populations of
  networks}.
\bjournal{Statistica Sinica}
\bvolume{34}
\bpages{87-110}.
\end{barticle}
\endbibitem

\bibitem{Chen:2023}
\begin{barticle}[author]
\bauthor{\bsnm{Chen},~\bfnm{Li}\binits{L.}},
  \bauthor{\bsnm{Zhou},~\bfnm{Jie}\binits{J.}} \AND
  \bauthor{\bsnm{Lin},~\bfnm{Lizhen}\binits{L.}}
(\byear{2023}).
\btitle{Hypothesis testing for populations of networks}.
\bjournal{Communications in Statistics - Theory and Methods}
\bvolume{52}
\bpages{3661--3684}.
\end{barticle}
\endbibitem

\bibitem{Dong:2020}
\begin{barticle}[author]
\bauthor{\bsnm{Dong},~\bfnm{Zhishan}\binits{Z.}},
  \bauthor{\bsnm{Wang},~\bfnm{Shuangshuang}\binits{S.}} \AND
  \bauthor{\bsnm{Liu},~\bfnm{Qun}\binits{Q.}}
(\byear{2020}).
\btitle{Spectral based hypothesis testing for community detection in complex
  networks}.
\bjournal{Information Sciences}
\bvolume{512}
\bpages{1360-1371}.
\end{barticle}
\endbibitem

\bibitem{Fu:2025a}
\begin{barticle}[author]
\bauthor{\bsnm{Fu},~\bfnm{Kang}\binits{K.}},
  \bauthor{\bsnm{Hu},~\bfnm{Jianwei}\binits{J.}},
  \bauthor{\bsnm{Keita},~\bfnm{Seydou}\binits{S.}} \AND
  \bauthor{\bsnm{Liu},~\bfnm{Hang}\binits{H.}}
(\byear{2025}).
\btitle{Two-sample test for stochastic block models via the largest singular
  value}.
\bjournal{Communications in Statistics - Theory and Methods}
\bvolume{54}
\bpages{1160-1179}.
\end{barticle}
\endbibitem

\bibitem{Fu:2025b}
\begin{barticle}[author]
\bauthor{\bsnm{Fu},~\bfnm{Kang}\binits{K.}},
  \bauthor{\bsnm{Hu},~\bfnm{Jianwei}\binits{J.}},
  \bauthor{\bsnm{Keita},~\bfnm{Seydou}\binits{S.}} \AND
  \bauthor{\bsnm{Liu},~\bfnm{Hao}\binits{H.}}
(\byear{2025}).
\btitle{Two-sample test for stochastic block models via maximum entry-wise
  deviation}.
\bjournal{Statistics and Its Interface}
\bvolume{18}
\bpages{299-313}.
\end{barticle}
\endbibitem

\bibitem{Gao:2017}
\begin{barticle}[author]
\bauthor{\bsnm{Gao},~\bfnm{Chao}\binits{C.}},
  \bauthor{\bsnm{Ma},~\bfnm{Zongming}\binits{Z.}},
  \bauthor{\bsnm{Zhang},~\bfnm{Anderson~Y.}\binits{A.~Y.}} \AND
  \bauthor{\bsnm{Zhou},~\bfnm{Harrison~H.}\binits{H.~H.}}
(\byear{2017}).
\btitle{Achieving optimal misclassification proportion in stochastic block
  models}.
\bjournal{Journal of Machine Learning Research}
\bvolume{18}
\bpages{1-45}.
\end{barticle}
\endbibitem

\bibitem{Gao:2018}
\begin{barticle}[author]
\bauthor{\bsnm{Gao},~\bfnm{Chao}\binits{C.}},
  \bauthor{\bsnm{Ma},~\bfnm{Zongming}\binits{Z.}},
  \bauthor{\bsnm{Zhang},~\bfnm{Anderson~Y.}\binits{A.~Y.}} \AND
  \bauthor{\bsnm{Zhou},~\bfnm{Harrison~H.}\binits{H.~H.}}
(\byear{2018}).
\btitle{Community detection in degree corrected block models}.
\bjournal{The Annals of Statistics}
\bvolume{46}
\bpages{2153-2185}.
\end{barticle}
\endbibitem

\bibitem{Ghoshdastidar:2020}
\begin{barticle}[author]
\bauthor{\bsnm{Ghoshdastidar},~\bfnm{D.}\binits{D.}},
  \bauthor{\bsnm{Gutzeit},~\bfnm{M.}\binits{M.}},
  \bauthor{\bsnm{Carpentier},~\bfnm{A.}\binits{A.}} \AND
  \bauthor{\bsnm{Luxburg},~\bfnm{U.~Von}\binits{U.~V.}}
(\byear{2020}).
\btitle{Two-sample hypothesis testing for inhomogeneous random graphs}.
\bjournal{The Annals of Statistics}
\bvolume{48}
\bpages{2208-2229}.
\end{barticle}
\endbibitem

\bibitem{Holland:1983}
\begin{barticle}[author]
\bauthor{\bsnm{Holland},~\bfnm{Paul~W.}\binits{P.~W.}},
  \bauthor{\bsnm{Laskey},~\bfnm{Kathryn~Blackmond}\binits{K.~B.}} \AND
  \bauthor{\bsnm{Leinhardt},~\bfnm{Samuel}\binits{S.}}
(\byear{1983}).
\btitle{Stochastic blockmodels: First steps}.
\bjournal{Social Networks}
\bvolume{5}
\bpages{109--137}.
\end{barticle}
\endbibitem

\bibitem{Hu:2020}
\begin{barticle}[author]
\bauthor{\bsnm{Hu},~\bfnm{Jianwei}\binits{J.}},
  \bauthor{\bsnm{Qin},~\bfnm{Hong}\binits{H.}},
  \bauthor{\bsnm{Yan},~\bfnm{Ting}\binits{T.}}, \bauthor{} \AND
  \bauthor{\bsnm{Zhao},~\bfnm{Yunpeng}\binits{Y.}}
(\byear{2020}).
\btitle{Corrected bayesian information criterion for stochastic block models}.
\bjournal{Journal of the American Statistical Association}
\bvolume{115}
\bpages{1771-1783}.
\end{barticle}
\endbibitem

\bibitem{Hu:2021}
\begin{barticle}[author]
\bauthor{\bsnm{Hu},~\bfnm{Jianwei}\binits{J.}},
  \bauthor{\bsnm{Zhang},~\bfnm{Jingfei}\binits{J.}},
  \bauthor{\bsnm{Qin},~\bfnm{Hong}\binits{H.}},
  \bauthor{\bsnm{Yan},~\bfnm{Ting}\binits{T.}} \AND
  \bauthor{\bsnm{Zhu},~\bfnm{Ji}\binits{J.}}
(\byear{2021}).
\btitle{Using maximum entry-wise deviation to test the goodness of fit for
  stochastic block models}.
\bjournal{Journal of The American Statistical Association}
\bvolume{116}
\bpages{1373-1382}.
\end{barticle}
\endbibitem

\bibitem{Hwang:2023}
\begin{barticle}[author]
\bauthor{\bsnm{Hwang},~\bfnm{Neil}\binits{N.}},
  \bauthor{\bsnm{Xu},~\bfnm{Jiarui}\binits{J.}},
  \bauthor{\bsnm{Chatterjee},~\bfnm{Shirshendu}\binits{S.}} \AND
  \bauthor{\bsnm{Bhattacharyya},~\bfnm{Sharmodeep}\binits{S.}}
(\byear{2024}).
\btitle{On the estimation of the number of communities for sparse networks}.
\bjournal{Journal of the American Statistical Association}
\bvolume{119}
\bpages{1895-1910}.
\end{barticle}
\endbibitem

\bibitem{Jin:2015}
\begin{barticle}[author]
\bauthor{\bsnm{Jin},~\bfnm{Jiashun}\binits{J.}}
(\byear{2015}).
\btitle{Fast community detection by SCORE}.
\bjournal{The Annals of Statistics}
\bvolume{43}
\bpages{57-89}.
\end{barticle}
\endbibitem

\bibitem{Jin:2025a}
\begin{barticle}[author]
\bauthor{\bsnm{Jin},~\bfnm{Jiashun}\binits{J.}},
  \bauthor{\bsnm{Ke},~\bfnm{Zheng~Tracy}\binits{Z.~T.}},
  \bauthor{\bsnm{Luo},~\bfnm{Shengming}\binits{S.}} \AND
  \bauthor{\bsnm{Ma},~\bfnm{Yucong}\binits{Y.}}
(\byear{2025}).
\btitle{Optimal network pairwise comparison}.
\bjournal{Journal of the American Statistical Association}
\bvolume{120}
\bpages{1048-1062}.
\end{barticle}
\endbibitem

\bibitem{Jin:2025b}
\begin{barticle}[author]
\bauthor{\bsnm{Jin},~\bfnm{Jiashun}\binits{J.}},
  \bauthor{\bsnm{Ke},~\bfnm{Zheng~Tracy}\binits{Z.~T.}},
  \bauthor{\bsnm{Tang},~\bfnm{Jiajun}\binits{J.}} \AND
  \bauthor{\bsnm{Wang},~\bfnm{Jingming}\binits{J.}}
(\byear{2025}).
\btitle{Network goodness-of-fit for the block-model family}.
\bjournal{Journal of the American Statistical Association (online)}.
\end{barticle}
\endbibitem

\bibitem{Le:2022}
\begin{barticle}[author]
\bauthor{\bsnm{Le},~\bfnm{Can~M.}\binits{C.~M.}} \AND
  \bauthor{\bsnm{Levina},~\bfnm{Elizaveta}\binits{E.}}
(\byear{2022}).
\btitle{Estimating the number of communities by spectral methods}.
\bjournal{Electronic Journal of Statistics}
\bvolume{16}
\bpages{3315-3342}.
\end{barticle}
\endbibitem

\bibitem{Lee:2014}
\begin{barticle}[author]
\bauthor{\bsnm{Lee},~\bfnm{Ji~Oon}\binits{J.~O.}} \AND
  \bauthor{\bsnm{Yin},~\bfnm{Jun}\binits{J.}}
(\byear{2014}).
\btitle{A necessary and sufficient condition for edge universality of Wigner
  matrices}.
\bjournal{Duke Mathematical Journal}
\bvolume{163}
\bpages{117-173}.
\end{barticle}
\endbibitem

\bibitem{Lei:2016}
\begin{barticle}[author]
\bauthor{\bsnm{Lei},~\bfnm{Jing}\binits{J.}}
(\byear{2016}).
\btitle{A goodness-of-fit test for stochastic block models}.
\bjournal{The Annals of Statistics}
\bvolume{44}
\bpages{401-424}.
\end{barticle}
\endbibitem

\bibitem{Li:2020}
\begin{barticle}[author]
\bauthor{\bsnm{Li},~\bfnm{Tianxi}\binits{T.}},
  \bauthor{\bsnm{Levina},~\bfnm{Elizaveta}\binits{E.}} \AND
  \bauthor{\bsnm{Zhu},~\bfnm{Ji}\binits{J.}}
(\byear{2020}).
\btitle{Network cross-validation by edge sampling}.
\bjournal{Biometrika}
\bvolume{107}
\bpages{257-276}.
\end{barticle}
\endbibitem

\bibitem{Rohe:2011}
\begin{barticle}[author]
\bauthor{\bsnm{Rohe},~\bfnm{Karl}\binits{K.}},
  \bauthor{\bsnm{Chatterjee},~\bfnm{Sourav}\binits{S.}} \AND
  \bauthor{\bsnm{Yu},~\bfnm{Bin}\binits{B.}}
(\byear{2011}).
\btitle{Spectral clustering and the high-dimensional stochastic blockmodel}.
\bjournal{The Annals of Statistics}
\bvolume{39}
\bpages{1878-1915}.
\end{barticle}
\endbibitem

\bibitem{Saldana:2017}
\begin{barticle}[author]
\bauthor{\bsnm{Saldana},~\bfnm{D.~Franco}\binits{D.~F.}},
  \bauthor{\bsnm{Yu},~\bfnm{Yi}\binits{Y.}} \AND
  \bauthor{\bsnm{Feng},~\bfnm{Yang}\binits{Y.}}
(\byear{2017}).
\btitle{How many communities are there?}
\bjournal{Journal of Computational and Graphical Statistics}
\bvolume{26}
\bpages{171-181}.
\end{barticle}
\endbibitem

\bibitem{Tracy:1994}
\begin{barticle}[author]
\bauthor{\bsnm{Tracy},~\bfnm{Craig~A.}\binits{C.~A.}} \AND
  \bauthor{\bsnm{Widom},~\bfnm{Harold}\binits{H.}}
(\byear{1994}).
\btitle{Level-spacing distributions and the {Airy} kernel}.
\bjournal{Communications in Mathematical Physics}
\bvolume{159}
\bpages{151-174}.
\end{barticle}
\endbibitem

\bibitem{Vershynin:2018}
\begin{bbook}[author]
\bauthor{\bsnm{Vershynin},~\bfnm{Roman}\binits{R.}}
(\byear{2018}).
\btitle{High-Dimensional Probability: An Introduction with Applications in Data
  Science}.
\bseries{Cambridge Series in Statistical and Probabilistic Mathematics}.
\bpublisher{Cambridge University Press}, \baddress{Cambridge}.
\end{bbook}
\endbibitem

\bibitem{Wang:2023}
\begin{barticle}[author]
\bauthor{\bsnm{Wang},~\bfnm{Jiangzhou}\binits{J.}},
  \bauthor{\bsnm{Zhang},~\bfnm{Jingfei}\binits{J.}},
  \bauthor{\bsnm{Liu},~\bfnm{Binghui}\binits{B.}},
  \bauthor{\bsnm{Zhu},~\bfnm{Ji}\binits{J.}} \AND
  \bauthor{\bsnm{Guo},~\bfnm{Jianhua}\binits{J.}}
(\byear{2023}).
\btitle{Fast network community detection with profile-pseudo likelihood
  methods}.
\bjournal{Journal of the American Statistical Association}
\bvolume{118}
\bpages{1359-1372}.
\end{barticle}
\endbibitem

\bibitem{Wang:2017}
\begin{barticle}[author]
\bauthor{\bsnm{Wang},~\bfnm{Y.~X.~Rachel}\binits{Y.~X.~R.}} \AND
  \bauthor{\bsnm{Bickel},~\bfnm{Peter~J.}\binits{P.~J.}}
(\byear{2017}).
\btitle{Likelihood-based model selection for stochastic block models}.
\bjournal{The Annals of Statistics}
\bvolume{45}
\bpages{500-528}.
\end{barticle}
\endbibitem

\bibitem{Wang:2021}
\begin{barticle}[author]
\bauthor{\bsnm{Wang},~\bfnm{Zhenggang}\binits{Z.}} \AND
  \bauthor{\bsnm{Yao},~\bfnm{Jianfeng}\binits{J.}}
(\byear{2021}).
\btitle{On a generalization of the CLT for linear eigenvalue statistics of
  Wigner matrices with inhomogeneous fourth moments}.
\bjournal{Random Matrices: Theory and Applications}
\bvolume{10}
\bpages{2150041}.
\end{barticle}
\endbibitem

\bibitem{Wu:2024}
\begin{barticle}[author]
\bauthor{\bsnm{Wu},~\bfnm{Qianyong}\binits{Q.}} \AND
  \bauthor{\bsnm{Hu},~\bfnm{Jiang}\binits{J.}}
(\byear{2024}).
\btitle{A spectral based goodness-of-fit test for stochastic block models}.
\bjournal{Statistics and Probability Letters}
\bvolume{209}
\bpages{110104}.
\end{barticle}
\endbibitem

\bibitem{Wu:2024a}
\begin{barticle}[author]
\bauthor{\bsnm{Wu},~\bfnm{Qianyong}\binits{Q.}} \AND
  \bauthor{\bsnm{Hu},~\bfnm{Jiang}\binits{J.}}
(\byear{2024}).
\btitle{Two-sample test of stochastic block models}.
\bjournal{Computational Statistics \& Data Analysis}
\bvolume{192}
\bpages{107903}.
\end{barticle}
\endbibitem

\bibitem{Wu:2024b}
\begin{barticle}[author]
\bauthor{\bsnm{Wu},~\bfnm{Qianyong}\binits{Q.}} \AND
  \bauthor{\bsnm{Hu},~\bfnm{Jiang}\binits{J.}}
(\byear{2024}).
\btitle{Two-sample test of stochastic block models via the maximum sampling
  entry-wise deviation}.
\bjournal{Journal of the Korean Statistical Society}
\bvolume{53}
\bpages{617-636}.
\end{barticle}
\endbibitem

\bibitem{Wu:2025}
\begin{bmisc}[author]
\bauthor{\bsnm{Wu},~\bfnm{Yujia}\binits{Y.}},
  \bauthor{\bsnm{Lan},~\bfnm{Wei}\binits{W.}},
  \bauthor{\bsnm{Feng},~\bfnm{Long}\binits{L.}} \AND
  \bauthor{\bsnm{Tsai},~\bfnm{Chih-Ling}\binits{C.-L.}}
(\byear{2025}).
\btitle{A goodness-of-fit test for sparse networks}.
\bhowpublished{arXiv: 2503.11990}.
\end{bmisc}
\endbibitem

\bibitem{Zhao:2012}
\begin{barticle}[author]
\bauthor{\bsnm{Zhao},~\bfnm{Yunpeng}\binits{Y.}},
  \bauthor{\bsnm{Levina},~\bfnm{Elizaveta}\binits{E.}} \AND
  \bauthor{\bsnm{Zhu},~\bfnm{Ji}\binits{J.}}
(\byear{2012}).
\btitle{Consistency of community detection in networks under degree-corrected
  stochastic block models}.
\bjournal{The Annals of Statistics}
\bvolume{40}
\bpages{2266-2292}.
\end{barticle}
\endbibitem

\bibitem{Zhu:2025}
\begin{barticle}[author]
\bauthor{\bsnm{Zhu},~\bfnm{Xiyue}\binits{X.}},
  \bauthor{\bsnm{Han},~\bfnm{Xiao}\binits{X.}} \AND
  \bauthor{\bsnm{Yang},~\bfnm{Qing}\binits{Q.}}
(\byear{2025}).
\btitle{Inference of community numbers in partial networks}.
\bjournal{Statistica Sinica (online)}.
\end{barticle}
\endbibitem

\end{thebibliography}

\end{document}